\documentclass[preprint,12pt]{elsarticle}
\usepackage[a4paper,left=0.70in,right=0.70in,top=0.75in,bottom=0.85in]{geometry}
\usepackage[utf8]{inputenc}
\usepackage[T1]{fontenc}
\usepackage{mathptmx}
\usepackage{setspace}
\usepackage{amsmath,mathtools}
\usepackage{bm}
\DeclareUnicodeCharacter{0394}{\ensuremath{\Delta}}
\DeclareUnicodeCharacter{03BB}{\ensuremath{\lambda}}
\DeclareUnicodeCharacter{03B1}{\ensuremath{\alpha}}
\DeclareUnicodeCharacter{03B2}{\ensuremath{\beta}}
\DeclareUnicodeCharacter{03B3}{\ensuremath{\gamma}}
\DeclareUnicodeCharacter{03B4}{\ensuremath{\delta}}
\DeclareUnicodeCharacter{03B5}{\ensuremath{\varepsilon}}
\DeclareUnicodeCharacter{03B7}{\ensuremath{\eta}}
\DeclareUnicodeCharacter{03B8}{\ensuremath{\theta}}
\DeclareUnicodeCharacter{03BA}{\ensuremath{\kappa}}
\DeclareUnicodeCharacter{03BC}{\ensuremath{\mu}}
\DeclareUnicodeCharacter{03C0}{\ensuremath{\pi}}
\DeclareUnicodeCharacter{03C1}{\ensuremath{\rho}}
\DeclareUnicodeCharacter{03C3}{\ensuremath{\sigma}}
\DeclareUnicodeCharacter{03C6}{\ensuremath{\phi}}
\DeclareUnicodeCharacter{03C7}{\ensuremath{\chi}}
\DeclareUnicodeCharacter{03C9}{\ensuremath{\omega}}
\DeclareUnicodeCharacter{2264}{\ensuremath{\leq}}
\DeclareUnicodeCharacter{2265}{\ensuremath{\geq}}
\DeclareUnicodeCharacter{00D7}{\ensuremath{\times}}
\DeclareUnicodeCharacter{2212}{\ensuremath{-}}
\DeclareUnicodeCharacter{2192}{\ensuremath{\rightarrow}}
\DeclareUnicodeCharacter{00B0}{\ensuremath{^\circ}}
\DeclareUnicodeCharacter{2013}{--}
\DeclareUnicodeCharacter{2014}{---}

\usepackage{graphicx}
\usepackage{booktabs,longtable,array,tabularx,multirow}
\usepackage{caption}
\usepackage{subcaption}
\usepackage{float}
\usepackage{placeins}
\usepackage{xcolor}
\usepackage{enumitem}
\setlist{nosep,leftmargin=*}
\usepackage{algorithm}
\usepackage{algpseudocode}
\floatname{algorithm}{Algorithm}
\usepackage[hidelinks]{hyperref}
\usepackage{xurl}
\providecommand{\tightlist}{\setlength{\itemsep}{0pt}\setlength{\parskip}{0pt}}
\providecommand{\RL}[1]{#1}

\makeatletter
\def\maxwidth{\ifdim\Gin@nat@width>\linewidth\linewidth\else\Gin@nat@width\fi}
\def\maxheight{\ifdim\Gin@nat@height>0.86\textheight 0.86\textheight\else\Gin@nat@height\fi}
\makeatother
\setkeys{Gin}{width=\maxwidth,height=0.86\textheight,keepaspectratio}
\newcommand{\centeredpaperfigure}[2][]{%
\begin{center}
\includegraphics[#1]{#2}
\end{center}%
}

\newcommand{\autopapercaption}[2]{%
\begin{center}
\parbox{0.94\linewidth}{\captionof{figure}{#2}\label{#1}}
\end{center}%
}
\journal{Journal of Computational Physics}
\biboptions{numbers,sort&compress}
\begin{document}
\begin{frontmatter}
\title{Resolving Cryogenic and Hypersonic Rarefied Flows via Deep Learning-Accelerated Lennard--Jones DSMC}
\author[fum]{Ahmad Shoja Sani}
\author[umass]{Ehsan Roohi\corref{cor1}}
\ead{roohie@umass.edu}
\author[bas,ict]{Stefan Stefanov}
\cortext[cor1]{Corresponding author.}
\affiliation[fum]{organization={High Performance Computing (HPC) Laboratory, Department of Mechanical Engineering, Ferdowsi University of Mashhad}, postcode={91775-1111}, city={Mashhad}, country={Iran}}
\affiliation[umass]{organization={Mechanical and Industrial Engineering, University of Massachusetts Amherst}, addressline={160 Governors Dr.}, city={Amherst}, postcode={MA 01003}, country={USA}}
\affiliation[bas]{organization={Institute of Mechanics, Bulgarian Academy of Sciences}, addressline={Acad. G. Bontchev Str.}, postcode={1113}, city={Sofia}, country={Bulgaria}}
\affiliation[ict]{organization={Centre of Excellence in Informatics and Information and Communication Technologies}, city={Sofia}, country={Bulgaria}}

\begin{abstract}
Integrating the physically realistic Lennard--Jones (LJ) potential into Direct Simulation Monte Carlo (DSMC) remains challenging because the long-range potential complicates collision-rate definition and makes repeated scattering-angle evaluation expensive. This study develops an LJ--DSMC framework built around two methodological advances and a transport-level validation of the resulting collision kernel. First, a generalized collision-selection treatment is formulated for Bird's DSMC algorithms (DSMC1, DSMC1S, and DS2V) through a Variable Effective Diameter (VED) model obtained from local Chapman--Enskog viscosity matching. This viscosity-consistent pair-selection model provides a finite DSMC collision-rate closure for the LJ potential and is validated in helium and argon normal shocks, cryogenic supersonic Couette flow, and hypersonic cylinder flows. The results show agreement with VHS in high-temperature repulsive regimes, but reveal clear LJ effects, including reduced shear stress and larger cryogenic wakes, when attractive forces become important. Second, the computational bottleneck of the accepted LJ binary-scattering step is removed by training a Deep Operator Network (DeepONet) to predict the LJ deflection angle from high-fidelity scattering data, replacing the numerical Matsumoto--Koura integral while preserving the standard elastic post-collision update. The surrogate gives a bulk mean wrapped-angle error of \(1.6\times10^{-3}\,\mathrm{rad}\) and a 99th-percentile error of \(9.9\times10^{-3}\,\mathrm{rad}\), accelerates the collision subroutine by 40\%, and reduces total wall time by 36\%. Finally, the same DeepONet--LJ scattering kernel is tested beyond viscosity-controlled flows through diffusion benchmarks. It recovers the Chapman--Enskog LJ diffusion coefficient within 1.10\% in an 80 K self-diffusion test and 1.03\% in an Ohr-style tracer-diffusion slab, confirming preservation of diffusion-relevant angular scattering statistics.
\end{abstract}

\end{frontmatter}

\section{Introduction}\label{sec:introduction}
Since its inception by Bird, the Direct Simulation Monte Carlo (DSMC) method has emerged as the preeminent numerical framework for simulating non-equilibrium and rarefied gas flows, where the breakdown of continuum mechanics necessitates a particle-based kinetic approach~\cite{bird1963approach}. By tracking a representative sample of particles through a stochastic treatment of movements and collisions, DSMC provides a robust pathway for modeling complex flows in high-altitude aerothermodynamics, vacuum technology, and microsystems \cite{roohi2025advances}.

The fidelity of a DSMC simulation is fundamentally determined by its collision procedure, which comprises two critical stages: selecting colliding pairs and implementing binary-collision dynamics. Accurate modeling of these stages requires a precise definition of the intermolecular interaction law, as the specified potential directly governs transport properties such as viscosity, thermal conductivity, and diffusion in the gas. Consequently, the choice of molecular model is not merely a technical detail but a primary factor influencing the simulation's overall predictive accuracy in capturing non-equilibrium phenomena. However, balancing the physical realism of these potentials with the associated computational overhead remains a challenge in high-fidelity kinetic modeling.

Historically, DSMC implementations relied on the simplistic Hard-Sphere (HS) model \cite{bird1963approach}, which fails to capture the realistic temperature dependence of viscosity. To improve physical fidelity, Bird \cite{bird1981montecarlo,bird1983meanfreepath} introduced the Variable Hard Sphere (VHS) model, establishing a generalized power-law relationship between viscosity and temperature based on the Inverse Power Law (IPL) potential \cite{chapman1970nonuniform}. While the computational efficiency of VHS makes it an attractive choice for engineering applications, it suffers from fundamental limitations: it neglects long-range attractive forces and assumes infinite repulsion at close molecular approaches. Furthermore, the VHS model is inherently incompatible with the IPL model's diffusion coefficients, leading to significant discrepancies in gas-mixture simulations where molecular diffusion is dominant \cite{koura1991equivalence}.

To bridge these gaps, various refinements, such as the Variable Soft Sphere (VSS) \cite{koura1991vss}, Generalized Hard Sphere (GHS) \cite{hassan1993ghs,hash1994direct}, and Generalized Soft Sphere (GSS) \cite{fan2002gss}, were developed. However, these models remain purely repulsive and rely on empirical parameters calibrated to match macroscopic transport properties rather than being derived from fundamental physical principles. In reality, intermolecular interactions are governed by a complex balance of repulsive and long-range attractive forces, such as dipole-dipole interactions, which become crucial in low-temperature or high-pressure regimes. The lack of a computationally efficient framework that incorporates such detailed collision dynamics is a critical bottleneck in high-fidelity rarefied-gas simulations.

Both repulsive and attractive components of the intermolecular forces are described comprehensively by the Lennard-Jones (LJ) potential (\(\phi(r)\)) \cite{hirschfelder1948transport}:

\begin{equation}
\phi(r) = 4\varepsilon_{LJ}\left\lbrack (d_{LJ}/r)^{12} - (d_{LJ}/r)^{6} \right\rbrack
\tag{1}
\end{equation}

, where \emph{ε\textsubscript{LJ}} denotes the well depth of the potential, \emph{d\textsubscript{LJ}} indicates the distance at which the potential energy is zero, and \emph{r} is the distance between the molecules. The potential well depth for a given gas species defines the relative importance of the attractive intermolecular forces. The LJ potential is particularly effective in representing the transport properties of gases across a wide range of temperatures. Its relevance to rarefied micro-device flows has also been demonstrated in Knudsen-pump simulations, where Lennard--Jones and \emph{ab initio} intermolecular potentials were compared in Boltzmann-equation calculations and shown to influence pumping and gas-separation characteristics~\cite{dodulad2014knudsen}. For common gases like nitrogen and argon, which possess relatively shallow potential wells, the attractive component of the intermolecular force becomes significant at low temperatures, such as those typically encountered in supersonic flow experiments \cite{pham1989nonequilibrium,lengrand1994flatplate,holden1995lens}. A similar situation is observed for metal vapors, which exhibit deeper potential wells at moderate temperatures. Additionally, many traditional molecular models derive their parameters by fitting to transport coefficients obtained using the LJ potential. However, direct implementation of this realistic potential in DSMC simulations remains a significant challenge due to the intensive computational cost of evaluating the exact scattering dynamics.

The LJ potential's capability to balance attractive and repulsive forces makes it an indispensable tool for simulating gas flows across diverse scientific and engineering disciplines. While it is widely utilized in molecular dynamics simulations to accurately describe intermolecular interactions, its integration into the DSMC framework has been the subject of several specialized studies~\cite{matsumoto1991argonshock,ayyaswamy2011ljdsmc,venkattraman2012binary,sharipov2012arbitrary}. For instance, Matsumoto and Koura \cite{matsumoto1991argonshock} employed the LJ potential in the null-collision DSMC method to calculate velocity distribution functions in argon normal shock waves. Subsequently, Ayyaswamy and Alexeenko \cite{ayyaswamy2011ljdsmc} and Venkattraman and Alexeenko \cite{venkattraman2012binary} proposed a more efficient implementation by representing the scattering angle as a polynomial approximation of the collision parameters. Furthermore, Sharipov et al. \cite{sharipov2012arbitrary} developed an \emph{ab initio} collision model capable of handling arbitrary potentials, including LJ and Stockmayer potentials, and applied it to study planar shock wave structures in helium, neon, and argon across various Mach numbers \cite{sharipov2017shock}. Despite these advancements, a common limitation of these implementations is their high degree of specificity to particular problems, which complicates their broader application to general and complex flow regimes.

This work presents a comprehensive strategy for incorporating LJ-based collision physics into the DSMC framework through two methodological advances, followed by a transport-level validation of the resulting collision kernel. While the implementation of the LJ potential was pioneered by Matsumoto and Koura~\cite{matsumoto1991argonshock}, their approach was specifically tailored to the null-collision DSMC scheme. In contrast, the first contribution of the present study is the development of a generalized LJ--DSMC collision-selection framework that is independently applicable across several collision algorithms, including No Time Counter (NTC)~\cite{bird1963approach}, Simplified Bernoulli Trials (SBT)~\cite{stefanov2011smallppc}, and Generalized Bernoulli Trials (GBT)~\cite{roohi2018gbt}. This first level of the framework addresses the collision-rate closure required to represent a long-range LJ potential within the finite-cross-section structure of DSMC. It includes both a baseline constant-diameter model and a variable-effective-diameter (VED) model, in which the effective diameter is determined by matching the local Chapman--Enskog LJ viscosity. Thus, this part of the method should be interpreted as a viscosity-consistent pair-selection model, primarily designed to recover LJ-consistent collision rates in regimes where shear viscosity is a leading transport mechanism.

The central challenge in any LJ--DSMC implementation arises from the long-range nature of the potential, which precludes a finite total collision cross-section in the classical hard-sphere sense. To resolve this difficulty at the pair-selection level, the proposed VED formulation assigns a temperature-dependent effective molecular diameter by equating the local LJ viscosity with the corresponding DSMC collision-rate closure. This allows the collision sampling process to adapt to the local gas temperature while remaining compatible with standard DSMC algorithms. Unlike conventional fixed-diameter approaches, which are accurate only within a limited thermal neighborhood of the reference state, the VED model incorporates both the well depth \(\varepsilon_{\mathrm{LJ}}\) and the characteristic distance \(d_{\mathrm{LJ}}\) through the LJ viscosity collision integral, thereby extending viscosity matching over a broad temperature range. This advancement addresses the limitations observed in earlier fixed-parameter models~\cite{bird1994molecular,weaver2015revised} and provides a rigorous foundation for viscosity-controlled LJ simulations in multiple collision-selection schemes.

The second methodological contribution is the introduction of a machine-learning strategy for accelerating the LJ binary-scattering step itself. Once a particle pair has been selected by the collision-rate model, the post-collision dynamics must still be determined by the LJ deflection angle. Direct numerical evaluation of the Matsumoto--Koura scattering integral~\cite{matsumoto1991argonshock} is computationally expensive and becomes a dominant bottleneck in large-scale DSMC calculations. To overcome this limitation, we train a Deep Operator Network (DeepONet) on high-fidelity LJ scattering data and use it as a surrogate for the deflection-angle evaluation. Importantly, the DeepONet does not learn the VED model and does not replace the DSMC collision structure; it learns the LJ scattering map defined by the reduced collision energy and impact parameter. The standard elastic velocity transformation is then retained, so momentum and kinetic energy conservation are preserved, while the expensive integral evaluation is avoided during runtime.

The landscape of rarefied gas dynamics (RGD) has been significantly reshaped by the emergence of scientific machine learning, primarily through the development of Physics-Informed Neural Networks (PINNs)~\cite{raissi2019pinn,cuomo2022pinnreview,farea2024pinnreview,zhao2024pinnfluids} and Deep Operator Networks (DeepONet)~\cite{lu2021deeponet,wang2021pideeponet}. Early contributions demonstrated the potential of neural networks to accelerate Boltzmann equation solvers by approximating collision operators~\cite{xiao2021boltzmannnn,deflorio2022rarefiedpinn,miller2022collisionoperators}. More recent studies have extended these ideas to non-equilibrium microflows~\cite{roohi2025microNozzleArxiv,zheng2025cfdeeponet,peyvan2026fusiondeeponet} and online optimization of learned collision models~\cite{ball2026onlinecollision}. Notably, recent works~\cite{roohi2026astdsmcsurrogate,roohi2026microstep,roohi2026neuralNetworksRGD} established neural-network surrogates for DSMC solutions at the macroscopic level. However, directly replacing the microscopic scattering evaluation associated with a realistic intermolecular potential remains much less explored. The present work addresses this gap by using DeepONet to emulate the LJ scattering dynamics at the collision level, rather than merely approximating macroscopic DSMC fields.

To validate these two methodological components, we perform DSMC simulations of several canonical and complex rarefied gas flows. We first investigate normal shock waves in helium and argon, where the contrast between a shallow and a deeper LJ potential well provides a sensitive test of the model's ability to capture species-dependent intermolecular effects. We then assess supersonic Couette flow in argon, emphasizing the influence of attractive--repulsive interactions on shear stress at cryogenic wall temperatures. Finally, to demonstrate robustness, stability, and scalability in multidimensional non-homogeneous flows, we simulate hypersonic flow over a two-dimensional cylinder. These test cases verify that the viscosity-consistent LJ collision-selection model and the DeepONet-accelerated scattering routine can be embedded in customized versions of Bird's DSMC codes (DSMC1, DSMC1S, and DS2V), providing a high-fidelity platform for rarefied-gas simulations across one- and two-dimensional configurations.

A separate transport-level validation is then carried out to address a question that is not resolved by viscosity matching alone. Because the VED model is constructed from the Chapman--Enskog viscosity, accurate diffusion is not automatically guaranteed by the pair-selection closure. Therefore, we first validate the DeepONet scattering-angle map directly against exact LJ scattering integrations on held-out collision samples, confirming that the learned angular kernel reproduces the LJ deflection angle with small wrapped-angle error. We then use the same DeepONet--LJ scattering kernel, without introducing a separate diffusion-tuned model, in two diffusion benchmarks: a homogeneous self-diffusion test at cryogenic temperature and a two-label tracer-diffusion slab inspired by the Ohr-style diffusion setup~\cite{ohr2023representative}. In these tests, the measured Fickian flux and local diffusion coefficient are compared directly with the Chapman--Enskog LJ diffusion target. These validations demonstrate that the DeepONet-accelerated LJ scattering map preserves diffusion-relevant angular statistics in addition to enabling viscosity-consistent DSMC simulations. We also compare DeepONet against a standard multilayer perceptron surrogate and a bilinear lookup table on the same held-out LJ scattering samples to isolate the benefit of the operator-based representation.

Together, the viscosity-matched VED model and the DeepONet scattering surrogate establish a coherent and computationally efficient pathway for using realistic molecular potentials within the DSMC framework. The VED model supplies a local Chapman--Enskog-consistent collision-rate closure for pair selection, while the DeepONet surrogate removes the dominant cost of LJ scattering-angle evaluation. The direct angular and diffusion benchmarks then verify that the learned scattering kernel preserves transport statistics beyond viscosity-controlled flows. This generalized approach overcomes the historical trade-off between physical accuracy and computational cost, extends LJ-based DSMC beyond specialized null-collision implementations, and provides a scalable framework for simulating complex rarefied flows in regimes where long-range attractive forces become macroscopically important.

The remainder of this paper is organized as follows. Section~\ref{sec:lj-reference-cross-section} presents the LJ--DSMC collision-rate formulation, including the viscosity-matched reference cross-section, the variable effective diameter (VED) model, and the dynamic bounding of the impact parameter used to remove the cut-off singularity while retaining LJ binary-scattering physics. Section~\ref{sec:ml-surrogate-scattering} describes the machine-learning acceleration strategy, including the DeepONet representation of the LJ scattering operator, the training dataset and sampling procedure, and the integration of the trained surrogate into the DSMC collision step. Section~\ref{sec:validation-results} reports the validation and performance assessment of the proposed framework, covering helium and argon normal shocks, cryogenic supersonic Couette flow, hypersonic flow over a circular cylinder, and diffusion benchmarks used to test transport statistics beyond viscosity matching. Finally, Section~\ref{sec:concluding-remarks} summarizes the main findings, discusses the implications of the DeepONet-accelerated LJ collision kernel, and outlines the resulting pathway for efficient high-fidelity DSMC simulations with realistic intermolecular potentials.

\section{Determination of a reference collision cross-section for the Lennard--Jones potential}\label{sec:lj-reference-cross-section}
\subsection{Detailed algorithm}\label{subsec:detailed-algorithm}
Consistent with the standard collision procedure of the selection-pair step, we developed an approach to implement the LJ potential in DSMC that is valid for all collision schemes. Our goal is to incorporate the new approach into our developed universal multilevel collision hybrid TAS method, which simultaneously uses transient adaptive subcell TAS and three collision schemes (NTC, SBT, and GBT) depending on the number of particles in the respective cells or subcells. To this aim, in the present paper, we utilized and tested the direct use of the no time counter (NTC), simplified Bernoulli trials (SBT) \cite{stefanov2011particle}, and generalized Bernoulli trials (GBT) \cite{roohi2018gbt}. In the NTC collision scheme, the total number of candidate pairs to be selected for collision checks within a single cell over the time step \(\Delta t\) is calculated as \cite{bird1994molecular}:

\begin{equation}
A_{\text{sel}} = \frac{N(N - 1)}{2}\frac{F_{\text{num}}(\sigma_{T}g_{r})_{\text{max}}\Delta t}{V_{c}}
\tag{2}
\end{equation}

, where \(N\) is the number of simulator particles currently in the cell, \(F_{\text{num}}\) is the number of real gas particles represented by each simulator particle, \(V_{c}\) is the volume of the cell, \(\Delta t\) is the simulation time step, \((\sigma_{T}g_{r})_{\text{max}}\) is the majorant product, representing the maximum value of the product of the total collision cross-section (\(\sigma_{T}\)) and the relative speed (\(g_{r}\)) between any two particles in the cell. In the NTC scheme, after randomly selecting a candidate pair of particles, the probability (\emph{p}) that this pair collides is determined by dividing the actual collision product by the majorant product.

\begin{equation}
p = \frac{(\sigma_{T}g_{r})_{\text{actual}}}{(\sigma_{T}g_{r})_{\text{max}}}
\tag{3}
\end{equation}

A critical challenge in implementing realistic intermolecular potentials, such as the LJ model, within the DSMC framework is efficiently selecting collision pairs. The standard NTC scheme calculates the number of candidate pairs based on the global maximum of the collision probability, \((\sigma_{T}g_{r})_{\text{max}}\), within a cell. However, the LJ potential poses a severe numerical vulnerability for the NTC scheme: at very low relative velocities, the attractive intermolecular forces cause the effective cross-section \((\sigma_{T}g_{r})_{\text{actual}}\) to increase dramatically. Consequently, minor macroscopic velocity fluctuations can induce anomalous, unbounded spikes in \((\sigma_{T}g_{r})_{\text{max}}\). This forces the NTC algorithm to select an astronomically large number of candidate pairs, most of which are subsequently rejected as dummy (null) collisions, thereby crippling computational efficiency. Furthermore, the NTC scheme is prone to selecting the same particle pair multiple times per time step when the number of simulated particles per cell (PPC) is strictly low.

To circumvent these physical and statistical vulnerabilities, the present solver also employs the SBT and GBT collision schemes \cite{stefanov2011particle,roohi2018gbt}. The SBT and GBT algorithms fundamentally decouple the collision-selection frequency from the global maximum \((\sigma_{T}g_{r})_{\text{max}}\). Instead, they formulate the selection process as a sequence of discrete combinatorial trials. This renders the SBT/GBT schemes inherently robust against the low-velocity singularities of attractive potentials and highly accurate even in extremely rarefied regions with low PPC.

In the SBT collision scheme, the particles are indexed in the cell (index \emph{i}), and each particle in the list is considered just with one of those located in the list after it, with the probability of:

\begin{equation}
p = \frac{(N - i)F_{num}g_{i,j}\sigma_{i,j}dt}{V}
\tag{4}
\end{equation}

In the GBT collision scheme, each particle in the list is considered just with one of those located in the list after it until \(N_{sel} - th\ \)particles (not until the last particle in the cell that was done in SBT), with the probability of:

\begin{equation}
p = \frac{N(N - 1)}{N_{sel}\left( 2N - N_{sel} - 1 \right)}(N - i)\frac{F_{num}g_{i,j}\sigma_{i,j}dt}{V}
\tag{5}
\end{equation}

By integrating the dynamic macroscopic cross-section (\(\sigma_{i,j}\)) directly into the exact combinatorial probability space of the SBT/GBT schemes, the present framework systematically eliminates artificial grazing collisions, bypasses the \((\sigma_{T}g_{r})_{\text{max}}\) singularity, and statistically prevents pair repetition. This ensures strict O(\emph{N}) computational complexity and provides an optimally clean numerical environment for high-fidelity simulations of strong shock waves.

The three partner selection schemes mentioned above are considered alongside three standard elastic collision models: HS, VHS, and VSS. These models were employed with a reference diameter recommended in the literature. The post-collision velocities are calculated using the deflection angle from the LJ procedure or VHS and VSS algorithms for elastic collisions. This approach does not yield perfect agreement with the experiment without a correct effective particle diameter. Bird suggests the best option, VHS with \(d=4.17\times10^{-10}\,\mathrm{m}\) and ω=0.816, α=1.0 for argon.

The first step in this approach is to estimate the effective molecular diameter based on the equalizing viscosity (\emph{η}) of VHS (called reference here) and LJ as follows:

\begin{equation}
\begin{aligned}
\eta &= \eta_{ref}\left(\frac{T}{T_{ref}}\right)^{\omega},\\
\eta_{ref} &= \frac{5(\alpha+1)(\alpha+2)\sqrt{\pi m k_B T_{ref}}}
{4\alpha(5-2\omega)(7-2\omega)\pi d_{ref}^{2}}.
\end{aligned}
\tag{6}
\end{equation}
\begin{equation}
\eta^{LJ}=\frac{5}{16d_{LJ}^{2}}\sqrt{\frac{m k_B T_1}{\pi}}\frac{V}{W^{(2)}(2)}.
\tag{7}
\end{equation}

Equations (6) and (7) present the viscosity of VHS and LJ, respectively. Here, \emph{T}, and α demonstrate the temperature and VSS scattering parameter. \emph{ω} is the viscosity-temperature index, and \emph{k\textsubscript{B}} denotes the Boltzmann constant. The molecular mass and diameter are presented by \emph{m} and \emph{d}, respectively.

Typically, for the VHS model, with \(\alpha = 1\) (for argon), we equalize the viscosities at the reference temperature:

\begin{equation}
\eta_{ref}=\frac{15\sqrt{\pi m k_B T_{ref}}}{2(5-2\omega)(7-2\omega)\pi d_{ref}^{2}}.
\tag{8}
\end{equation}
\begin{equation}
\begin{aligned}
\eta_{1}^{VHS} &= \eta_{1}^{LJ},\qquad T_1=T_{ref},\\
\frac{15\sqrt{\pi m k_B T_{ref}}}{2(5-2\omega)(7-2\omega)\pi d_{ref}^{2}}
&=\frac{5}{16d_{LJ}^{2}}\sqrt{\frac{m k_B T_1}{\pi}}\frac{V}{W^{(2)}(2)},\\
\frac{3}{(5-2\omega)(7-2\omega)d_{ref}^{2}}
&=\frac{1}{4d_{LJ}^{2}}\frac{1}{2}\frac{V}{W^{(2)}(2)}.
\end{aligned}
\tag{9}
\end{equation}

As a result, we obtain the basic formula for the reference diameter depending on the viscosity parameter ω and the ratio of functions \(V\), \(W^{(2)}(2)\), that is related to the Chapman-Enskog \RL{}expansion.

\begin{equation}
d_{ref}=2d_{LJ}\sqrt{\frac{1}{V/W^{(2)}(2)}\frac{6}{(5-2\omega)(7-2\omega)}}.
\tag{10}
\end{equation}

, where \emph{d\textsubscript{LJ}} is the Lennard-Jones diameter for which the LJ potential equals zero.

It is worth noting that we are free to choose reference collision parameters that differ from the standard recommended for VHS. It turned out that the best results of the shock wave simulations were obtained when the reference collision diameter was selected by using the upstream flow temperature as a reference temperature and computing the reference diameter by using Eq. (10). In Eq. (6), \emph{V} and \emph{W\textsuperscript{(2)}}(2) are functions of \(\varepsilon_{LJ}/k_{B}T\) tabulated in Ref \cite{hirschfelder1948transport} or approximated by a series of \(\varepsilon^{*} = 1/T^{*}{= \varepsilon}_{LJ}/k_{B}T\) with high accuracy. Within a wide range of \emph{T}, the function \emph{V}=1 and the expression \emph{W \textsuperscript{(2)}}(2) for \(\varepsilon^{*} = x = \varepsilon_{LJ}/k_{B}T\) is computed as follows:

\begin{equation}
\begin{aligned}
W^{(2)}(2;x) &= \frac{1}{6}x^{4}\Big[6.88155x^{-3.855}
+2.89488(x+0.190)^{-3.211}\\
&\quad +7.33242(x+0.67742)^{-3.303}
-660079(x+6.7461)^{-7.01}\\
&\quad +0.03950e^{-0.9274x}+0.03732e^{-0.8291x}
+0.03888e^{0.6161x}-0.009625e^{-0.2211x}\Big].
\end{aligned}
\tag{11}
\end{equation}

For argon, the source mentioned above gives \(x = \varepsilon_{LJ}/k_{B}T = \frac{124}{T}\). Other sources give different values. Some of them (Ref. \cite{weaver2015revised}, Table 4) were also used in the present investigation. In the implementation, Eq. (11) is evaluated for the local reduced temperature and stored in a one-dimensional table; arbitrary intermediate temperatures are obtained by interpolation in the reduced variable \(x=\varepsilon_{LJ}/k_B T\). This avoids repeated evaluation of the fitting expression while preserving the same Chapman--Enskog collision-integral relation used in the analytical derivation.

To quantify the temperature-dependent validity of the fixed-diameter approximation, we compared the viscosity implied by a fixed reference diameter with the Chapman--Enskog LJ viscosity over the temperature range relevant to the present simulations. For a fixed reference temperature \(T_{\mathrm{ref}}\), the reference diameter is chosen so that
\begin{equation}
\eta_{\mathrm{fixed}}(T_{\mathrm{ref}})=\eta_{\mathrm{LJ}}(T_{\mathrm{ref}})
\tag{11a}
\end{equation}
Away from this reference state, the fixed-diameter model follows the VHS-type power-law viscosity,
\begin{equation}
\eta_{\mathrm{fixed}}(T;T_{\mathrm{ref}})
=
\eta_{\mathrm{LJ}}(T_{\mathrm{ref}})
\left(\frac{T}{T_{\mathrm{ref}}}\right)^{\omega}
\tag{11b}
\end{equation}
whereas the LJ reference viscosity is evaluated from the same Chapman--Enskog collision-integral relation used in Eq.~(11). The relative viscosity-closure error is then defined as
\begin{equation}
E_{\eta}(T;T_{\mathrm{ref}})
=
100\left[
\frac{\eta_{\mathrm{fixed}}(T;T_{\mathrm{ref}})}
{\eta_{\mathrm{LJ}}(T)}
-1
\right]
\tag{11c}
\end{equation}
This diagnostic provides a direct measure of the temperature interval over which a single fixed effective diameter can be used without significant transport inconsistency.

\begin{figure}[!htbp]
\centering
\includegraphics[width=0.78\linewidth]{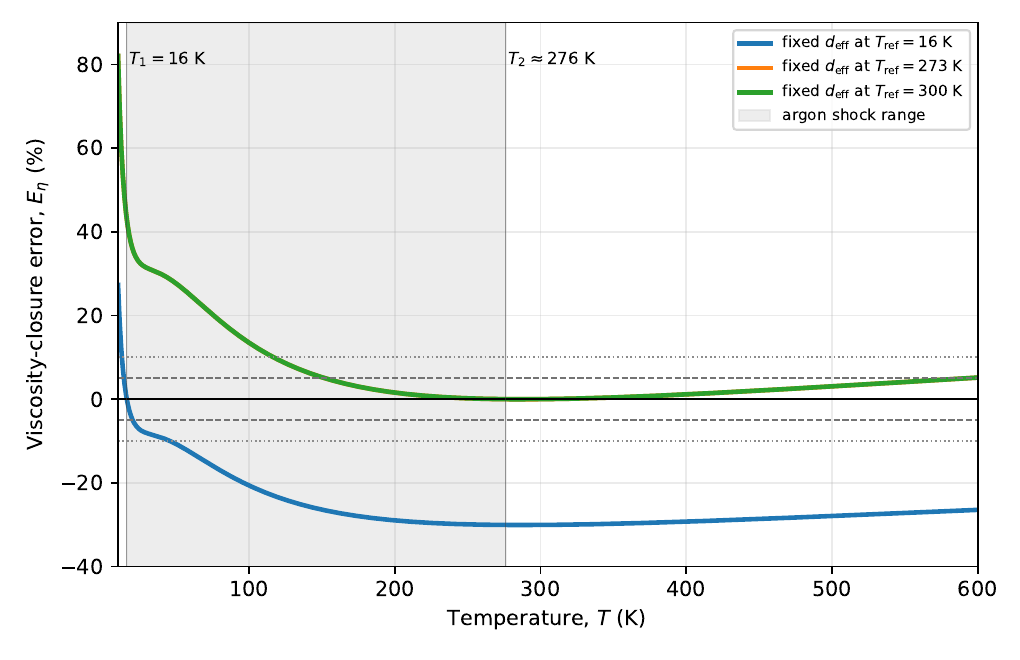}
\caption{Temperature-dependent viscosity-closure error of the fixed-diameter LJ--DSMC approximation for argon. The fixed effective diameter is calibrated at \(T_{\mathrm{ref}}\), and the error is measured relative to the Chapman--Enskog LJ viscosity evaluated with the same collision-integral relation used in the VED model. The shaded region denotes the temperature range between the upstream and downstream states of the argon shock-wave case. The fixed-diameter closure is accurate only near its calibration temperature, whereas the VED formulation removes this viscosity-closure error by recomputing the effective diameter from the local temperature.}
\label{fig:fixed_diameter_validity}
\end{figure}

Figure~\ref{fig:fixed_diameter_validity} shows that the fixed-diameter approximation is accurate only in a neighborhood of the calibration temperature. For the argon shock-wave case, calibrating the diameter at the cryogenic upstream temperature gives the correct viscosity at \(T_{\mathrm{ref}}=16\,\mathrm{K}\), but produces a large error as the gas heats through the shock. Conversely, calibrating near room temperature improves the agreement around \(273\)--\(300\,\mathrm{K}\), but leads to a large error at the cryogenic upstream state. Thus, a single fixed effective diameter cannot simultaneously represent the upstream and downstream regions of the cryogenic shock. This temperature-dependent loss of validity motivates the VED formulation, in which the effective diameter is recomputed locally from \(T(x)\) using the same LJ collision-integral relation. By construction, the VED model removes the viscosity-closure error associated with the fixed-diameter approximation while retaining the LJ scattering dynamics for accepted collisions.

\FloatBarrier

Another subtle point in the algorithm is the definition of the number of selected pairs for collisions. The NTC selection algorithm remains unchanged with the VHS acceptance-rejection procedure, where each selected pair is accepted for collision with probability (\emph{p}):

\begin{equation}
p = \frac{c_{r}\sigma(T_{ref})F_{num}(2kT_{ref}/mc_{r}^{2})^{(\omega - 0.5)}/\Gamma(5/2 - \omega)}{(c_{r}\sigma)_{\max}}
\tag{12}
\end{equation}

, where \emph{T(x)} is the temperature in a cell with coordinate \emph{x}. The temperature \emph{T(x)} is updated alongside the macroscopic properties. \emph{c\textsubscript{r}} and \emph{σ} are the relative velocity and collision cross-section of the colliding particles, respectively.

In the SBT collision model, the number of selected pairs equals \(N_{sel} = N^{(l)} - 1\), where \emph{N\textsuperscript{(l)}} is the instantaneous number of particles per cell. To obtain the correct number of collisions, the selected pairs were checked for collision with a probability, i.e.,

\begin{equation}
p = (N - i).c_{r}\sigma\left( T_{ref} \right)F_{num}\frac{1}{\Gamma(5/2 - \omega)}\left( \frac{2kT_{ref}}{mc_{r}^{2}} \right)^{(\omega - 0.5)}\frac{\Delta t}{V}
\tag{13}
\end{equation}

A similar approach is followed in the GBT collision model. It is worth noting that, in the procedure described above, the reference diameter (Eq. 10) is kept constant throughout the simulations.

It is important to clarify the microscopic interpretation of the temperature-dependent effective diameter in the VED extension. The VED diameter is used only in the stochastic pair-selection probability, i.e., in the macroscopic collision-rate closure that enforces the local Chapman--Enskog LJ viscosity. It does not alter the accepted binary-collision dynamics. Once a pair is selected, the scattering angle is still evaluated from the LJ collision parameters, namely the relative collision energy and the impact parameter, either by direct numerical integration or by the trained DeepONet surrogate. The post-collision velocity transformation remains the standard elastic binary-collision update and therefore conserves momentum and kinetic energy for each accepted pair.

This separation between collision-rate sampling and accepted scattering dynamics is consistent with Bird's reference-diameter formulation for VHS/VSS gases~\cite{bird1994molecular}. In Bird's treatment, the reference diameter can be written in terms of the reference viscosity as
\begin{equation}
d_{\mathrm{ref}}
=
\left[
\frac{
5(\alpha+1)(\alpha+2)(m k_B T_{\mathrm{ref}}/\pi)^{1/2}
}{
4\alpha(5-2\omega)(7-2\omega)\mu_{\mathrm{ref}}
}
\right]^{1/2}
\tag{13a}
\end{equation}
while the corresponding velocity-dependent effective diameter can be expressed as
\begin{equation}
d
=
d_{\mathrm{ref}}
\left[
\frac{
(2k_B T_{\mathrm{ref}}/m_r c_r^2)^{\omega-1/2}
}{
\Gamma(5/2-\omega)
}
\right]^{1/2}
\tag{13b}
\end{equation}
The reference diameter is therefore a transport closure used to normalize the collision frequency rather than a new microscopic force law. Equivalently, the equilibrium collision frequency and mean free path can be written as
\begin{equation}
\nu_0
=
4d_{\mathrm{ref}}^2 n
\left(\frac{\pi k_B T_{\mathrm{ref}}}{m}\right)^{1/2}
\left(\frac{T}{T_{\mathrm{ref}}}\right)^{1-\omega}
\tag{13c}
\end{equation}
and
\begin{equation}
\lambda_0
=
\left[
2^{1/2}\pi d_{\mathrm{ref}}^2 n
\left(\frac{T_{\mathrm{ref}}}{T}\right)^{\omega-1/2}
\right]^{-1}
\tag{13d}
\end{equation}
These relations show that the local temperature enters the collision-rate normalization through the local ensemble-averaged relative speed and transport coefficient, not through a modification of the microscopic scattering law.

In DSMC, each collision cell represents a local point in phase space in the small-cell limit. Therefore, two pairs with the same instantaneous relative speed but embedded in cells with different local temperatures may sample different local collision-rate ensembles, because the surrounding local distributions are different. However, once a collision is accepted, both pairs still undergo the same LJ angular scattering dynamics for the same collision energy and impact parameter. The locality of the VED construction therefore does not introduce a new temperature-dependent force law; it provides a local Chapman--Enskog-consistent sampling frequency for viscosity-controlled flows while preserving the LJ microscopic scattering statistics of accepted collisions.

As previously stated, the VHS and VSS collision models cannot equalize the viscosity of the LJ potential across a wide range of temperatures. Therefore, the fixed reference diameters used in the VHS or VSS models match the viscosity distribution of the LJ potential only within a narrow range around the standard temperature, typically between 150 and 500 K. This discrepancy is especially pronounced at lower temperatures. To address this issue, we propose a novel universal algorithm that calculates a variable effective diameter based on the local temperature. This method ensures that the viscosity of the VHS or VSS models matches that of the LJ potential at the specified local temperature, thereby improving simulation accuracy across a broader temperature range.

The following parts of the novel, combined with the VHS/VSS-LJ procedure, are implemented into the standard DSMC algorithm.

\begin{enumerate}
\def\labelenumi{\arabic{enumi}.}
\item
  Initialization of the reference collision diameter \(d_{ref}(x)\), following the initial reference temperature profile \(T_{ref}(x)\) by using Eqs. (6) and (7). Here, \emph{x} denotes the location in the simulation domain. In our calculation, a constant upstream temperature \(T_{ref} = T_{1}\) was used for the initial temperature profile and correspondingly \(d_{ref}\left( T_{1} \right) = \text{const}\).
\item
  During the simulation, the local reference diameter \(d_{ref}(x) = d_{ref}(l)\) in each cell \(l = 1,N_{c}\) is updated periodically by using Eqs. (10) and (11) together with the macroscopic properties update, including temperature \(T(x) = T(l)\).
\item
  In the NTC algorithm, the number of selected pairs \emph{N\textsubscript{sel}} is calculated in a standard manner by using the maximum product of cross-section and relative velocity. Each selected pair \((i,j),\mspace{6mu} 1 \leq i \neq j \leq N^{(l)}\) is accepted for collision with a probability:
\end{enumerate}

\begin{equation}
p = \frac{c_{r}\sigma(x)F_{num}\left\lbrack \left( 2k_{B}T(x)/m \right)/c_{r}^{2} \right\rbrack^{(\omega - 0.5)}/\Gamma(5/2 - \omega)}{\left\lbrack c_{r}\sigma(x)\left\lbrack \left( 2k_{B}T(x)/m \right)/c_{r}^{2} \right\rbrack^{(\omega - 0.5)}/\Gamma(5/2 - \omega) \right\rbrack_{\max}}
\tag{14}
\end{equation}

in conformity with the acceptance-rejection procedure.

\begin{enumerate}
\def\labelenumi{\arabic{enumi}.}
\setcounter{enumi}{3}
\item
  In the SBT algorithm, the number of selected pairs is \(N_{sel} = N - 1\). The collision probability of each selected pair \((i,j),\mspace{6mu}\left( 1 \leq i \leq N^{(l)} - 1 \right) \cap \left( i < j \leq N^{(l)} \right)\) is equal to
\end{enumerate}

\begin{equation}
p = (N - i).c_{r}\sigma\left( T(x) \right)F_{num}\frac{1}{\Gamma(5/2 - \omega)}\left( \frac{2k_{B}T(x)}{mc_{r}^{2}} \right)^{(\omega - 0.5)}\frac{\Delta t}{V}
\tag{15}
\end{equation}

The implementation of the GBT algorithm approach is similar to that of the SBT.

\begin{enumerate}
\def\labelenumi{\arabic{enumi}.}
\setcounter{enumi}{4}
\item
  The post-collision velocities of particles of each accepted collision pair are computed using the LJ binary collision dynamics. Matsumoto and Koura \cite{matsumoto1991argonshock} suggested the algorithm, which we describe below. Points 3 to 5 are implemented and used at every time step Δt, and point 2 is applied after several steps, when the macroscopic variables are sampled, and their average values are updated.
\end{enumerate}

The next step in a typical implementation is to determine the deflection angle based on LJ intermolecular potential. The single parameter that fully characterizes an elastic collision between two atoms or molecules, relating the pre-collisional and post-collisional velocities, is the deflection angle \emph{χ}. The nature of the intermolecular forces between the colliding molecules highly influences this angle. Following the methodology of Matsumoto and Koura \cite{matsumoto1991argonshock}, we detail the approach for calculating the scattering angle based on the LJ molecular interaction potentials. The LJ potential is represented as Eq. (1). Based on Eq. (1), the deflection angle is given by:

\begin{equation}
\begin{aligned}
\chi &= \pi - 2\left[1 + cz - (1 + c)z^{2}\right]^{1/2} \\
&\quad \times \int_{0}^{1}\left\{ 1 - \left[1 + cz - (1 + c)z^{2}\right]u^{2} + czu^{6} - (1 + c)z^{2}u^{12} \right\}^{-1/2}\,du .
\end{aligned}
\tag{16}
\end{equation}

, with \(c = (2/\varepsilon^{*})\left\lbrack 1 + (1 + \varepsilon^{*})^{1/2} \right\rbrack\), \(z = (4/c\varepsilon^{*})(d_{LJ}/r_{0})^{6}\), and \(u = r_{0}/r\) where \emph{r\textsubscript{0}} is the closest distance between the two molecules. The \(\varepsilon^{*} = \varepsilon/\varepsilon_{LJ}\) and \(b^{*} = b/d_{LJ}\) are the reduced relative energy and impact parameters, respectively. For given values of \emph{ε\textsuperscript{*}} and \emph{b\textsuperscript{*}}, the value of \emph{z} is obtained by solving the implicit equation below:

\begin{equation}
b^{*} = (4/c\varepsilon^{*})\sqrt{1 + cz - (1 + c)z^{2}}z^{- 1/6}
\tag{17}
\end{equation}

The solution for \emph{z} is used to compute the \emph{χ} by numerical integration of Eq. (16). Using the Gauss-Chebyshev quadrature method, the integral in Eq. (16) is written as:

\begin{equation}
\int_{0}^{1}{I(u)du = \sum_{k = 1}^{M}{\frac{1}{2}w_{k}I(\frac{y_{k} + 1}{2})}}\sqrt{1 - {y_{k}}^{2}}
\tag{18}
\end{equation}

, where \emph{w\textsubscript{k}} are the weights for Gauss-Chebyshev quadrature, \emph{y\textsubscript{k}} are zeros of the \emph{M\textsuperscript{th}} degree Chebyshev polynomial, and \emph{I} is the integral in Eq. (16) given by

\begin{equation}
I(u) = \left\{ 1 - \left\lbrack 1 + cz - (1 + c)z^{2} \right\rbrack u^{2} + czu^{6} - (1 + c)z^{2}u^{12} \right\}^{- 1/2}
\tag{19}
\end{equation}

The zeros, \emph{y\textsubscript{k}}, of \emph{M\textsuperscript{th}} degree Chebyshev polynomial are given by

\begin{equation}
y_{k} = \cos(\frac{(2k + 1)\pi}{2M})
\tag{20}
\end{equation}

and the weights are all equal and are given by

\begin{equation}
w_{k} = \frac{\pi}{M}
\tag{21}
\end{equation}

In the classical approach by Matsumoto and Koura \cite{matsumoto1991argonshock}, molecular collisions are not calculated for deflection angles below the small cutoff angle \(\chi_{\min}\). This leads to the finite total collision cross section \emph{σ\textsubscript{t}} for the long-range force of the LJ potential. With specified values for \emph{ε\textsuperscript{*}} and \(\chi_{\min}\), the effective maximum value of \emph{b\textsuperscript{*}}, \(b_{\max}^{*}\), is estimated from the small-angle approximation \cite{mason1962atomic} as:

\begin{equation}
b_{\max}^{*}{= max}\left\lbrack {(\frac{4\pi}{\varepsilon^{*}\chi_{\min}})}^{1/6},{(\frac{6\pi}{\varepsilon^{*}\chi_{\min}})}^{1/12} \right\rbrack
\tag{22}
\end{equation}

and the reduced total cross-section \(\sigma_{t}^{*} = \sigma_{t}/\sigma_{LJ}\) is given by:

\begin{equation}
\sigma_{t}^{*} = b_{\max}^{*^{2}}
\tag{23}
\end{equation}

, where \(\sigma_{LJ} = \pi d_{LJ}^{2}\) is the constant cross-section of the LJ potential. The value of the reduced impact parameter \emph{b\textsuperscript{*}} is chosen with a uniform random number \emph{R} in the range (0, 1) as:

\begin{equation}
b^{*} = b_{\max}^{*^{}}R^{1/2}
\tag{24}
\end{equation}

It is confirmed that for \(b^{*} = b_{\max}^{*}\), the value of the deflection angle is smaller than \emph{χ}\textsubscript{min}, and in this situation, the collision is ignored as a null collision \cite{koura1986null,koura1990sensitive}.

To generate the high-fidelity scattering data required for training the surrogate models (and for the exact LJ-DSMC benchmarks), the deflection angle is computed by numerically solving the scattering integral (Eq. 16). This process involves two critical numerical steps implemented in the standard code. First, the distance of closest approach (\emph{r\textsubscript{0}}), which appears as the lower limit of the scattering integral, is determined by finding the root of the denominator in Eq. (16). This root-finding problem is handled using the Bisection method, chosen for its guaranteed convergence, with a tolerance threshold of 5 × 10\textsuperscript{-5} for the reduced coordinate \emph{z}.

Once the root is located, the singularity at the lower integration limit is removed by a variable transformation, and the resulting definite integral is evaluated using the Gauss-Chebyshev quadrature rule. Based on a sensitivity analysis performed in the solver, varying quadrature points from 6 to 1600 in the source code, it was determined that 100 quadrature points provide a sufficient balance between numerical precision and computational speed for generating the large-scale training dataset. This numerical treatment makes the reference scattering data sufficiently accurate for surrogate training, with quadrature and root-finding errors kept well below the observed surrogate error.

\subsection{Dynamic bounding of the impact parameter: resolving the cut-off singularity}\label{subsec:dynamic-impact-parameter-bound}
A fundamental challenge in implementing realistic intermolecular potentials with infinite interaction ranges, such as the Lennard-Jones (LJ) model, within the Direct Simulation Monte Carlo (DSMC) framework is the mathematical singularity of the total collision cross-section. Theoretically, the infinite tail of the attractive force yields an unbounded cross-section, dictating that the microscopic impact parameter, b, must be truncated at a finite maximum value, bmax, to enable stochastic sampling. The formulation of bmax critically affects both the physical fidelity and the computational efficiency of the solver.

Traditionally, the DSMC community has relied on two distinct methodologies to address this singularity

\begin{enumerate}
\def\labelenumi{\arabic{enumi}.}
\item
  Arbitrary Deflection Angle Cut-off (e.g., Matsumoto and Koura \cite{matsumoto1991argonshock}): In this classical approach, the infinite range is bypassed by prescribing an arbitrarily small minimum cutoff deflection angle (e.g., $\chi_{\min} \approx 0.1^{\circ}$). While mathematically permissible within a formal ``null-collision'' framework, substituting this infinitesimal $\chi_{\min}$ yields an excessively large maximum impact parameter. Consequently, the algorithm is forced to select a large number of candidate collision pairs per cell, most of which exhibit grazing trajectories with extreme distance ($\chi \to 0$). These \emph{grazing} interactions exchange negligible momentum and are eventually discarded as null collisions. This drastically increases computational overhead, rendering the routine prohibitively expensive and highly inefficient for modern data-driven or Machine-Learning (ML)-accelerated solvers.
\item
  Empirical Profile Truncation (e.g., Venkattraman and Alexeenko \cite{venkattraman2012binary}): Recognizing the severe computational inefficiency of the pure null-collision approach, other researchers have determined $b_{\max}$ empirically by visually analyzing the deterministic scattering profile ($\chi$ vs. $b$). For the LJ potential, the deflection angle exhibits a distinct negative well, often referred to as rainbow scattering, caused by the attractive intermolecular forces. Empirically, \emph{$b_{\max}$} is chosen just past the depth of this minimum, where the negative angle asymptotically approaches the zero axis. While this effectively avoids simulating excess-grazing collisions and guarantees that the essential attractive-scattering kinematics are captured, it remains an ad hoc, static estimate that lacks a formal, dynamic coupling to the gas's local macroscopic thermodynamic state.
\end{enumerate}

To establish strict mathematical consistency between the macroscopic and microscopic scales, the present hybrid algorithm entirely eliminates the need for arbitrary cutoff angles or empirical geometric truncations. In our proposed methodology, the maximum impact parameter for the microscopic scattering phase is dynamically constrained to be strictly identical to the effective cross-section utilized for the collision pair selection in the macroscopic phase:

\begin{equation}
\pi b_{\max}^{2} = \sigma_{cell}\ \ \overset{yields}{\rightarrow}\ b_{\max}^{*} = \frac{b_{\max}}{d_{LJ}} = \frac{1}{d_{LJ}}\sqrt{\frac{\sigma_{cell}}{\pi}}
\tag{25}
\end{equation}

, where \(\sigma_{cell}\) is rigorously updated at each cell based on the local macroscopic temperature.

A critical physical requirement raised by this dynamic bounding is ensuring that the derived \(b_{\max}^{*}\) does not prematurely truncate the attractive potential well (i.e., the region of significant negative scattering angles must not be artificially excluded). This physical fidelity is inherently guaranteed by our approach. The local effective cross-section σcell is rigorously derived from the Chapman-Enskog collision integral \emph{W\textsuperscript{(2)}}(2) (\(T_{inv}^{*}\)). In kinetic theory, this integral evaluates the momentum transfer contributions over the entire domain of impact parameters (b $\in$ {[}0, $\infty$)), weighted by a momentum-transfer efficiency factor of sin\textsuperscript{2} χ (or equivalently, 1 − cos\textsuperscript{2} χ).

Because of this specific transport weighting, the \emph{W\textsuperscript{(2)}}(2) integral heavily and naturally encapsulates the deep negative deflection region (the attractive tail) while intrinsically and analytically filtering out the zero-momentum-transfer contributions of extreme-distance grazing collisions (where χ → 0 and sin\textsuperscript{2} χ → 0). Consequently, equating the microscopic scattering area directly to this macroscopic integral guarantees that the dynamically derived \(b_{\max}^{*}\)fully encompasses the physically critical negative scattering regime. This self-consistent constraint avoids cutting off important attractive interactions and eliminates computationally wasteful null collisions. Exact momentum and kinetic-energy conservation are then retained by the standard elastic binary-collision velocity update. Ultimately, this optimal reduction of the collision space provides a highly efficient, theoretically sound foundation that is a crucial prerequisite for in-situ neural network inference, as discussed in subsequent sections.

The detailed algorithm of the variable effective diameter LJ is summarized below.

\textbf{\hfill\break
}

\begin{algorithm}[!htbp]
\caption{Hybrid Lennard-Jones Scattering Algorithm with Variable Effective Diameter (VED)}
\label{alg:ved-lj}
\begin{algorithmic}[1]
\Require Reduced collision energy $\varepsilon^*=E_c/\varepsilon_{LJ}$, local cell temperature $T_{cell}$, random number $R\in[0,1]$.
\Ensure Deflection angle $\chi$.
\State \textbf{Step 1: Dynamic Bounding of Impact Parameter (VED Model)} \Comment{See Section~\ref{subsec:dynamic-impact-parameter-bound}}
\State Calculate local effective cross-section $\sigma_{cell}(T_{cell})$ using viscosity matching (Eq. 25).
\State Determine maximum reduced impact parameter: $b^{*}_{max}=d_{LJ}^{-1}\sqrt{\sigma_{cell}/\pi}$.
\State Sample reduced impact parameter: $b^*=b^*_{max}R^{1/2}$.
\State \textbf{Step 2: Root Finding for Closest Approach ($r_0$)} \Comment{Subroutine BISECTION}
\State Initialize search interval for reduced coordinate $z$: $[x_0,x_1]\leftarrow[10^{-20},1.0]$.
\State Set tolerance $\delta\leftarrow5\times10^{-5}$.
\State Define function $F(z)$ based on the denominator of the scattering integral (Eq. 17).
\While{$|x_1-x_0|>\delta$}
  \State $x_{mid}\leftarrow(x_0+x_1)/2$.
  \If{$F(x_{mid})F(x_0)<0$}
    \State $x_1\leftarrow x_{mid}$.
  \Else
    \State $x_0\leftarrow x_{mid}$.
  \EndIf
\EndWhile
\State $z_{root}\leftarrow x_{mid}$.
\State \textbf{Step 3: Numerical Integration of Scattering Angle} \Comment{Subroutine ELASTIC\_LJ}
\State Set quadrature nodes $M\leftarrow100$.
\State Initialize integral sum $S\leftarrow0$.
\For{$k=0$ to $M-1$}
  \State $y_k\leftarrow\cos((2k+1)\pi/(2M))$.
  \State $u_k\leftarrow(y_k+1)/2$.
  \State Evaluate integrand $I(u_k)$ from Eq. 19.
  \State $S\leftarrow S+I(u_k)\sqrt{1-y_k^2}$.
\EndFor
\State Calculate final integration weight $W\leftarrow\pi/(2M)$.
\State Compute deflection angle $\chi$.
\State \Return $\chi$.
\end{algorithmic}
\end{algorithm}
\FloatBarrier

\section{Machine learning-based surrogate models for scattering dynamics}\label{sec:ml-surrogate-scattering}
As established in the previous section, the deflection angle χ uniquely determines the post-collision velocities in elastic binary collisions governed by the Lennard--Jones intermolecular potential. In the classical formulation introduced by Matsumoto and Koura \cite{matsumoto1991argonshock}, χ is obtained by numerically evaluating an implicit scattering integral that depends on the relative collision energy and the impact parameter. This formulation provides an exact representation of the LJ collision physics and is employed in the present study as the reference solution.

Despite its physical rigor, repeatedly evaluating the LJ scattering integral for each collision event is the dominant computational bottleneck in DSMC simulations that incorporate realistic intermolecular potentials. This limitation becomes particularly restrictive in multidimensional and large-scale simulations, where a vast number of collision events must be processed.

To alleviate this computational burden without compromising physical fidelity, the LJ scattering operator is approximated with a DeepONet-based surrogate. Importantly, the underlying intermolecular interaction remains the Lennard--Jones potential; the surrogate serves solely as an efficient numerical realization of the scattering solution described in Section~\ref{sec:lj-reference-cross-section}.

Mathematically, the LJ scattering process defines a deterministic operator mapping the collision parameters to the deflection angle,

\begin{equation}
S_{LJ}(E,b) = \chi
\tag{26}
\end{equation}

, where \emph{E} denotes the relative collision energy and \emph{b} represents the impact parameter. The operator \emph{S\textsubscript{LJ}} is implicitly defined through the classical scattering formulation and does not admit a closed-form analytical expression. In this work, the operator is primarily approximated by a DeepONet surrogate, while simpler MLP and interpolation-based models are used only as baselines for comparison.

\subsection{Deep Operator Network (DeepONet) formulation}\label{subsec:deeponet-formulation}
To more faithfully capture the operator nature of the LJ scattering process, a surrogate model based on Deep Operator Networks (DeepONet) is employed. DeepONet is specifically designed to approximate nonlinear operators by decomposing the input dependence into separate functional and coordinate representations \cite{lu2021deeponet}.

In this framework, the LJ scattering operator is represented through the angular embedding
\begin{equation}
\mathbf{y}_{\chi}(E,b)=\left(\cos\chi,\sin\chi\right)
\tag{26a}
\end{equation}
and each output component is approximated by a DeepONet expansion,
\begin{equation}
\hat{y}_{q}(E,b)=\sum_{k=1}^{p} B_{k}^{(q)}(E)T_{k}^{(q)}(b), \qquad q\in\{\cos\chi,\sin\chi\}
\tag{26b}
\end{equation}
The scattering angle is then reconstructed as
\begin{equation}
\hat{\chi}=\mathrm{atan2}(\widehat{\sin\chi},\widehat{\cos\chi})
\tag{26c}
\end{equation}

, where \emph{B\textsubscript{k}} and \emph{T\textsubscript{k}} denote the outputs of the branch and trunk networks, respectively, and \emph{p} is the dimension of the latent representation. The branch network encodes the dependence of the scattering process on the collision energy, while the trunk network captures the dependence on the impact parameter. This operator-based formulation enables the DeepONet model to represent the intrinsic structure of the LJ scattering mechanism more effectively than standard deep neural networks (DNNs), resulting in improved accuracy and robustness across a broad range of collision conditions.

To capture the continuous mapping between the collision parameters and the resulting scattering angle, a DeepONet architecture is employed. As illustrated in Fig.~\ref{fig:fig01}, unlike standard feed-forward networks, DeepONet consists of two parallel sub-networks: the branch net and the trunk net. In this framework, the branch net processes the reduced collision energy \(\epsilon^*\), while the trunk net takes the normalized impact parameter \(b/b_{\max}\) as the evaluation coordinate. The DeepONet used in the DSMC calculations has a latent dimension \(p=128\). Both the branch and trunk sub-networks have width 128 and four hidden layers, followed by a dot-product decoder. The outputs of these two networks are combined through a dot product, augmented by a bias term, to predict the angular embedding from which \(\chi\) is reconstructed. This decomposition allows the model to learn the underlying integral operator of the LJ potential and to generalize across the wide range of collision energies encountered in DSMC simulations. Training details are given in Section~\ref{subsec:training-dataset-sampling}.

\includegraphics[width=\linewidth,height=0.32\textheight,keepaspectratio]{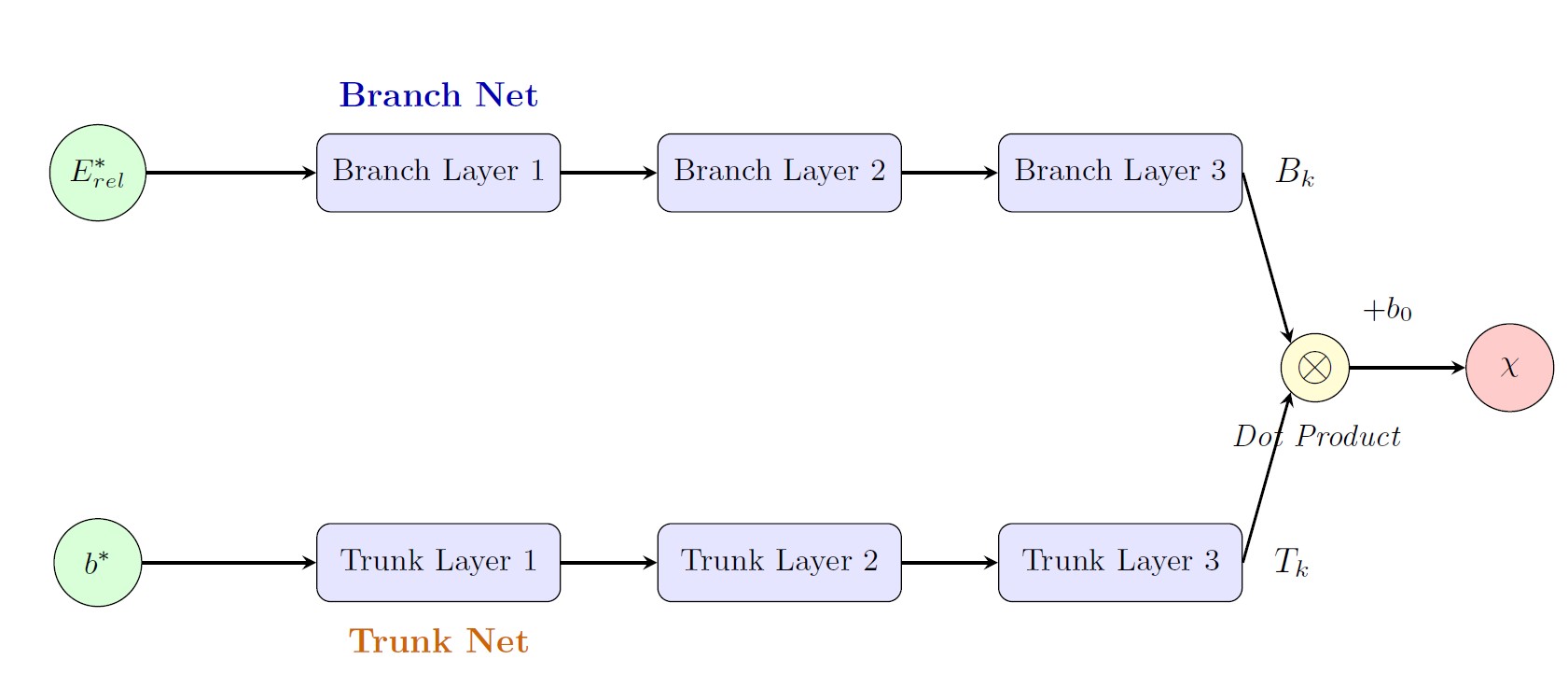}

\autopapercaption{fig:fig01}{Schematic of the DeepONet architecture.}

\subsection{Training dataset and sampling strategy}\label{subsec:training-dataset-sampling}
The DeepONet surrogate model is trained on high-fidelity scattering data generated from the exact Lennard--Jones formulation described in Section~\ref{sec:lj-reference-cross-section}. The input variables are the reduced collision energy, \(\epsilon^*\), and the normalized impact parameter, \(b/b_{\max}\), and the output is represented by \((\cos\chi,\sin\chi)\) rather than \(\chi\) itself. This representation avoids artificial discontinuities associated with angular periodicity and improves stability in the low-angle tail. The final dataset contains \(1.28\times10^7\) valid LJ scattering samples, split into \(1.152\times10^7\) training samples and \(1.28\times10^6\) validation samples. The reduced collision energy spans \(10^{-3}\le \epsilon^* \le 40\), while \(b/b_{\max}\) spans the interval \([0,1]\), where \(b_{\max}\) is the upper impact-parameter bound used to normalize each sampled collision state. The sampling is deliberately nonuniform: log-uniform sampling covers the full energy range, additional log-uniform enrichment is applied for \(\epsilon^*<1\), and uniformly distributed samples are included to avoid biasing the surrogate exclusively toward the low-energy tail.

The DeepONet used in the DSMC calculations has a latent dimension \(p=128\), fully connected branch and trunk sub-networks with width 128 and four hidden layers, and a dot-product decoder. Training uses the Adam optimizer with learning rate \(10^{-3}\), mini-batch size 8192, and a 100-epoch maximum with validation monitoring. The cost of training is paid once offline; only the trained weights are used during DSMC. To determine whether the operator architecture is necessary, we also trained and evaluated two baseline surrogates on the same held-out collision set: a standard multilayer perceptron (MLP) with comparable input--output variables and a bilinear interpolation table constructed over the same reduced variables. The comparative results are reported in Section~\ref{subsec:diffusion-validation-baselines}. These baseline models are used only for offline accuracy comparison and are not used in the DSMC production simulations.

\begin{figure}[!htbp]
\centering
\includegraphics[width=0.99\textwidth,trim={0 0 0 36pt},clip]{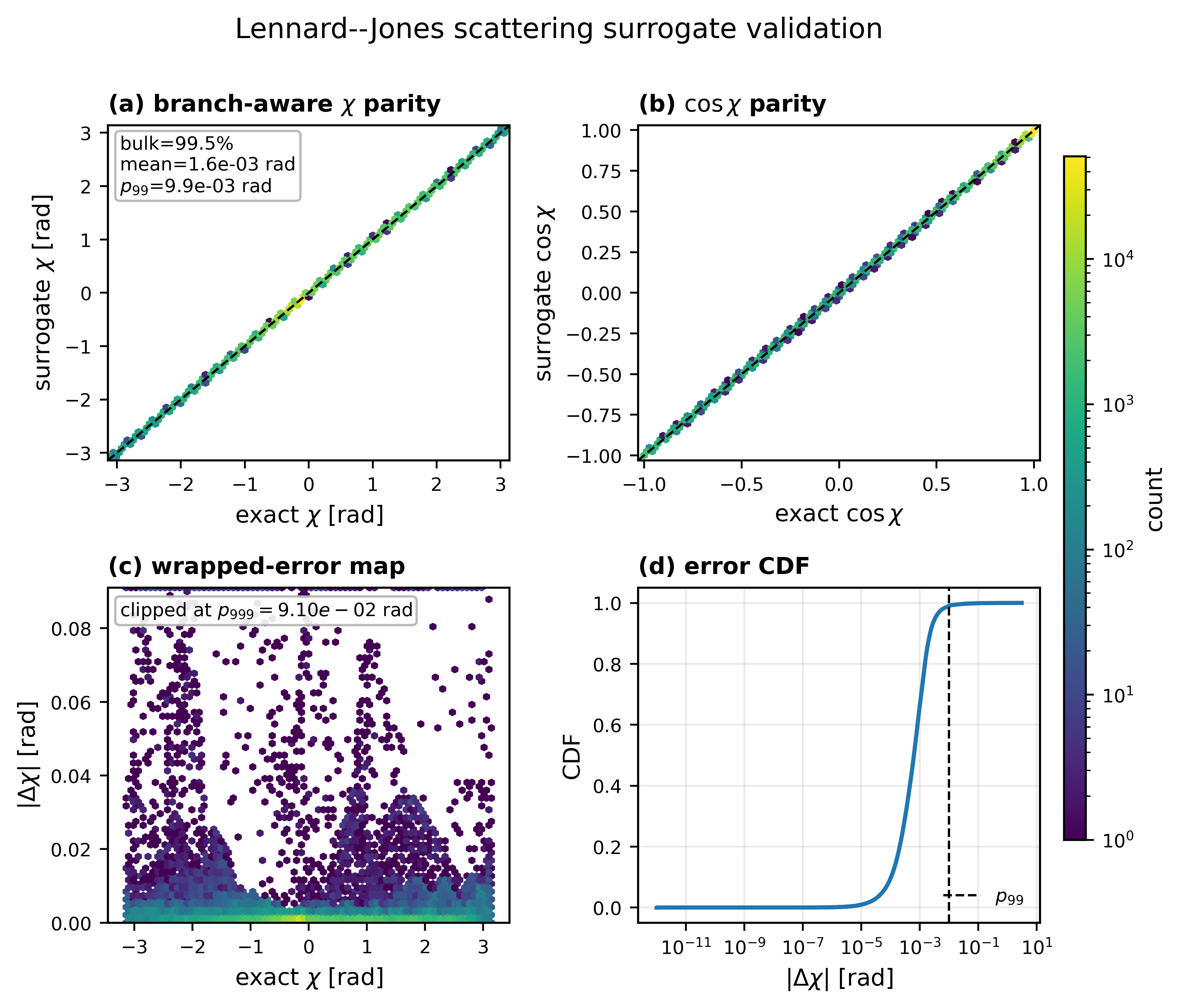}
\caption{Direct validation of the DeepONet--LJ scattering-angle surrogate against exact numerical integration on held-out collision samples. Panel (a) shows the branch-aware reconstructed scattering angle \(\chi\), panel (b) shows the parity of \(\cos\chi\), panel (c) maps the wrapped angular error, and panel (d) reports the error cumulative distribution. The bulk mean wrapped error is approximately \(1.6\times10^{-3}\,\mathrm{rad}\), and the 99th-percentile error is approximately \(9.9\times10^{-3}\,\mathrm{rad}\).}
\label{fig:scattering_angle_surrogate_validation}
\end{figure}

To quantitatively assess the predictive performance of the DeepONet-based surrogate model, a parity plot comparing the predicted scattering angles (\(\chi_{predicted}\)) against the exact numerical solutions (\(\chi_{true}\)) is presented in the first frame of Fig.~\ref{fig:scattering_angle_surrogate_validation}. The testing dataset, which was not seen by the network during training, spans a wide range of collision energies and impact parameters. As illustrated, the data points are tightly clustered around the 45-degree dashed red line, which represents a perfect fit (\emph{y} = \emph{x}). The high degree of linearity across the full range of scattering angles (from 0 to π radians) demonstrates that DeepONet effectively captures the complex, nonlinear scattering dynamics of the Lennard--Jones potential. This perfect correlation validates the reliability of the neural operator as a high-fidelity and computationally efficient substitute for the exact scattering integral in DSMC simulations.

The trained DeepONet scattering-angle map is further validated against exact LJ scattering integrations using the same branch-aware angular reconstruction employed in the DSMC collision step. Frames b-d in Figure~\ref{fig:scattering_angle_surrogate_validation} provide a more detailed diagnostic of the surrogate error: it compares the reconstructed \(\cos\chi\), the wrapped angular error over the sampled collision space, and the cumulative distribution of the error. This additional validation is important because the same angular kernel is used later in the diffusion tests; therefore, the surrogate must reproduce not only the overall parity trend but also the angular statistics that enter transport coefficients. The bulk region, containing 99.5\% of the held-out validation samples, gives a mean wrapped-angle error of \(1.6\times10^{-3}\,\mathrm{rad}\), while the 99th-percentile error is \(9.9\times10^{-3}\,\mathrm{rad}\). These results confirm that the DeepONet surrogate provides an accurate and computationally efficient realization of the LJ scattering map before it is embedded in the full DSMC transport calculations.

\FloatBarrier

\subsection{Integration into the DSMC collision procedure}\label{subsec:dsmc-integration}
Once trained, the DeepONet surrogate is integrated directly into the DSMC collision algorithm by replacing the numerical evaluation of the LJ scattering integral during the velocity-transformation step. The pair selection procedure and the overall structure of the DSMC algorithm are described in Section~\ref{sec:lj-reference-cross-section}, and it remains unchanged. As a result, the DSMC solver supports two interchangeable numerical realizations of the LJ scattering process: an exact formulation based on numerical integration and a machine learning-accelerated surrogate formulation. This modular structure enables a direct assessment of the trade-off between computational efficiency and numerical accuracy in the subsequent validation and results sections.

The resulting inference call is deterministic and enters only through the deflection-angle evaluation; the velocity transformation remains the standard elastic binary-collision rotation and therefore conserves linear momentum and kinetic energy exactly for each accepted pair. The in-situ DeepONet inference algorithm used to predict the scattering angle is detailed below.

\textbf{\hfill\break
}

\begin{algorithm}[!htbp]
\caption{In-Situ ML Inference for Deflection Angle Prediction}
\label{alg:deeponet-inference}
\begin{algorithmic}[1]
\Require Pre-collision relative velocity $v_r$, dimensionless impact parameter $b^*$.
\Require Extracted scaler parameters $\mu=[\mu_{v_r},\mu_{b^*}]^T$, $\sigma=[\sigma_{v_r},\sigma_{b^*}]^T$.
\Require Network weights and biases $W^{(l)}_{branch}$, $b^{(l)}_{branch}$ and $W^{(l)}_{trunk}$, $b^{(l)}_{trunk}$ for all trained layers.
\State \textbf{Step 1: Input Z-score Normalization}
\State $\tilde{v}_r\leftarrow(v_r-\mu_{v_r})/\sigma_{v_r}$.
\State $\tilde{b}^*\leftarrow(b^*-\mu_{b^*})/\sigma_{b^*}$.
\State \textbf{Step 2: Forward Pass - Branch Network (Kinematics)}
\State $z_b^{(1)}\leftarrow\tilde{v}_rW_{branch}^{(1)}+b_{branch}^{(1)}$.
\State $a_b^{(1)}\leftarrow\max(0,z_b^{(1)})$.
\State $z_b^{(2)}\leftarrow a_b^{(1)}W_{branch}^{(2)}+b_{branch}^{(2)}$.
\State $a_b^{(2)}\leftarrow\max(0,z_b^{(2)})$.
\State $h_{branch}\leftarrow a_b^{(2)}W_{branch}^{(3)}+b_{branch}^{(3)}$.
\State \textbf{Step 3: Forward Pass - Trunk Network (Geometry)}
\State $z_t^{(1)}\leftarrow\tilde{b}^*W_{trunk}^{(1)}+b_{trunk}^{(1)}$.
\State $a_t^{(1)}\leftarrow\max(0,z_t^{(1)})$.
\State $z_t^{(2)}\leftarrow a_t^{(1)}W_{trunk}^{(2)}+b_{trunk}^{(2)}$.
\State $a_t^{(2)}\leftarrow\max(0,z_t^{(2)})$.
\State $h_{trunk}\leftarrow a_t^{(2)}W_{trunk}^{(3)}+b_{trunk}^{(3)}$.
\State \textbf{Step 4: Output Evaluation}
\State \(\widehat{\cos\chi}\leftarrow h_{branch}^{(c)}\cdot h_{trunk}^{(c)}\).
\State \(\widehat{\sin\chi}\leftarrow h_{branch}^{(s)}\cdot h_{trunk}^{(s)}\).
\State Normalize the angular pair if needed:
\begin{equation}
(\widehat{\cos\chi},\widehat{\sin\chi}) \leftarrow 
\frac{(\widehat{\cos\chi},\widehat{\sin\chi})}
{\sqrt{\widehat{\cos\chi}^{\,2}+\widehat{\sin\chi}^{\,2}}}
\tag{26d}
\end{equation}
\State \(\chi_{pred}\leftarrow \mathrm{atan2}(\widehat{\sin\chi},\widehat{\cos\chi})\).
\State \Return \(\chi_{pred}\).
\end{algorithmic}
\end{algorithm}
\FloatBarrier

\section{Validation and results}\label{sec:validation-results}

The proposed frameworks for incorporating the Lennard-Jones (LJ) potential into the DSMC method are evaluated through a series of demanding test cases. This validation strategy assesses both physical fidelity and computational performance across diverse flow regimes. The investigations comprise normal shock waves in monatomic gases, supersonic Couette flow, and hypersonic flow over a 2D circular cylinder, collectively probing strong translational non-equilibrium, effects of attractive forces, and algorithmic robustness in complex geometries.

For normal shock waves, results are compared against the experimental benchmarks of Muntz and Harnett \cite{muntz1969shockvdf} for helium and Holtz and Muntz \cite{holtz1983argonshock} for argon. These comparisons facilitate a quantitative assessment of the LJ-based formulations against traditional VHS models. Furthermore, the performance of machine learning-accelerated models, specifically DeepONet, is systematically benchmarked against the exact LJ implementation to evaluate their accuracy and efficiency gains.

Simulations are conducted using customized versions of Bird's DSMC suite. The 1D codes (DSMC1 and DSMC1S) are employed for shock and Couette flows, while a modified version of the two-dimensional DS2V environment is adapted to accommodate the LJ-based scattering models. For the ML-accelerated approach, surrogate networks are trained offline in Python using high-fidelity scattering data. The optimized weights and biases are exported to external data structures, which are subsequently integrated into the Fortran-based collision subroutines. This modular implementation enables seamless runtime switching between exact LJ and ML-based scattering models, preserving the core DSMC algorithmic structure while significantly reducing the computational cost of the collision transformation stage.

\subsection{Normal shock wave of helium and argon}\label{subsec:normal-shock-he-ar}
In this section, we examine the structure of normal shock waves in helium and argon to evaluate the accuracy of the proposed LJ-based DSMC framework. These cases cover a range of Mach numbers, providing a robust test for both the physical fidelity of the potential models and the predictive power of the neural operator surrogates. The helium normal shock wave at an upstream Mach number of 1.59 and the temperature of T=160 K is simulated. The upstream number density is 2.88982×10\textsuperscript{21} m\textsuperscript{-3}, and the velocity is 1184.87 m/s. The helium gas, with a molecular mass of 6.65×10-27 kg and a viscosity-temperature index of 0.647, is considered. Accordingly, the upstream speed of sound is calculated to be \emph{a\textsubscript{1}} = 744.065 m/s. The upstream and downstream flow conditions are summarized in Table~\ref{tab:helium-shock-conditions}.

\setcounter{table}{0}

\begin{table}[!htbp]
\centering
\caption{Conditions of the simulated helium shock wave at \(Ma=1.59\), chosen to match the experimental conditions of Muntz and Harnett~\cite{muntz1969shockvdf}.}
\label{tab:helium-shock-conditions}
\begin{tabular}{lcc}
\toprule
\textbf{Quantity} & \textbf{Free stream} & \textbf{Behind shock} \\
\midrule
Mach number & 1.59 & 0.7034 \\
Number density \((\mathrm{m}^{-3})\) & \(2.88982\times10^{21}\) & \(5.2808\times10^{21}\) \\
Temperature \((\mathrm{K})\) & 160 & 244.92 \\
Velocity \((\mathrm{m\,s}^{-1})\) & 1184.87 & 648.09 \\
\bottomrule
\end{tabular}
\end{table}

The normal shock wave of argon at a relatively high Mach number of 7.183 and a low temperature of 16 K is considered. The argon gas with a molecular mass of \(6.63\times10^{-26}\,\mathrm{kg}\) with a viscosity-temperature index equal to 0.816 and VSS scattering parameter α=1 at a reference temperature of 16 K is used. The upstream number density and the velocity are set to 1.144×10\textsuperscript{21} m\textsuperscript{-3} and 539.58 m/s, respectively. For the temperature and molecular mass considered here, the upstream speed of sound is calculated as 75.119 m/s. The upstream and downstream flow conditions for the argon shock-wave case are summarized in Table~\ref{tab:argon-shock-conditions}.

\begin{table}[!htbp]
\centering
\caption{Conditions of the simulated argon shock wave at \(Ma=7.183\), chosen to match the experimental conditions of Holtz and Muntz~\cite{holtz1983argonshock}.}
\label{tab:argon-shock-conditions}
\begin{tabular}{lcc}
\toprule
\textbf{Quantity} & \textbf{Free stream} & \textbf{Behind shock} \\
\midrule
Mach number & 7.183 & 0.461 \\
Number density \((\mathrm{m}^{-3})\) & \(1.144\times10^{21}\) & \(4.324\times10^{21}\) \\
Temperature \((\mathrm{K})\) & 16 & 276 \\
Velocity \((\mathrm{m\,s}^{-1})\) & 539.58 & 142.74 \\
\bottomrule
\end{tabular}
\end{table}

Before presenting the physical results, a rigorous numerical sensitivity analysis is conducted to ensure that the solutions are independent of the spatial and statistical discretizations. Because of the significant role of attractive forces in argon, grid- and particle-per-cell (PPC) independence studies are performed for this gas.

\subsubsection{Grid and particle independence study}\label{subsubsec:grid-particle-independence}
\begin{figure}[!htbp]
\centering

\begin{subfigure}[t]{0.48\linewidth}
\centering
\includegraphics[width=\linewidth,height=0.30\textheight,keepaspectratio]{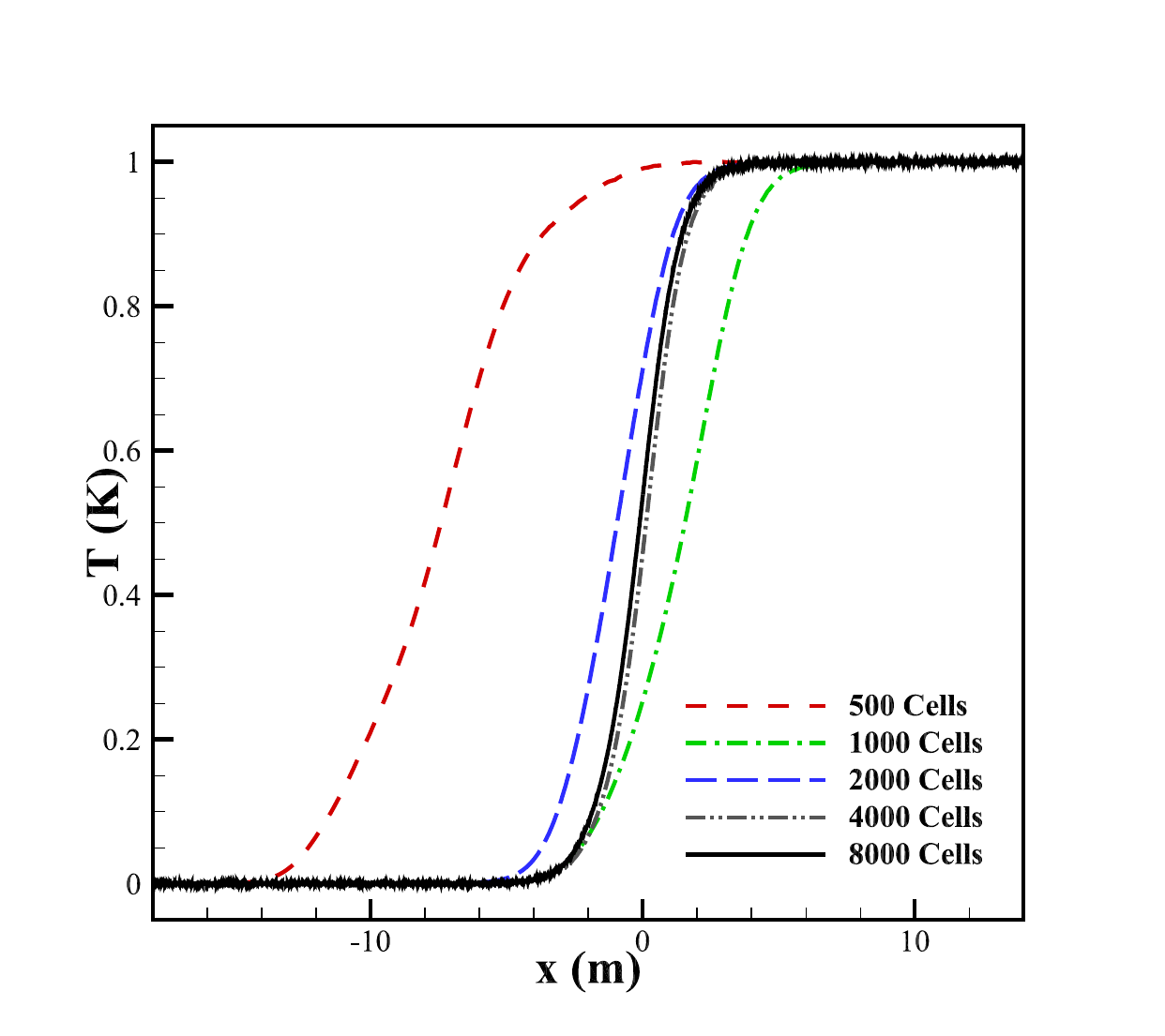}
\caption{}
\label{fig:fig03a}
\end{subfigure}
\hfill
\begin{subfigure}[t]{0.48\linewidth}
\centering
\includegraphics[width=\linewidth,height=0.30\textheight,keepaspectratio]{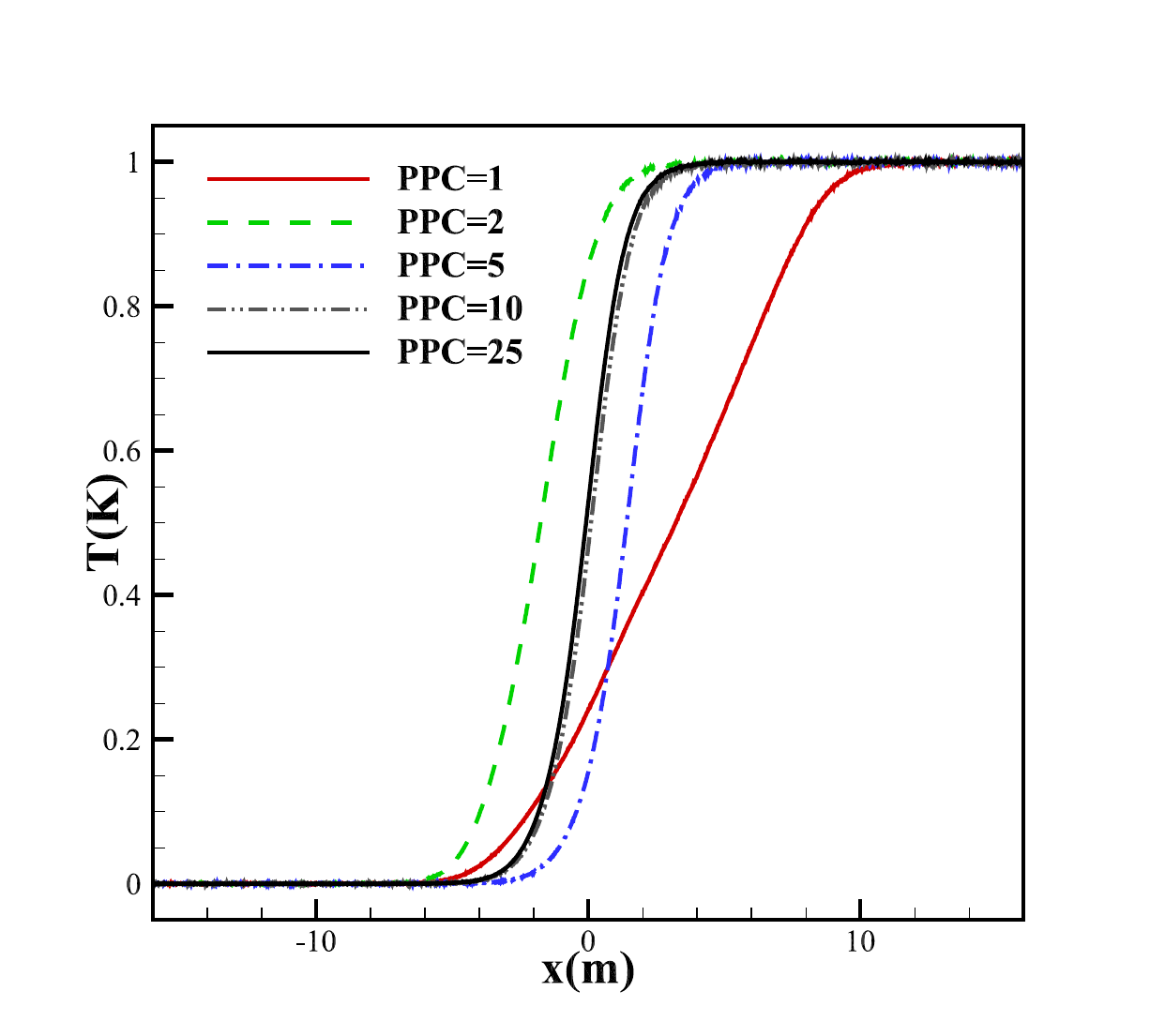}
\caption{}
\label{fig:fig03b}
\end{subfigure}

\caption{Normalized density profile for the independence tests: (a) grid-size independence and (b) particles-per-cell (PPC) independence, using the NTC collision scheme.}
\label{fig:fig03}
\end{figure}

To ensure that the numerical results are independent of the spatial discretization and the number of simulated particles, a rigorous sensitivity analysis was performed for the Argon normal shock wave using the Lennard-Jones (LJ) potential. First, the effect of grid resolution was investigated by testing five different cell counts: 500, 1000, 2000, 4000, and 8000. As shown in Fig.~\ref{fig:fig03}, the density and temperature profiles converge as the number of cells increases. The results for 4000 and 8000 cells show no significant discrepancies, indicating that 4000 cells provide sufficient resolution to capture the sharp gradients within the shock structure while maintaining the \(\Delta x < \lambda/3\) criterion. Consequently, a grid of 4000 cells was selected for all subsequent shock wave simulations.

Simultaneously, the influence of the number of particles per cell (PPC) was evaluated. Five different upstream PPC values (1, 2, 5, 10, and 25) were examined. The comparison in Fig.~\ref{fig:fig03} shows that increasing the PPC from 10 to 25 yields negligible changes in flow properties, suggesting that statistical noise is effectively minimized at 10 PPC. Based on these findings, an upstream value of 10 PPC was adopted for the remainder of the study to balance statistical accuracy and computational efficiency.

\subsubsection{Physical validation: helium and argon shock structures}\label{subsubsec:physical-validation-shocks}
The physical accuracy of the developed LJ-DSMC framework is first evaluated by comparing the normalized density and temperature profiles against the experimental data of Muntz and Harnett \cite{muntz1969shockvdf} for helium and Holtz and Muntz \cite{holtz1983argonshock} for argon. The results for gas parameters are reported in the normalized form, for instance, for density, i.e.: \begin{equation}
\hat{\rho}=\frac{\rho-\rho_1}{\rho_2-\rho_1}
\tag{26e}
\end{equation}, where ρ denotes the density across the shock wave, and \emph{ρ\textsubscript{1}} and \emph{ρ\textsubscript{2}} represent the densities at the upstream and downstream flow conditions, respectively.

Starting with helium, the simulations were performed at Mach 1.59. For the initial validation, the LJ parameters were taken as \emph{ε/k}=6.03 K and \emph{d\textsubscript{LJ}}=2.7×10\textsuperscript{-10} m \cite{hirschfelder1948transport}. Fig.~\ref{fig:fig04} shows the shock-wave structure of helium, illustrating excellent agreement between the simulation results and experimental measurements. Specifically, Panel (a) shows the normalized density profile, where the results obtained using the LJ potential---implemented via NTC, SBT, and GBT collision schemes---are compared with the VHS model and experimental data \cite{muntz1969shockvdf}. The results of the VHS model were obtained using the NTC scheme and a molecular diameter of 2.33×10\textsuperscript{-10} m \cite{bird1994molecular}. The overlap between the LJ and VHS results is expected for helium because its very shallow potential well makes the attractive forces nearly negligible at these temperatures.

Furthermore, Fig.~\ref{fig:fig04}(b) depicts the temperature component profiles in the parallel and perpendicular directions. The DSMC results for these translational temperature components closely align with the experimental data, capturing the characteristic non-equilibrium behavior in the shock layer. These results verify the core implementation of LJ scattering dynamics and confirm that the choice of collision partner selection scheme (NTC, SBT, or GBT) does not introduce numerical artifacts when the effective molecular diameter is properly calibrated.

\begin{longtable}[]{@{}
  >{\raggedright\arraybackslash}p{(\columnwidth - 2\tabcolsep) * \real{0.4937}}
  >{\raggedright\arraybackslash}p{(\columnwidth - 2\tabcolsep) * \real{0.5063}}@{}}

\begin{minipage}[b]{\linewidth}\raggedright
\includegraphics[width=\linewidth,height=0.32\textheight,keepaspectratio]{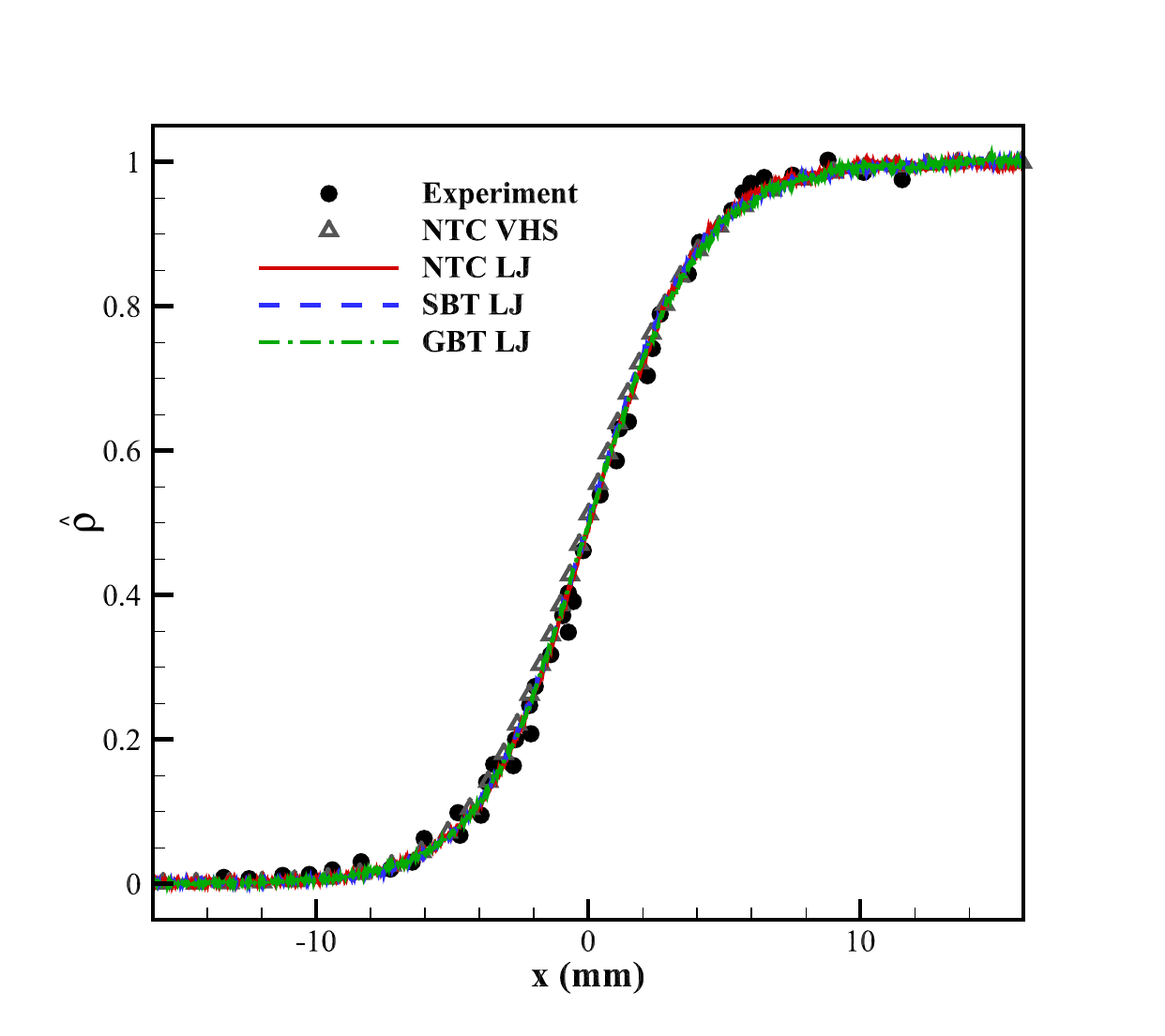}

(a)
\end{minipage} & \begin{minipage}[b]{\linewidth}\raggedright
\includegraphics[width=\linewidth,height=0.32\textheight,keepaspectratio]{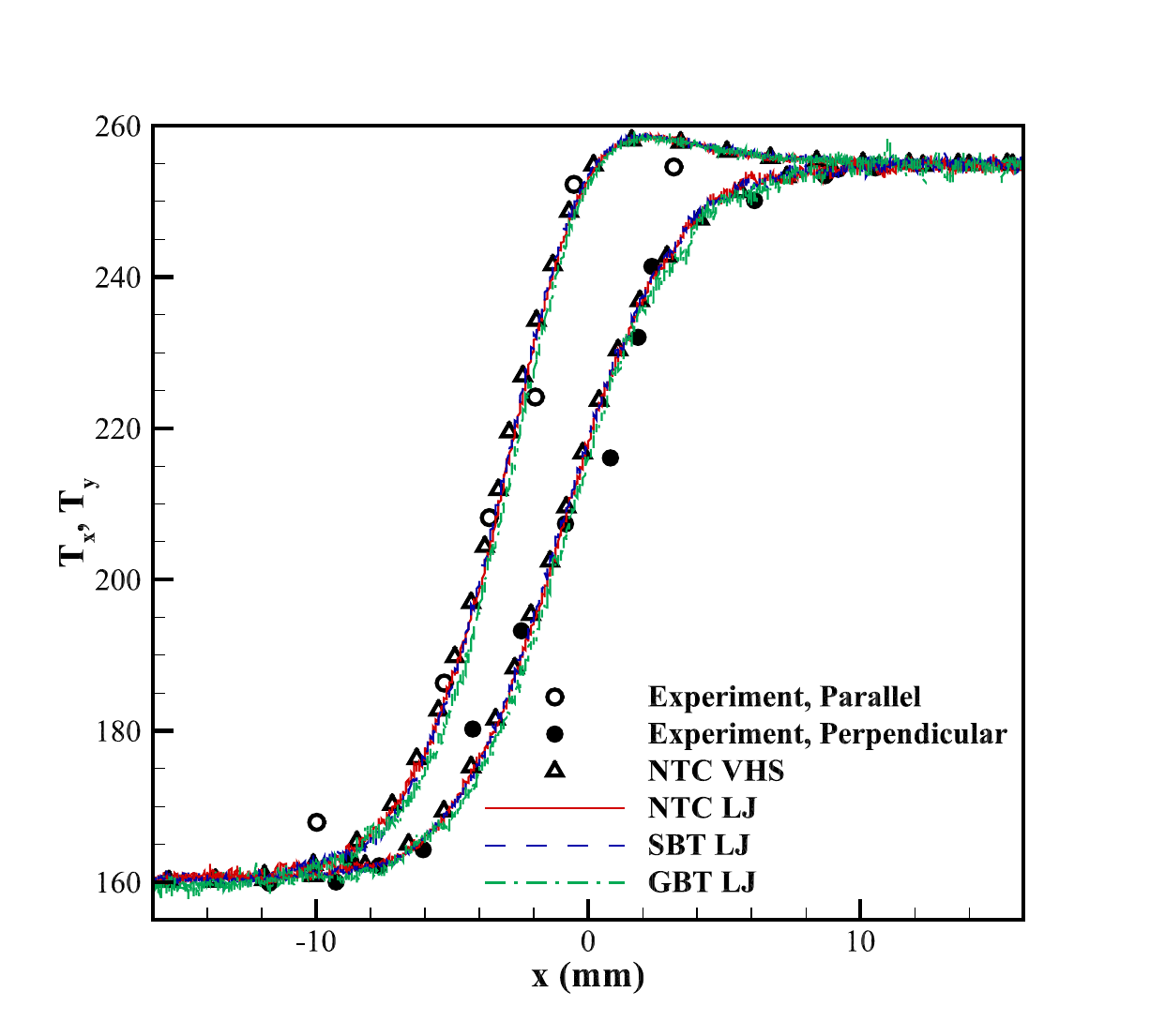}

(b)
\end{minipage} \\

\end{longtable}

\autopapercaption{fig:fig04}{Comparison of the LJ model (constant effective molecular diameter approach) results for Helium with that of the experimental data from Ref.~\cite{muntz1969shockvdf} and VHS simulation: (a) normalized density and (b) temperature in parallel and perpendicular directions.}

The proposed variable-effective-molecular-diameter approach is further validated through a comparative study against experimental benchmarks and the conventional VHS model for the helium normal shock wave. As illustrated in Fig.~\ref{fig:fig05}, the variable-diameter formulation accurately captures the normalized density and temperature profiles and agrees well with the experimental data from Ref. \cite{muntz1969shockvdf}. For these simulations, the LJ parameters for helium were consistently maintained as as \emph{ε/k}=6.03 K and \emph{d\textsubscript{LJ}}=2.7×10\textsuperscript{-10} m \cite{hirschfelder1948transport}. The precise alignment of the results confirms that the variable diameter framework---despite its ability to handle complex attractive-repulsive regimes---remains robust and physically consistent in the weak-attraction limit of helium, effectively bridging the gap between simplified collision models and realistic molecular physics.

\begin{longtable}[]{@{}
  >{\raggedright\arraybackslash}p{(\columnwidth - 2\tabcolsep) * \real{0.4958}}
  >{\raggedright\arraybackslash}p{(\columnwidth - 2\tabcolsep) * \real{0.5042}}@{}}

\begin{minipage}[b]{\linewidth}\raggedright
\includegraphics[width=\linewidth,height=0.32\textheight,keepaspectratio]{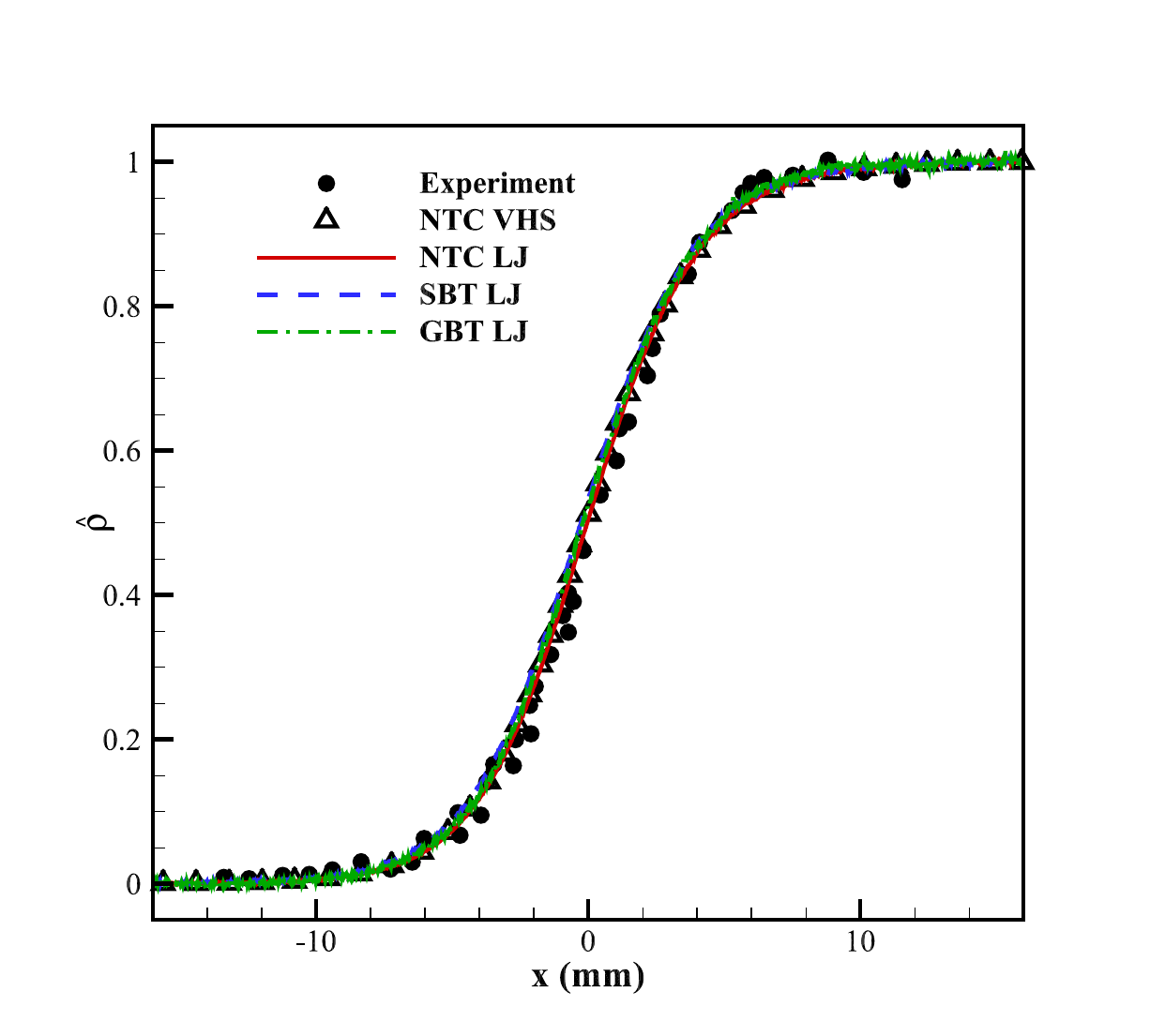}

(a)
\end{minipage} & \begin{minipage}[b]{\linewidth}\raggedright
\includegraphics[width=\linewidth,height=0.32\textheight,keepaspectratio]{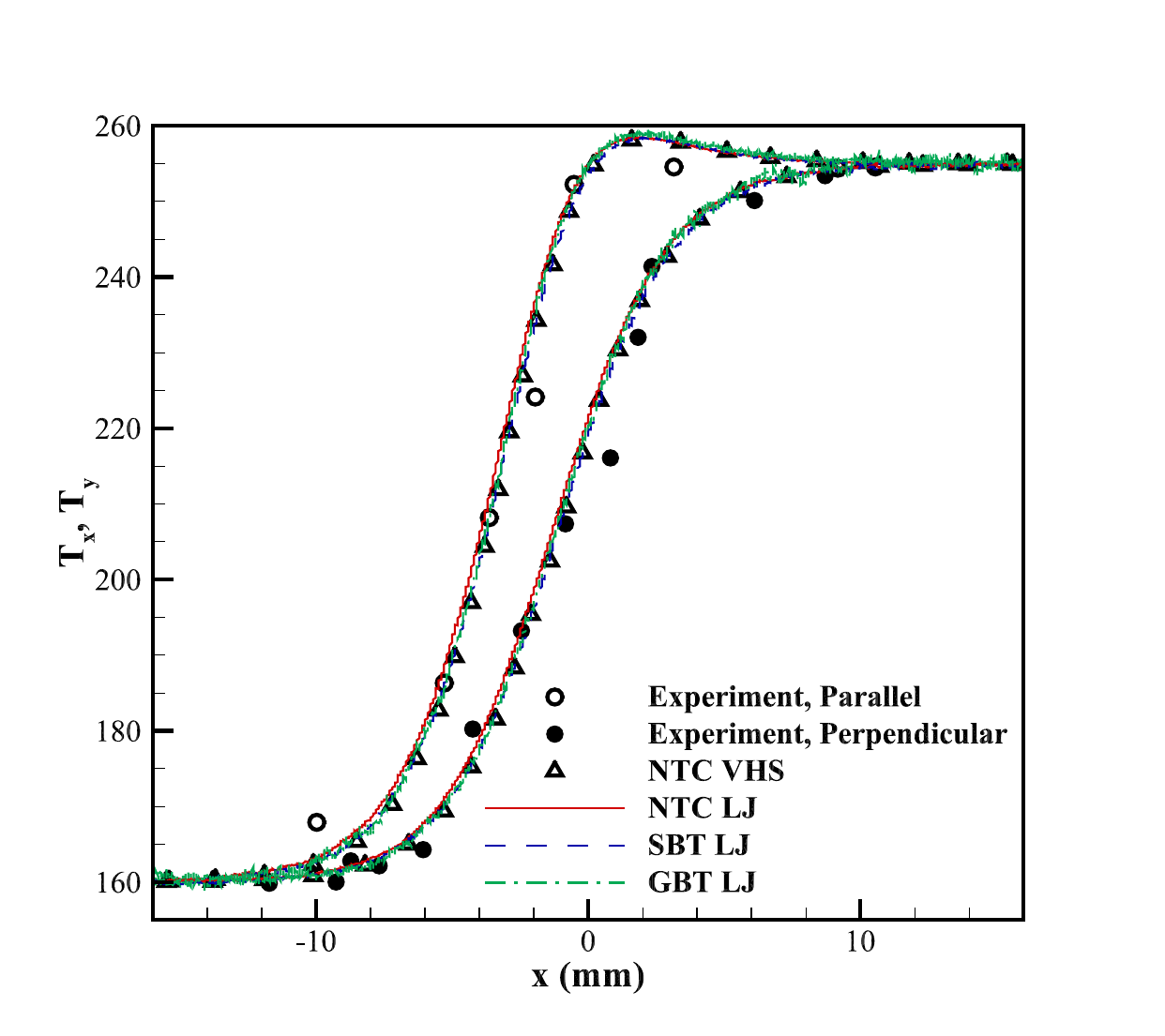}

(b)
\end{minipage} \\

\end{longtable}

\autopapercaption{fig:fig05}{Comparison of the LJ model (variable effective diameter approach) results for Helium with that of the experimental data from Ref.~\cite{muntz1969shockvdf} and VHS simulation: (a) normalized density and (b) temperature in parallel and perpendicular directions.}

In the preceding cases where the local temperature significantly exceeds the potential well depth (\(T \gg \frac{\varepsilon}{k_B}\)), the intermolecular interactions primarily occur within the repulsive regime. This explains the strong agreement observed between the LJ and VHS models, both of which effectively capture the experimental trends for helium. However, for heavier noble gases, such as argon, which have relatively deeper potential wells, the attractive component of the intermolecular force becomes dominant at low temperatures---conditions frequently encountered in hypersonic expansion and shock-wave experiments.

To evaluate the capability of our variable-diameter and deep-learning-based approaches under such challenging conditions, we investigate the normal shock wave of argon at Mach 7.183 and an upstream temperature of 16 K, utilizing the experimental benchmarks provided by Holtz and Muntz \cite{holtz1983argonshock}. For these simulations, the LJ parameters are set to \(\varepsilon/k_B\)=124 K and \emph{d\textsubscript{LJ}}=3.418×10\textsuperscript{-10} m \cite{hirschfelder1948transport}. This specific case provides critical validation of the model's ability to handle the complex attractive-repulsive balance that governs flow physics in high-gradient, low-temperature regimes.

The normalized density profiles for the argon normal shock wave at Mach 7.183 are presented in Fig.~\ref{fig:fig06}, comparing the constant (Panel a) and variable (Panel b) effective molecular diameter approaches. Unlike the helium case, the deeper potential well of argon \(\varepsilon/k_B\)=124 K renders the attractive component of the intermolecular force dominant at the low-temperature upstream condition of 16 K.

As demonstrated in Fig.~\ref{fig:fig06}(a), the conventional VHS model (using \(d_{\mathrm{ref}}=4.17\times10^{-10}\,\mathrm{m}\)) fails to accurately capture the shock layer structure, predicting a steeper gradient that deviates from the experimental data. This underscores the limitation of models that consider only the repulsive force in cryogenic or high-Mach regimes. In contrast, both LJ-based implementations effectively recover the experimental density profile, with the variable effective diameter approach (Fig.~\ref{fig:fig06}(b)) showing remarkable precision in bridging the repulsive and attractive regimes. These results validate the proposed framework's ability to handle challenging non-equilibrium flows in which the intricate balance of molecular forces dictates macroscopic flow properties.

\begin{longtable}[]{@{}
  >{\raggedright\arraybackslash}p{(\columnwidth - 2\tabcolsep) * \real{0.4984}}
  >{\raggedright\arraybackslash}p{(\columnwidth - 2\tabcolsep) * \real{0.5016}}@{}}

\begin{minipage}[b]{\linewidth}\raggedright
\includegraphics[width=\linewidth,height=0.32\textheight,keepaspectratio]{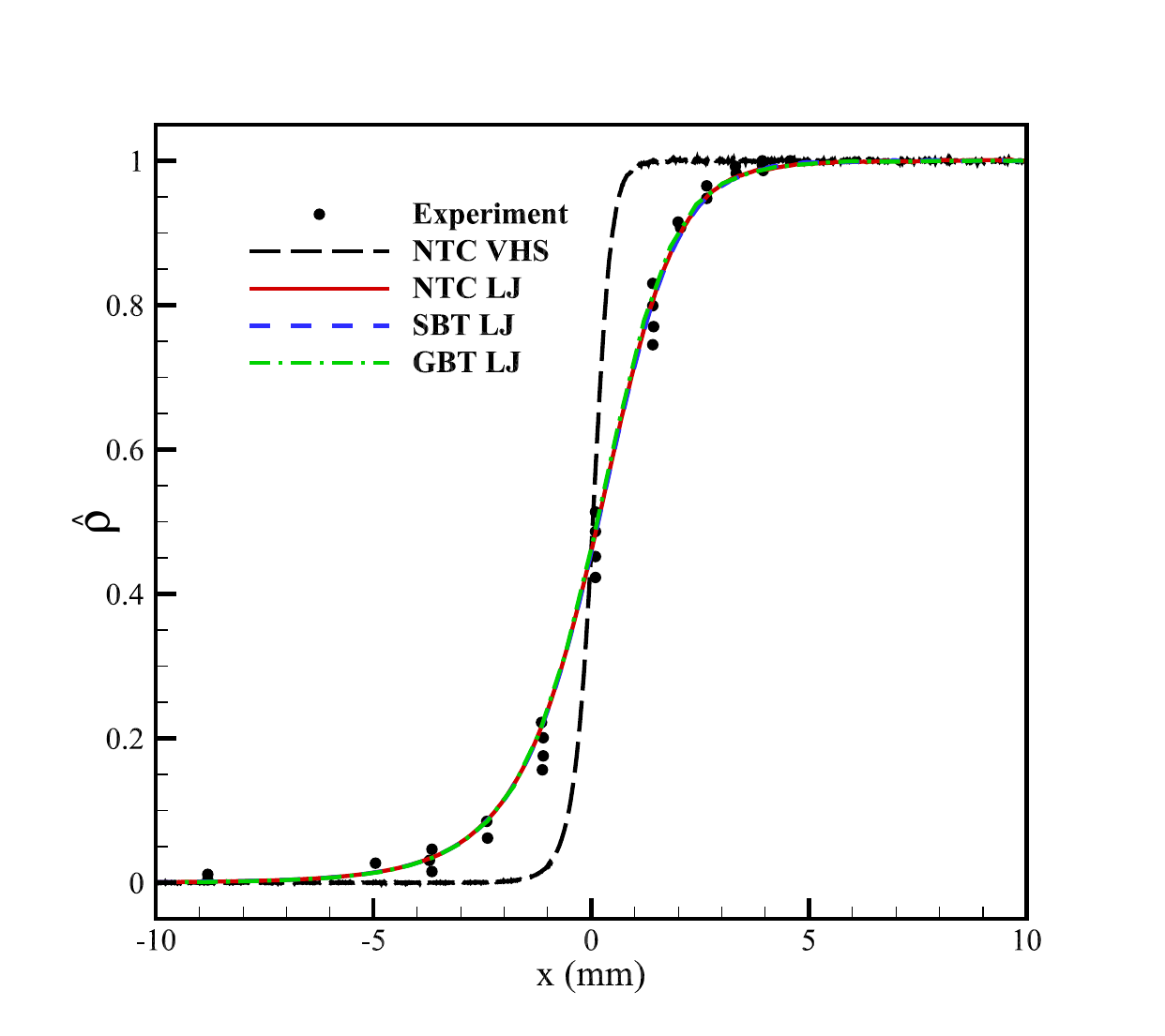}

(a)
\end{minipage} & \begin{minipage}[b]{\linewidth}\raggedright
\includegraphics[width=\linewidth,height=0.32\textheight,keepaspectratio]{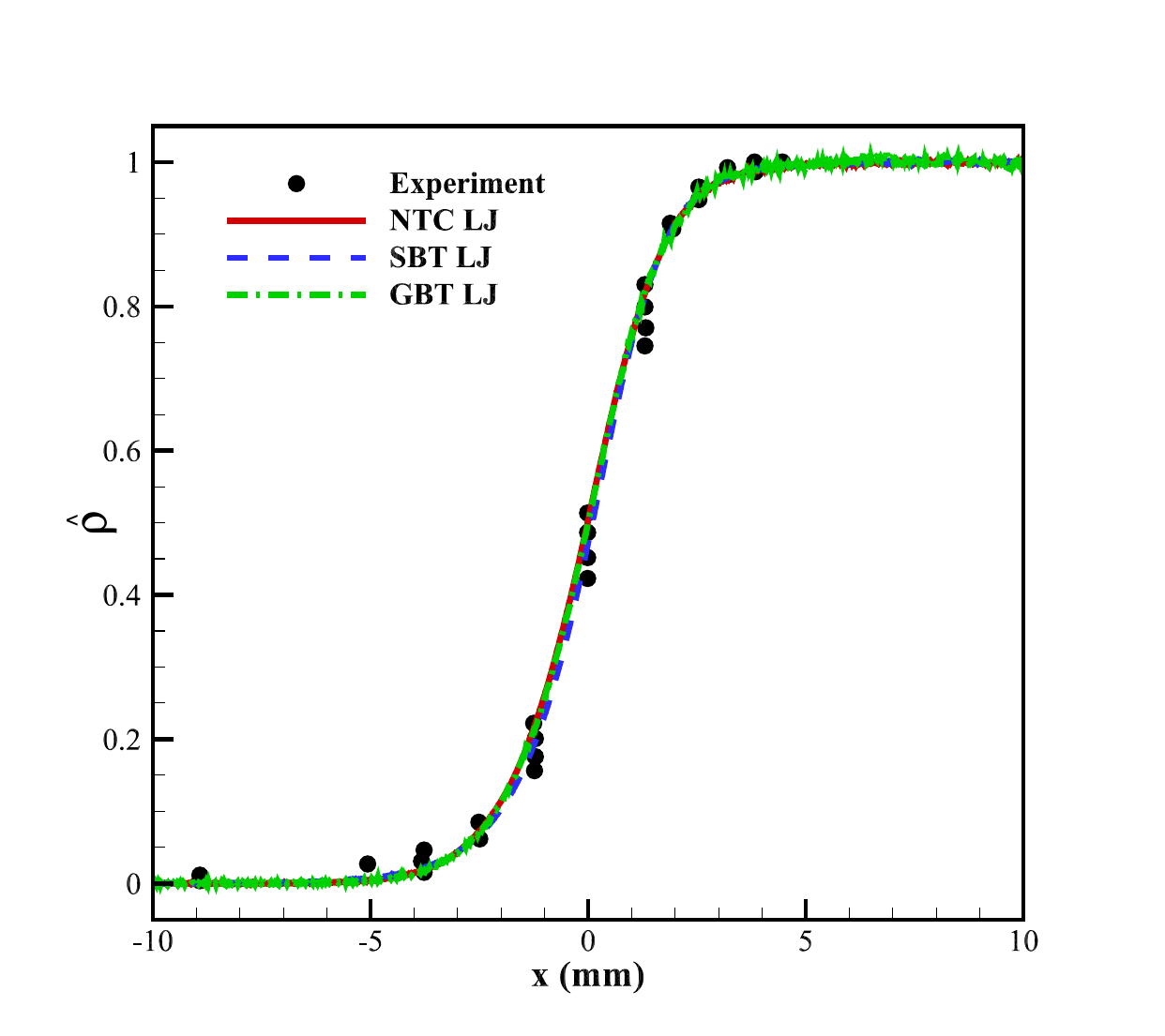}

(b)
\end{minipage} \\

\end{longtable}

\autopapercaption{fig:fig06}{Normalized density profile for argon normal shock wave simulated using the LJ and VHS models compared with the experimental data from Ref.~\cite{holtz1983argonshock}: (a) constant effective molecular diameter and (b) variable effective molecular diameter.}

The Lennard-Jones (LJ) potential is highly effective for modeling gas transport properties over broad temperature ranges. However, the accuracy of DSMC simulations is significantly sensitive to the selection of model parameters. While traditional LJ parameters are widely cited, Weaver and Alexeenko \cite{weaver2015revised} proposed an optimized set based on a comprehensive analysis of shear viscosity and thermal conductivity. In this study, we evaluate our models\textquotesingle{} performance using these distinct parameter sets by comparing simulated helium temperature profiles with experimental data from Refs. \cite{muntz1969shockvdf,erwin1991nonequilibrium}.

Fig.~\ref{fig:fig07} illustrates the normalized temperature profiles, \(\widehat{T} = \frac{T - T_{1}}{T_{2} - T_{1}}\), across the shock wave for various potential well depths and LJ diameters. The results demonstrate that all simulation cases align closely with the experimental benchmarks. A key observation from the close-up view in Fig.~\ref{fig:fig07} is that the constant and variable effective molecular diameters yield nearly identical results for helium, confirming the consistency of the proposed variable-diameter framework. Notably, the simulation employing the parameters suggested by Weaver and Alexeenko \cite{weaver2015revised} (\(\varepsilon/k_B\)=9.8725 K and \emph{d\textsubscript{LJ}}=2.5238×10\textsuperscript{-10} m) exhibits enhanced predictive accuracy, particularly in capturing non-equilibrium behavior in the upstream shock-foot region.

\centeredpaperfigure[width=0.92\linewidth,height=0.50\textheight,keepaspectratio]{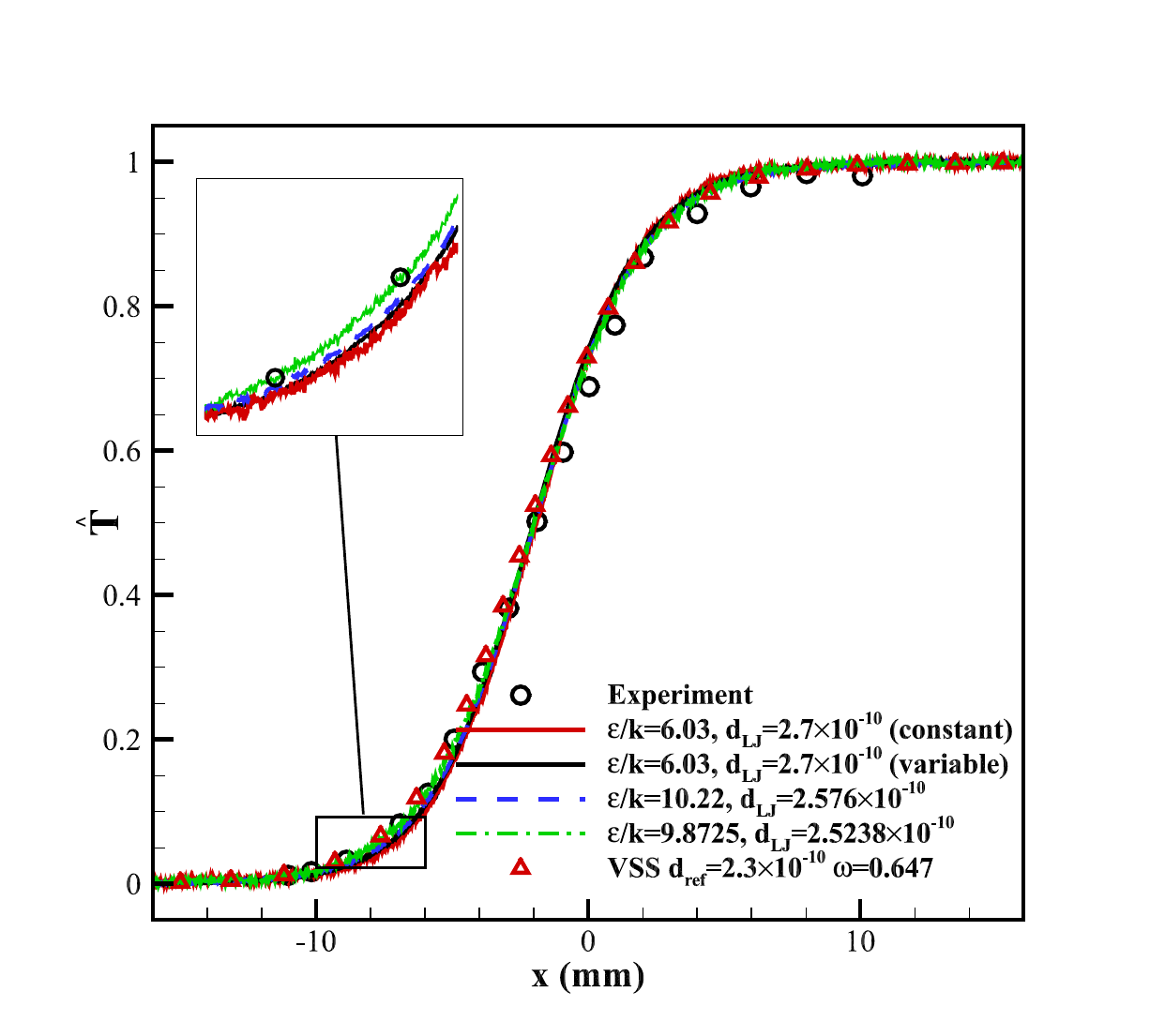}

\autopapercaption{fig:fig07}{Comparison of different sets of LJ parameters: normalized temperature profile for helium normal shock wave simulated using the LJ and VSS models compared with the experimental data from Refs.~\cite{muntz1969shockvdf,erwin1991nonequilibrium}.}

\subsubsection{Internal shock structure: the spatial--kinetic paradox and bimodal non-equilibrium}\label{subsubsec:internal-shock-structure}
To investigate fundamental translational non-equilibrium collision mechanics independent of multidimensional aerodynamic boundary effects, the one-dimensional normal shock wave in argon at Mach 5 was revisited. A profound kinetic paradox emerges when comparing the microscopic velocity distributions with the macroscopic density profile (shown in Fig.~\ref{fig:fig06}a). While the macroscopic density profiles differ markedly between the VHS and LJ models, the microscopic Velocity Distribution Functions (VDFs) sampled at identical internal stations match with remarkable precision (see Fig.~\ref{fig:fig08}).

The spatial discrepancy observed in the density profiles across the physical x-coordinate is a direct consequence of the cryogenic upstream conditions. At 16 K, the attractive tail of the LJ potential profoundly governs the scattering kinematics, drastically inflating the collision cross-section due to grazing and orbiting trajectories. The phenomenological VHS model is purely repulsive and inherently devoid of an attractive tail. Consequently, it artificially underestimates the collision cross-section in the cold upstream region, computing a locally larger mean free path. This discrepancy in spatial transport properties inherently stretches and distorts the macroscopic density gradient of the shock along the physical x-axis compared to the physically complete LJ potential.

To examine the internal microscopic state, the VDFs for the stream-wise (parallel, \emph{V\textsubscript{x}}) and cross-stream (perpendicular, \emph{V\textsubscript{y}}) molecular velocity components were extracted at five distinct internal locations. As shown in the VDF plots in Fig.~\ref{fig:fig08}, the parallel velocity distributions exhibit highly complex, non-Maxwellian behavior with a bimodal (two-peaked) profile. This structural anomaly aligns well with the classical Mott-Smith kinetic theory for strong shock waves and with the observations reported in Refs. \cite{pham1989nonequilibrium,matsumoto1991argonshock}. Because the physical thickness of a hypersonic shock spans merely a few mean free paths, the gas experiences extreme translational non-equilibrium. Within this narrow transition layer, the hyper-velocity, ultra-cold unshocked molecules from the upstream spatially interpenetrate the heavily decelerated, highly thermalized downstream molecules. The lack of sufficient intermolecular collisions prevents these two distinct populations from immediately merging into a single equilibrium Maxwellian distribution. Consequently, they form a bimodal kinetic signature: a sharp, high-speed peak representing the surviving upstream molecules, and a broad, low-speed plateau representing the shock-heated downstream gas. The perpendicular VDFs remain unimodal, simply demonstrating pronounced thermal broadening as directed kinetic energy is violently dissipated into transverse random thermal motion.

Despite the spatial divergence of the macroscopic profiles, the microscopic VDFs of the VHS and LJ models collapse onto each other flawlessly. The resolution to this apparent paradox lies in two fundamental principles:

First, the VDFs were deliberately extracted not at absolute spatial coordinates (x-locations), but at exact matching mentioned in the caption of the figure. Evaluating the VDFs in the \(\widehat{\rho}\)-space elegantly factors out the spatial stretching error induced by the VHS model. The normalized density acts as a thermodynamic progress variable that dictates the exact "mixing ratio" of the upstream and downstream molecular populations. Because both simulations are constrained by the exact same macroscopic Rankine-Hugoniot jump relations, a specific \(\widehat{\rho}\) value mandates an identical proportional blend of fast and slow particles, regardless of the collision model's spatial mapping.

Second, while the upstream gas is cryogenic, the interior of a Mach 5 shock wave is a region of violent, high-speed molecular impacts. The relative kinetic energy \emph{E\textsubscript{K}} of colliding pairs, particularly between the fast freestream and hot downstream molecules, vastly exceeds the LJ potential well depth (\emph{E\textsubscript{K}} $\gg$ \emph{ε}). Under these hyper-energetic impacts, the attractive tail becomes kinematically irrelevant, and molecular scattering is overwhelmingly governed by the steep, short-range Pauli repulsive core. Because the dynamically shrinking diameter of the VHS model implemented in this solver (calibrated via the viscosity-temperature index ω=0.816) is specifically designed to replicate this exact inverse-power-law repulsive behavior at high temperatures, the microscopic scattering mechanics of VHS and LJ become virtually identical within the hot shock core.

Ultimately, this analysis reconfirms a physical duality: while a purely repulsive variable-diameter VHS model fails to accurately predict where in physical space \emph{x} a specific thermodynamic state occurs for an ultra-cold gas, it perfectly resolves the highly non-equilibrium bimodal phase-space transition, yielding internal velocity distributions that phenomenologically harmonize with the Lennard--Jones potential when mapped against the localized thermodynamic density fractions.

\begin{longtable}[]{@{}
  >{\raggedright\arraybackslash}p{(\columnwidth - 2\tabcolsep) * \real{0.5000}}
  >{\raggedright\arraybackslash}p{(\columnwidth - 2\tabcolsep) * \real{0.5000}}@{}}

\begin{minipage}[b]{\linewidth}\raggedright
\includegraphics[width=\linewidth,height=0.32\textheight,keepaspectratio]{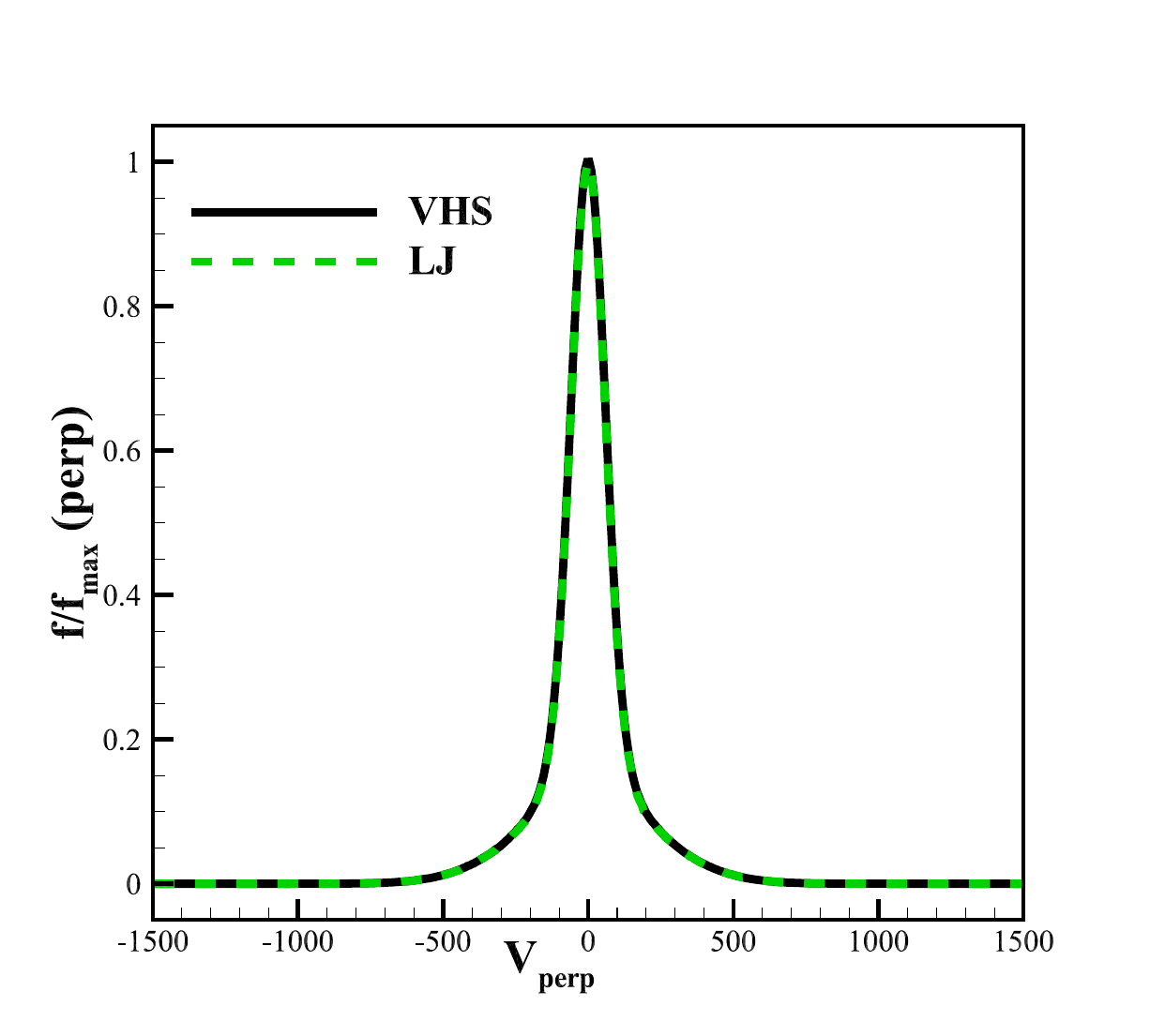}
\end{minipage} & \begin{minipage}[b]{\linewidth}\raggedright
\includegraphics[width=\linewidth,height=0.32\textheight,keepaspectratio]{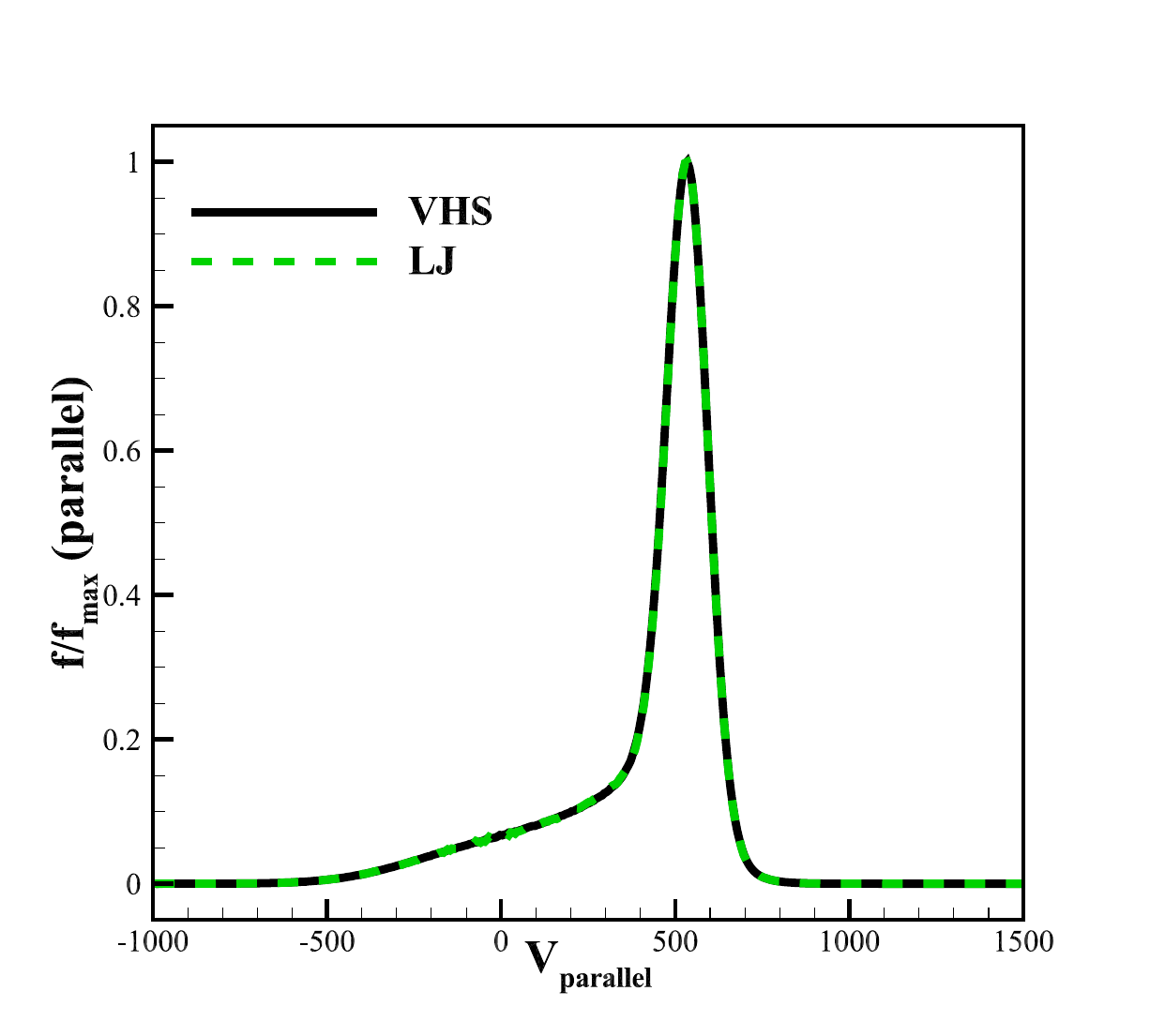}
\end{minipage} \\

\multicolumn{2}{@{}>{\raggedright\arraybackslash}p{(\columnwidth - 2\tabcolsep) * \real{1.0000} + 2\tabcolsep}@{}}{%
\begin{minipage}[t]{\linewidth}\raggedright
\begin{enumerate}
\def\labelenumi{\alph{enumi})}
\item
  \(\widehat{\rho}\)= 0.0993
\end{enumerate}
\end{minipage}} \\
\includegraphics[width=\linewidth,height=0.32\textheight,keepaspectratio]{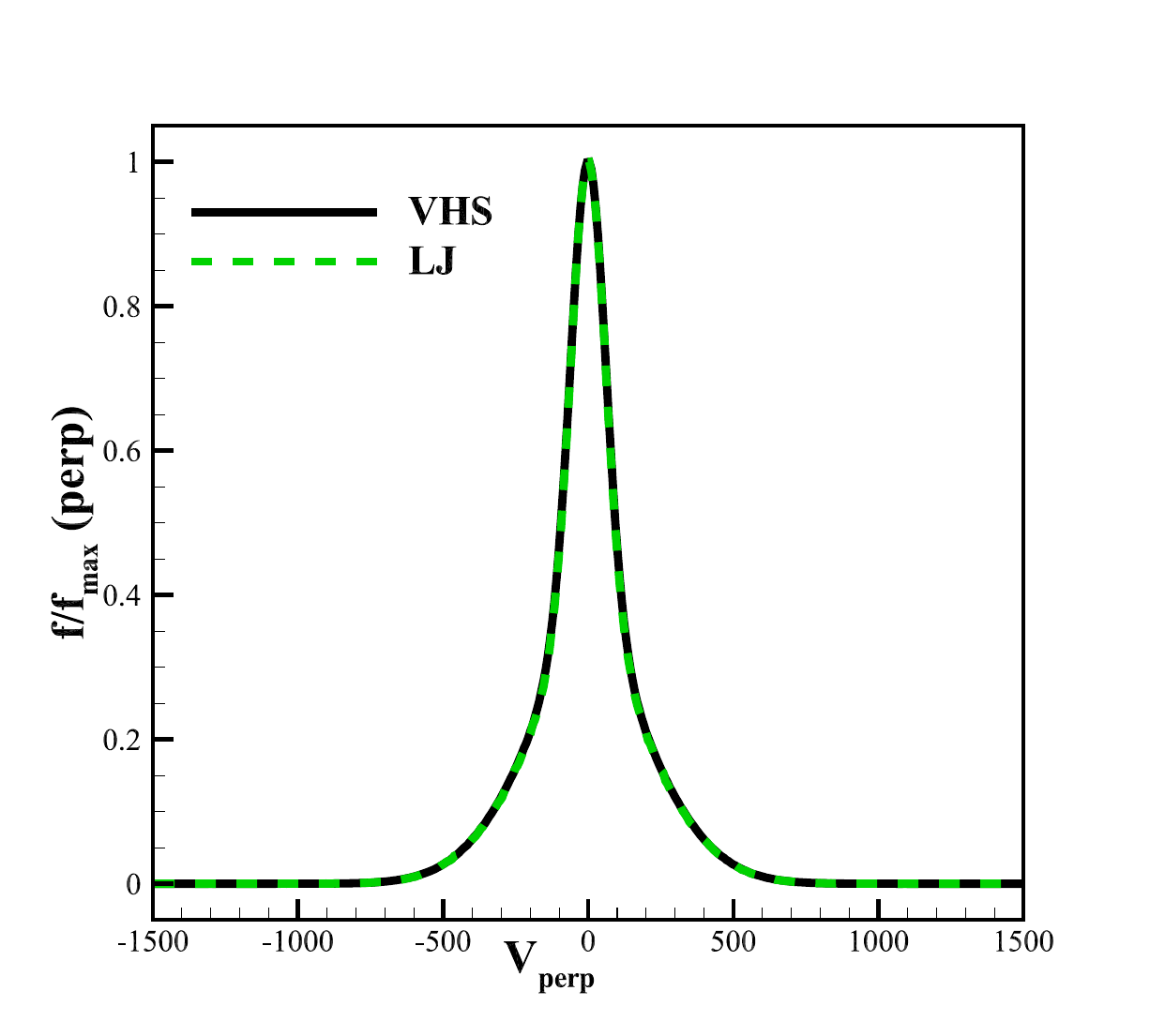} & \includegraphics[width=\linewidth,height=0.32\textheight,keepaspectratio]{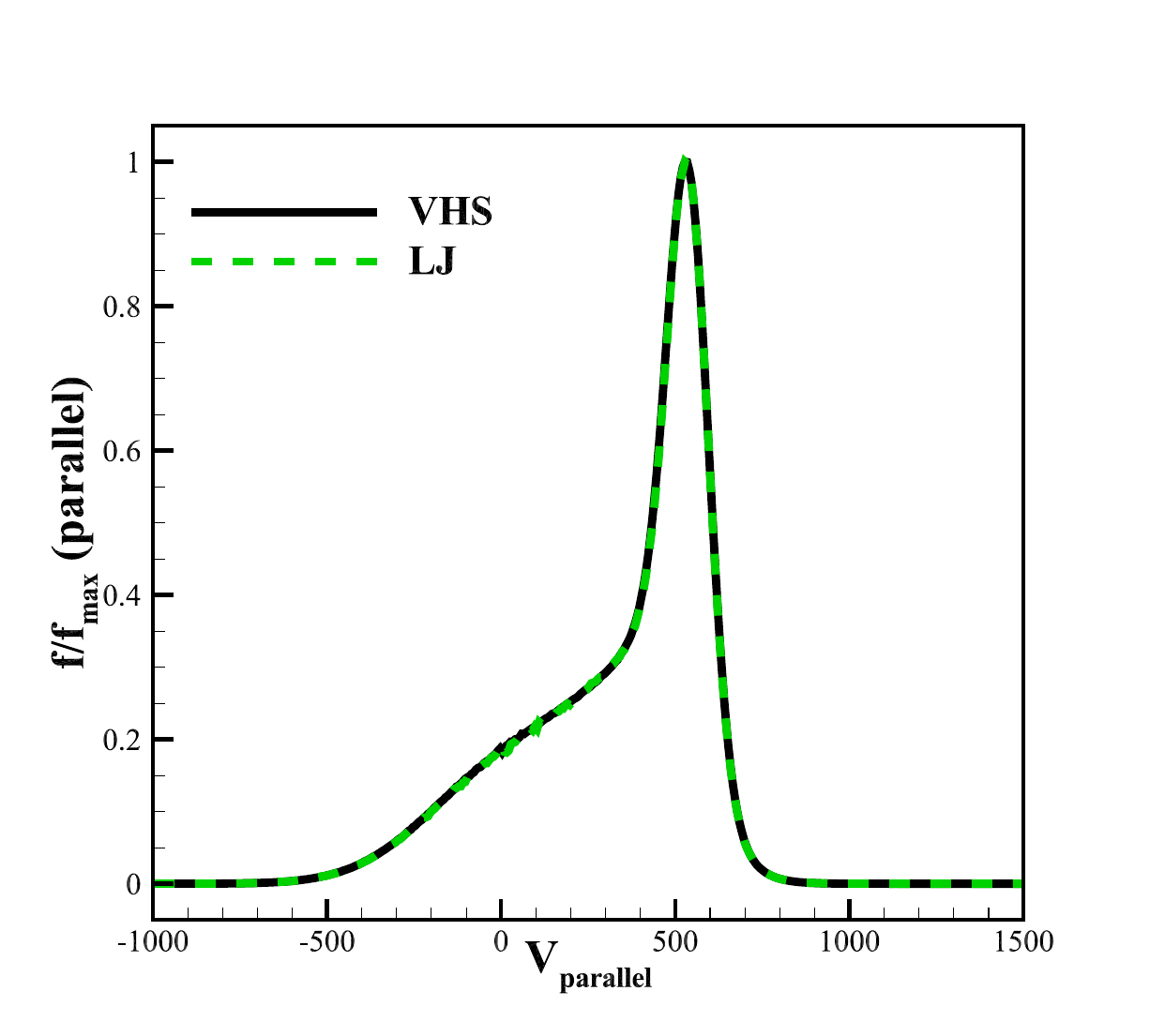} \\
\multicolumn{2}{@{}>{\raggedright\arraybackslash}p{(\columnwidth - 2\tabcolsep) * \real{1.0000} + 2\tabcolsep}@{}}{%
\begin{minipage}[t]{\linewidth}\raggedright
\begin{enumerate}
\def\labelenumi{\alph{enumi})}
\setcounter{enumi}{1}
\item
  \(\widehat{\rho}\) = 0.212
\end{enumerate}
\end{minipage}} \\
\includegraphics[width=\linewidth,height=0.32\textheight,keepaspectratio]{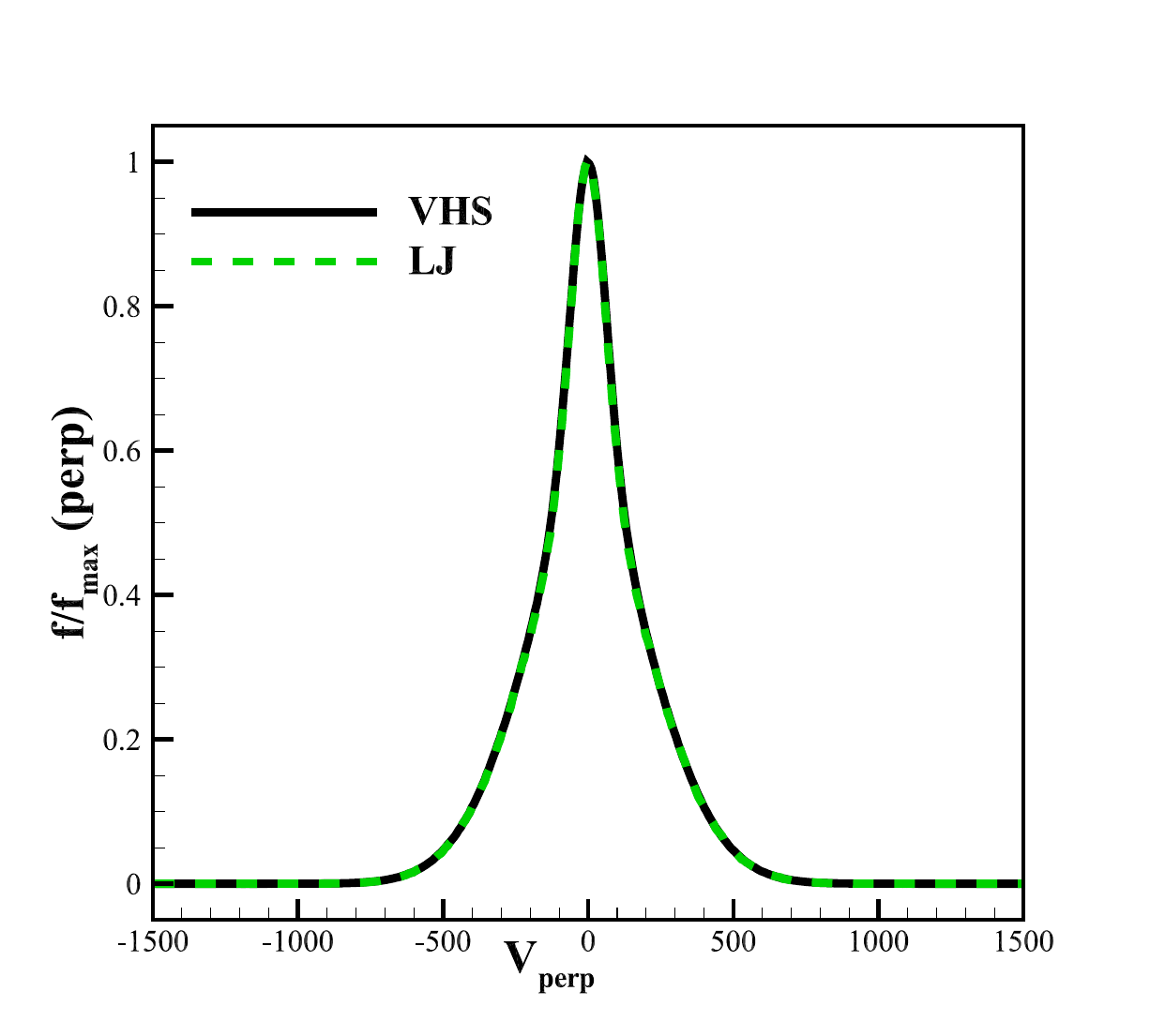} & \includegraphics[width=\linewidth,height=0.32\textheight,keepaspectratio]{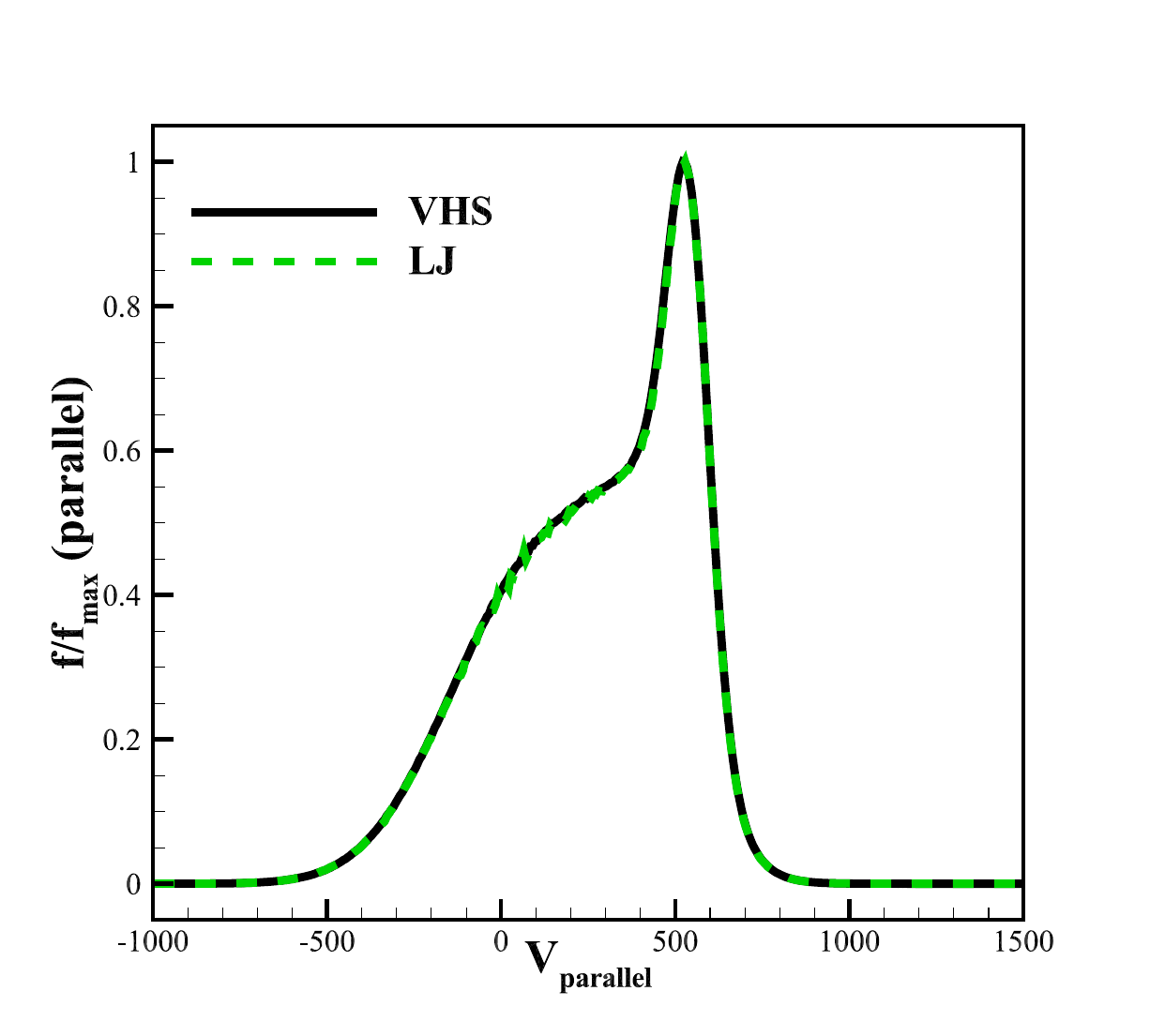} \\
\multicolumn{2}{@{}>{\raggedright\arraybackslash}p{(\columnwidth - 2\tabcolsep) * \real{1.0000} + 2\tabcolsep}@{}}{%
\begin{minipage}[t]{\linewidth}\raggedright
\begin{enumerate}
\def\labelenumi{\alph{enumi})}
\setcounter{enumi}{2}
\item
  \(\widehat{\rho}\) = 0.3568
\end{enumerate}
\end{minipage}} \\
\includegraphics[width=\linewidth,height=0.32\textheight,keepaspectratio]{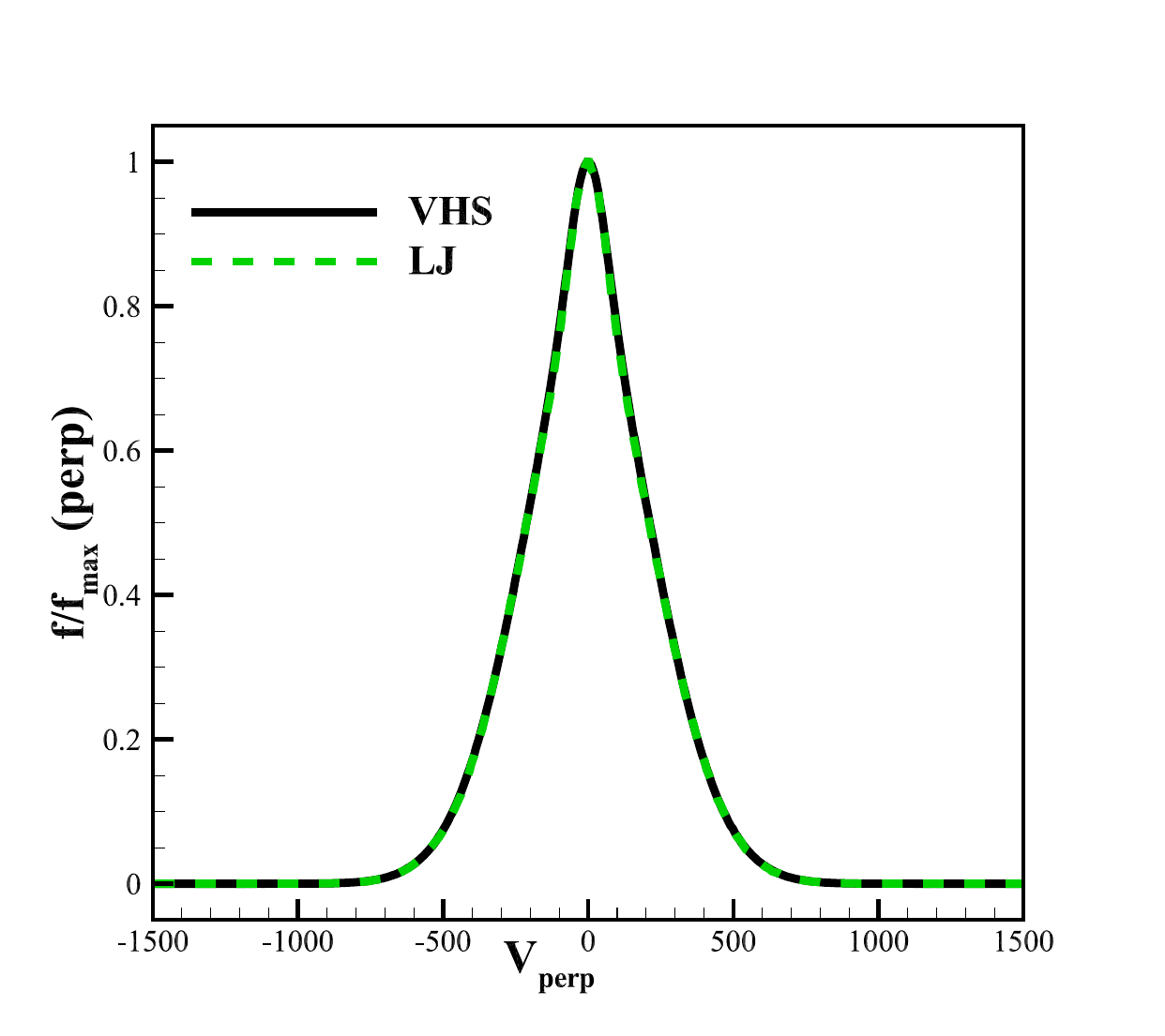} & \includegraphics[width=\linewidth,height=0.32\textheight,keepaspectratio]{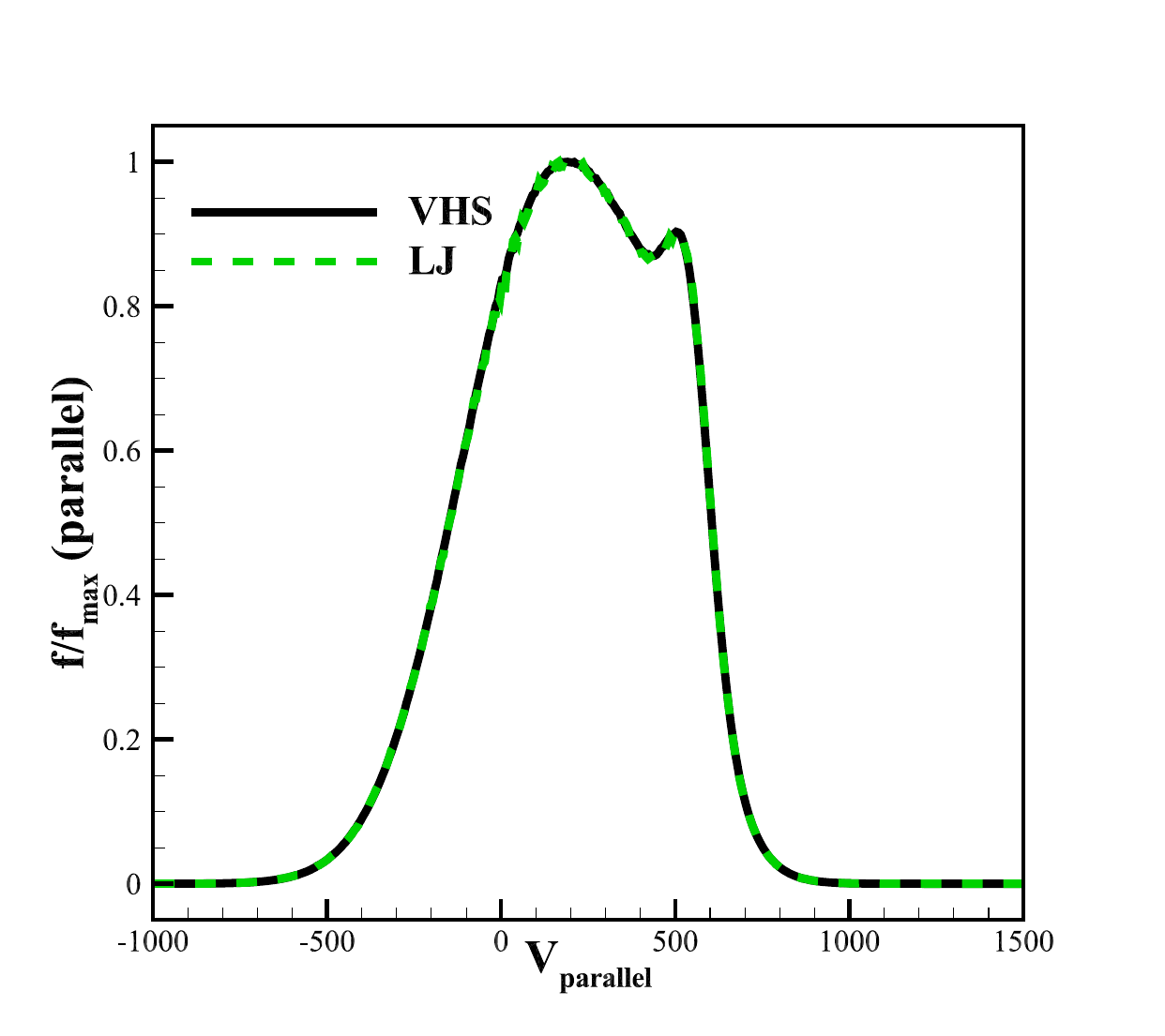} \\
\multicolumn{2}{@{}>{\raggedright\arraybackslash}p{(\columnwidth - 2\tabcolsep) * \real{1.0000} + 2\tabcolsep}@{}}{%
\begin{minipage}[t]{\linewidth}\raggedright
\begin{enumerate}
\def\labelenumi{\alph{enumi})}
\setcounter{enumi}{3}
\item
  \(\widehat{\rho}\ \)= 0.577
\end{enumerate}
\end{minipage}} \\
\includegraphics[width=\linewidth,height=0.32\textheight,keepaspectratio]{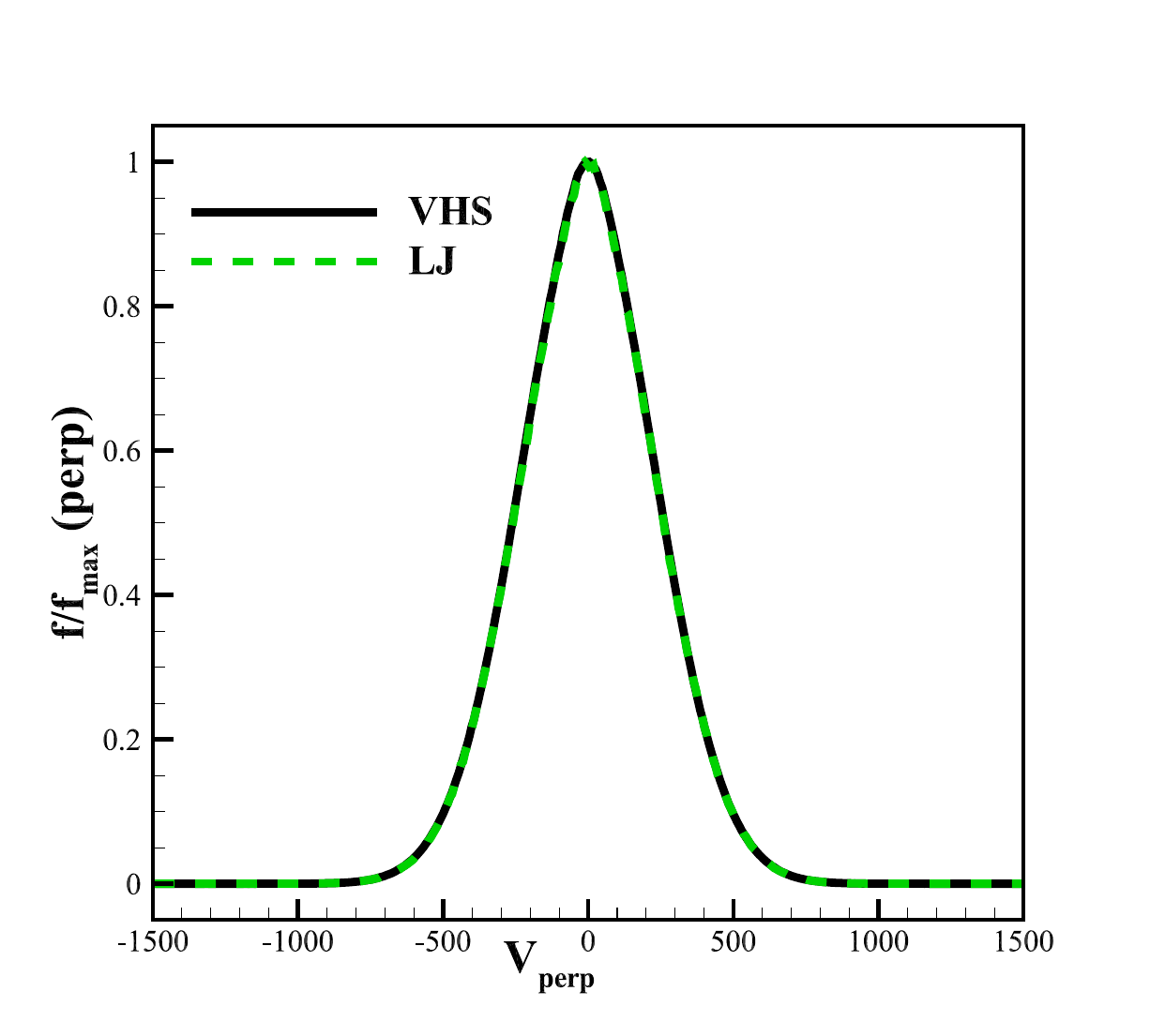} & \includegraphics[width=\linewidth,height=0.32\textheight,keepaspectratio]{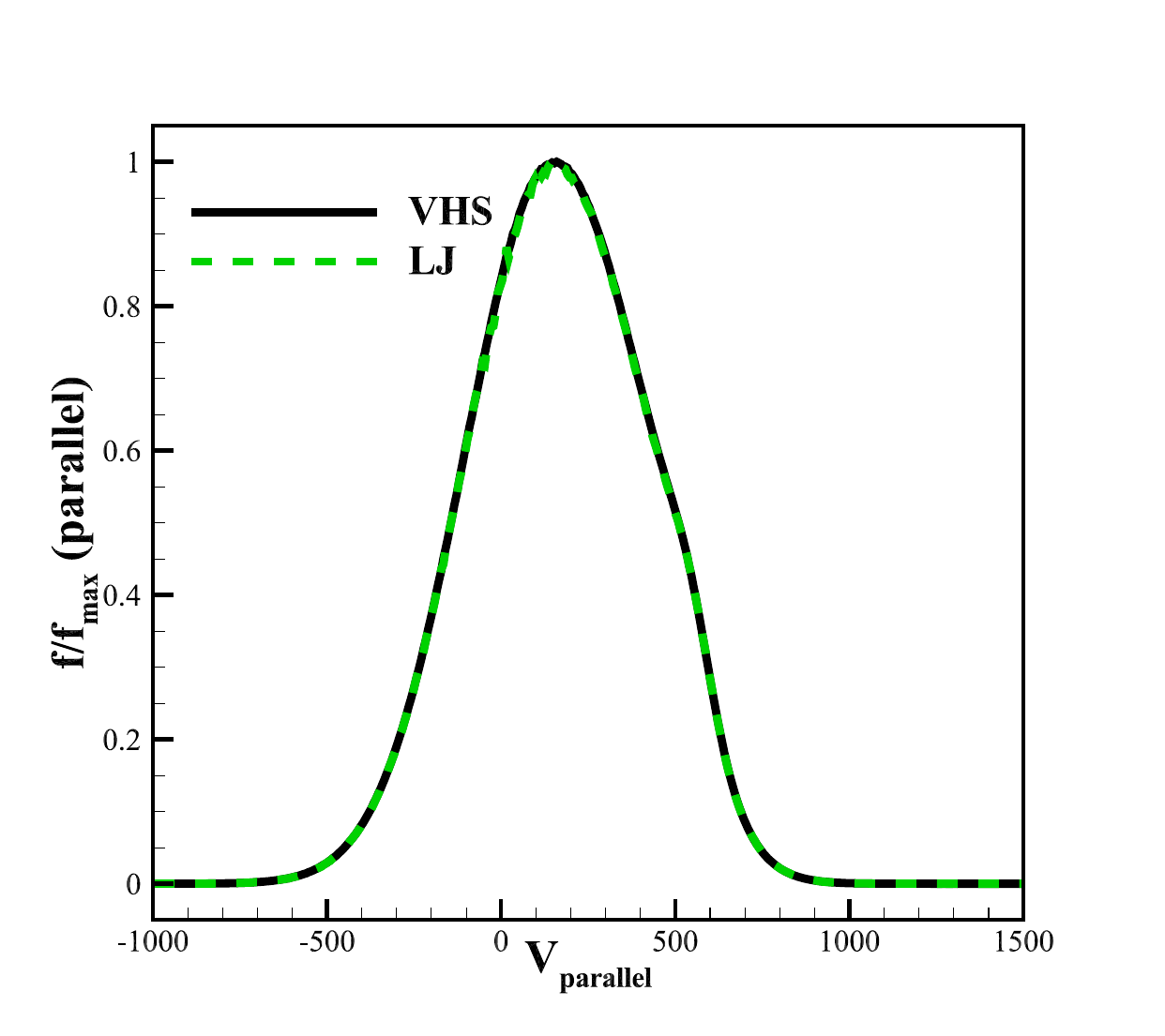} \\
\multicolumn{2}{@{}>{\raggedright\arraybackslash}p{(\columnwidth - 2\tabcolsep) * \real{1.0000} + 2\tabcolsep}@{}}{%
\begin{minipage}[t]{\linewidth}\raggedright
\begin{enumerate}
\def\labelenumi{\alph{enumi})}
\setcounter{enumi}{4}
\item
  \(\widehat{\rho}\) = 0.7916
\end{enumerate}
\end{minipage}} \\
\end{longtable}

\autopapercaption{fig:fig08}{Perpendicular and parallel velocity distribution functions at various normalized density locations for argon shock wave.}

\subsubsection{Machine-learning assessment for normal shock waves}\label{subsubsec:ml-assessment-shocks}
Following the physical validation of the LJ-DSMC framework, this section evaluates the ability of the deep learning-based surrogate models to reproduce the exact shock structures of helium and argon. The primary objective is to verify whether the trained DeepONet architecture can replicate the high-fidelity results presented in Section~\ref{subsubsec:physical-validation-shocks} with a precision indistinguishable from that of the exact numerical implementation.

\begin{figure}[!htbp]
\centering

\begin{subfigure}[t]{0.48\linewidth}
\centering
\includegraphics[width=\linewidth,height=0.28\textheight,keepaspectratio]{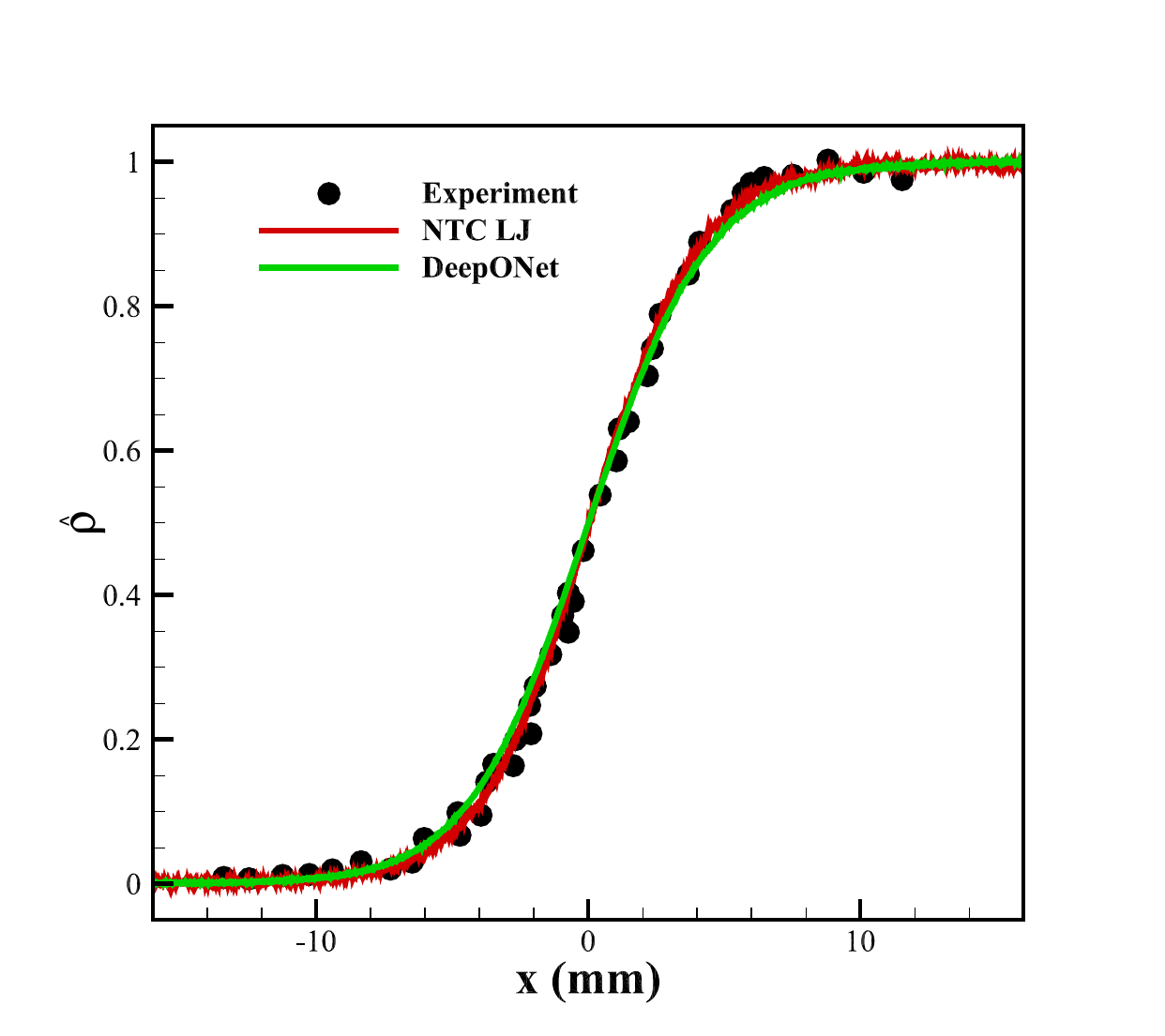}
\caption{}
\label{fig:fig09a}
\end{subfigure}
\hfill
\begin{subfigure}[t]{0.48\linewidth}
\centering
\includegraphics[width=\linewidth,height=0.28\textheight,keepaspectratio]{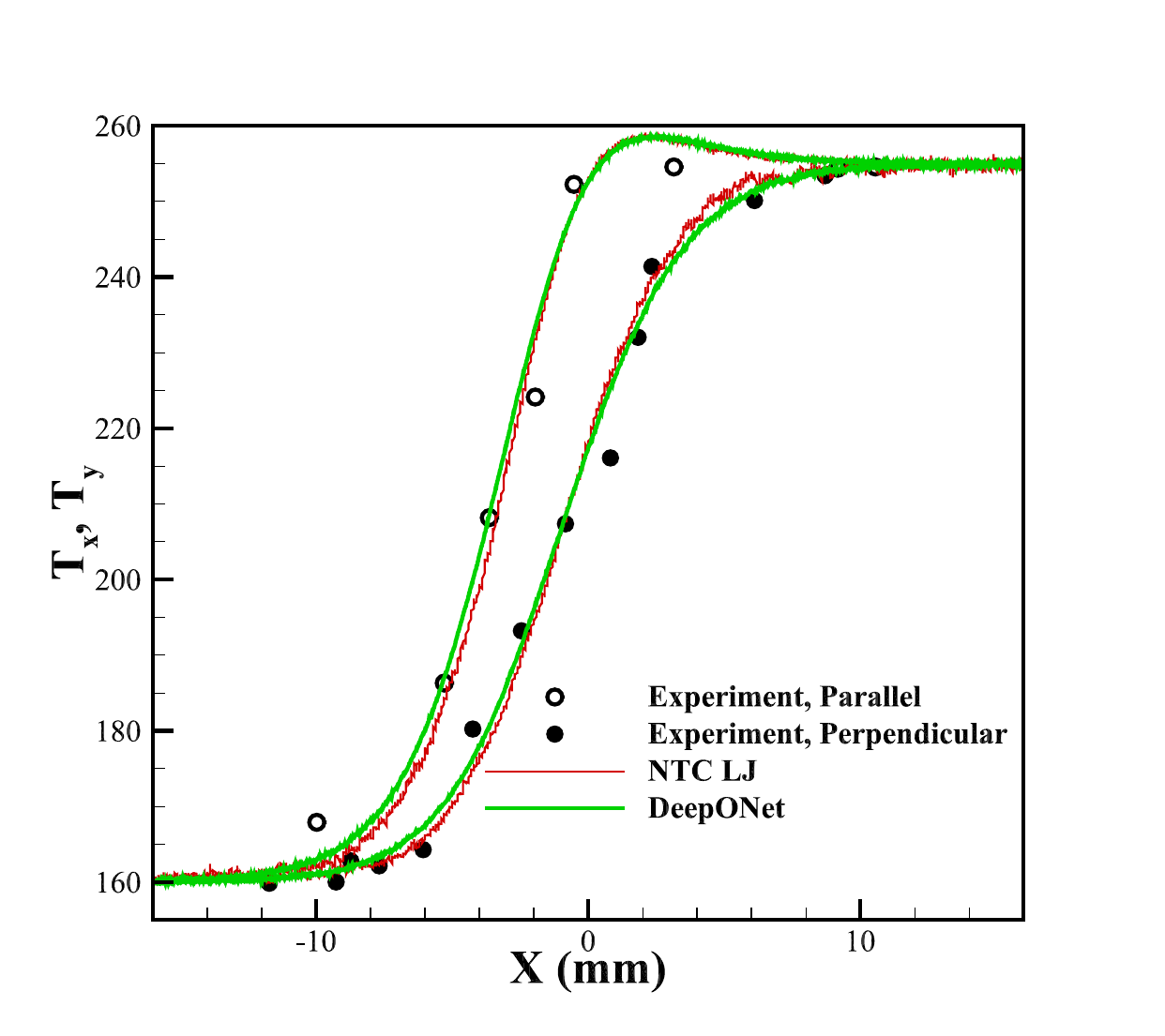}
\caption{}
\label{fig:fig09b}
\end{subfigure}

\vspace{0.4em}

\begin{subfigure}[t]{0.62\linewidth}
\centering
\includegraphics[width=\linewidth,height=0.28\textheight,keepaspectratio]{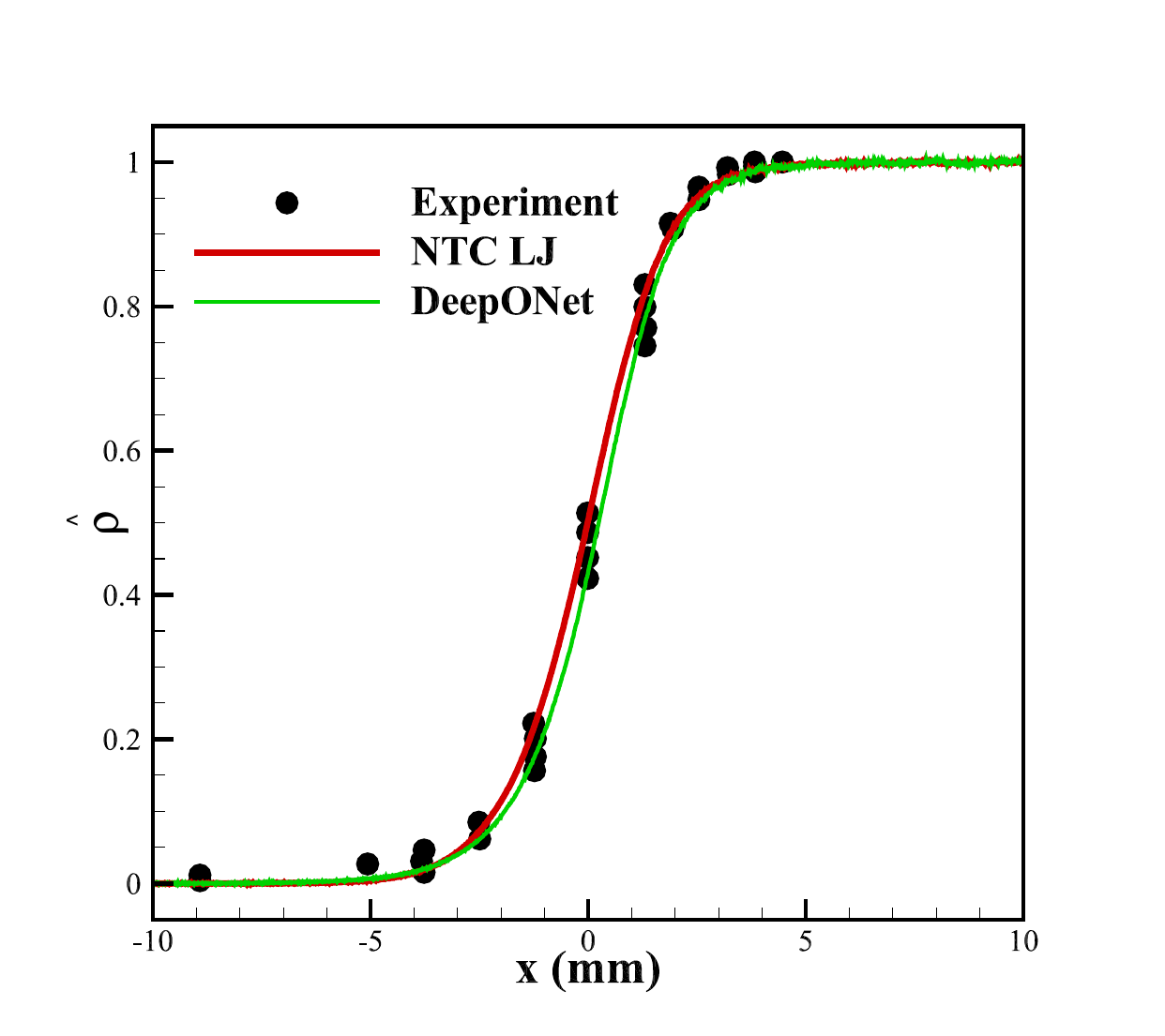}
\caption{}
\label{fig:fig09c}
\end{subfigure}

\caption{Comparison of the ML results with the experimental data and LJ simulation: (a) normalized density and (b) parallel and perpendicular temperature components for helium, and (c) normalized density for argon.}
\label{fig:fig09}
\end{figure}

The predictive accuracy of the trained DeepONet model is further scrutinized by simulating the internal structure of normal shock waves for both helium and argon. Fig.~\ref{fig:fig09} presents a comprehensive comparison between the machine learning-based results and experimental benchmarks \cite{muntz1969shockvdf,holtz1983argonshock}, and the exact LJ numerical solutions. Specifically, Fig.~\ref{fig:fig09}(a) illustrates the normalized density profile for helium at Mach 1.59. The DeepONet results demonstrate an excellent agreement with both the experimental data \cite{muntz1969shockvdf} and the exact LJ implementation, confirming the model's precision in capturing the macroscopic flow gradients. To evaluate the model's fidelity in resolving thermal non-equilibrium, Fig.~\ref{fig:fig09}(b) depicts the parallel and perpendicular temperature components across the helium shock wave. The neural operator successfully reproduces directional temperature profiles, matching experimental trends and high-fidelity LJ results with remarkable accuracy.

Furthermore, the framework's robustness is tested with argon, a gas with significantly stronger attractive interactions. Fig.~\ref{fig:fig09}(c) shows the normalized density profile for the argon shock wave at Mach 7.18. Despite the challenging low-temperature and high-gradient conditions, the DeepONet surrogate yields results that are virtually identical to the exact LJ-NTC solutions and align perfectly with the experimental data \cite{holtz1983argonshock}. Note that the VHS results are omitted from these comparisons, as their inability to accurately capture these specific shock structures was established in Section~\ref{subsubsec:physical-validation-shocks}. The consistent performance across both species and varying flow properties confirms that the deep learning approach preserves the essential molecular physics of the Lennard--Jones potential while offering a computationally efficient alternative to DSMC simulations.

To provide a deeper insight into the microscopic fidelity of the surrogate model, the probability density function (PDF) of the deflection angles (χ) is examined. Fig.~\ref{fig:fig10} illustrates the comparison of \emph{χ} (ranging from 0 to π radians) between the exact LJ potential and the DeepONet model for both helium and argon. The distribution curves for both gases overlap perfectly, confirming that the deflection angle exhibits universal behavior independent of the gas species when governed by the same potential model. The DeepONet surrogate successfully captures this fundamental scattering characteristic, ensuring that the macroscopic accuracy observed in the shock profiles is rooted in a rigorous representation of the inter-particle collision dynamics.

\centeredpaperfigure[width=0.92\linewidth,height=0.50\textheight,keepaspectratio]{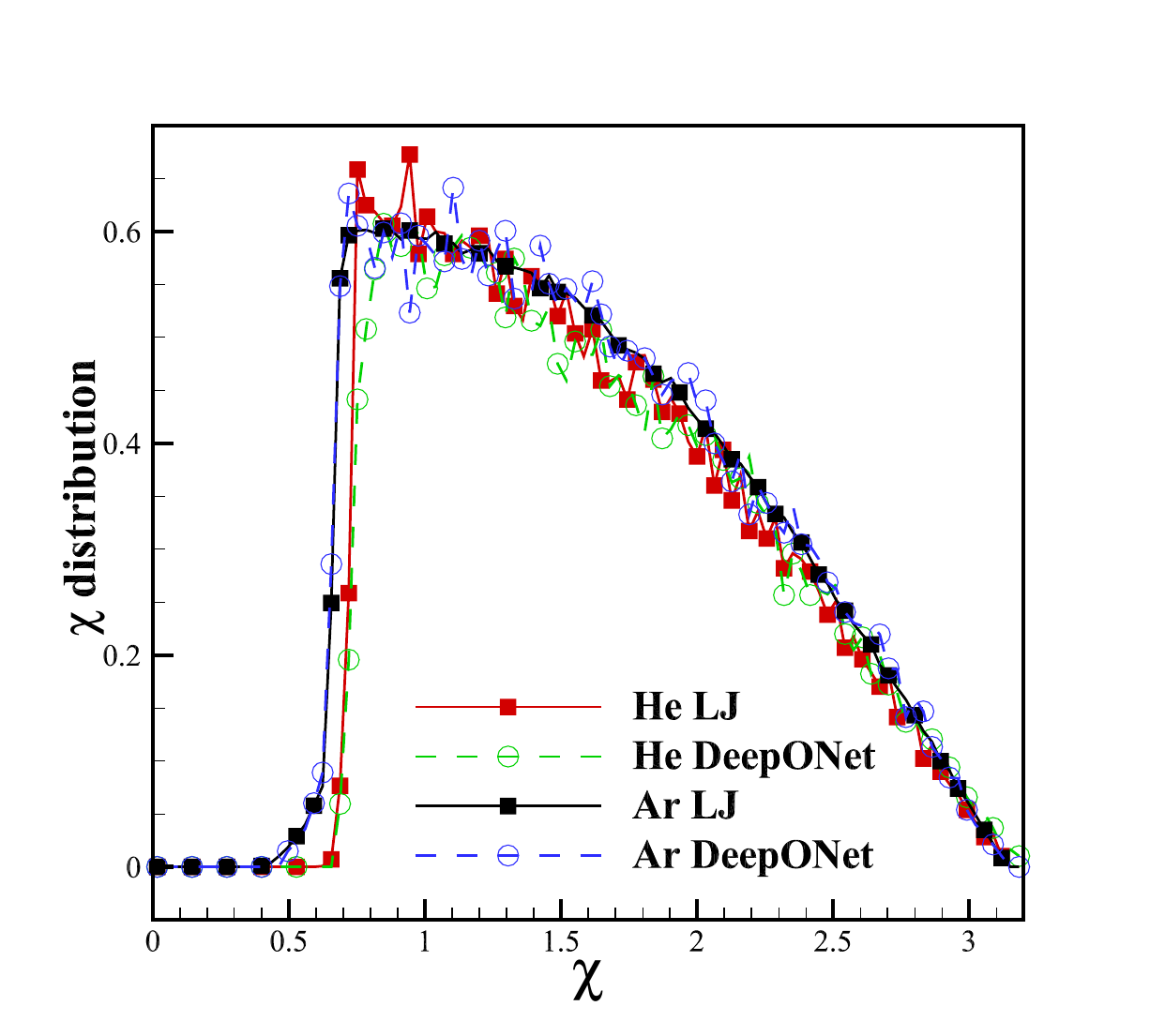}

\autopapercaption{fig:fig10}{Comparison of the deflection angle (\emph{χ}) probability distribution for helium and argon obtained from the exact LJ potential and the DeepONet architecture.}

Finally, the computational advantage of integrating deep learning surrogates is quantified. Table~\ref{tab:computational-performance-argon} summarizes the execution time for 2,000 iterations of the argon shock wave simulation, measured on a standard CPU. The results indicate that replacing the intensive numerical integration of the LJ scattering kernel with the DeepONet surrogate reduces total simulation time by 31\%. More specifically, the profiling analysis reveals that the collision subroutine itself is accelerated by 40\%. These performance gains demonstrate that the proposed framework not only preserves the rigorous physics of the Lennard--Jones potential but also significantly enhances the computational feasibility of DSMC for complex flow regimes.

\setcounter{table}{2}
\begin{table}[!htbp]
\centering
\caption{Comparison of computational performance between the exact LJ--DSMC and the DeepONet-accelerated framework for the argon shock-wave simulation over 2000 iterations.}
\label{tab:computational-performance-argon}
\begin{tabularx}{\linewidth}{lccc}
\toprule
\textbf{Model} &
\textbf{Normalized total time} &
\textbf{Collision time fraction (\%)} &
\textbf{Normalized collision time} \\
\midrule
Exact LJ--DSMC & 1.00 & 57.50 & 1.00 \\
DeepONet--LJ--DSMC & 0.69 & 34.76 & 0.60 \\
\bottomrule
\end{tabularx}
\end{table}

\subsection{Supersonic Couette flow of argon}\label{subsec:supersonic-couette-argon}
The second case study focuses on the compressible supersonic Couette flow of argon. This configuration is selected to evaluate the performance of the developed LJ-DSMC framework in shear-driven flows, where the transport properties, specifically viscosity, are the primary determinants of the flow field. Given that the gas in the vicinity of the walls operates at a cryogenic temperature (\emph{T\textsubscript{w}} = 40 K), the influence of the attractive intermolecular forces in the Lennard--Jones potential is expected to be more pronounced compared to the repulsive-only VHS model.

In this configuration, the argon gas is confined between two parallel plates with the wall temperature maintained at 40 K. The initial gas number density is set to 1.4×10\textsuperscript{20} \RL{}m\textsuperscript{-3}. The upper wall moves at a constant velocity of 300 \RL{}m/s, which corresponds to a Mach number of 2.55 and a global Knudsen number of 0.0051. The argon species is modeled with a molecular mass of 6.63×10\textsuperscript{-26} kg, a viscosity-temperature index of ω=0.816, and a VSS scattering parameter of α=1 at the reference temperature. For the numerical discretization, the domain is divided into 500 uniform computational cells without subcells, with an initial population of 20 particles per cell to ensure statistical stability. These specific flow and numerical parameters are summarized in Table~\ref{tab:couette-flow-conditions}. This setup enables a rigorous comparison between the VHS model and the LJ potential, particularly in capturing viscous dissipation and the resulting non-isothermal profiles at low temperatures.

\setcounter{table}{3}
\begin{table}[!htbp]
\centering
\caption{Summary of flow conditions used for the supersonic Couette-flow problems.}
\label{tab:couette-flow-conditions}
\begin{tabular}{lc}
\toprule
\textbf{Parameter} & \textbf{Value} \\
\midrule
Wall temperature, \(T_{\mathrm{wall}}\) \((\mathrm{K})\) & 40 \\
Initial number density, \(n\) \((\mathrm{m}^{-3})\) & \(1.4\times10^{20}\) \\
Moving-wall velocity, \(v_{\mathrm{wall}}\) \((\mathrm{m\,s}^{-1})\) & 300 \\
Mach number, \(Ma\) & 2.55 \\
Knudsen number, \(Kn\) & 0.0051 \\
\bottomrule
\end{tabular}
\end{table}

\subsubsection{Physical analysis: rarefied shear-flow structures}\label{subsubsec:couette-physical-analysis}
The macroscopic flow properties of supersonic argon Couette flow are illustrated in Fig.~\ref{fig:fig11}, highlighting the impact of the intermolecular potential on transport phenomena in the cryogenic regime. In the central region of the channel, the VHS potential predicts a very slightly higher temperature peak (\emph{T/T\textsubscript{wall} $\approx$ 1.365}) but almost identical density profile. The velocity profiles of both models were identical and are not shown here. The slightly higher temperature of the VHS model is due to its higher viscosity, which is almost unphysical in low-temperature regimes. Thus, VHS predicts more viscous heating. The profile of shear stress distribution shown in Fig.~\ref{fig:fig11}(c) confirms higher shear stress of the VHS. The same trend appears in the hypersonic-cylinder simulations discussed in Section~\ref{subsec:hypersonic-cylinder}.

\begin{longtable}[]{@{}
  >{\raggedright\arraybackslash}p{(\columnwidth - 2\tabcolsep) * \real{0.5003}}
  >{\raggedright\arraybackslash}p{(\columnwidth - 2\tabcolsep) * \real{0.4997}}@{}}

\begin{minipage}[b]{\linewidth}\raggedright
\includegraphics[width=\linewidth,height=0.32\textheight,keepaspectratio]{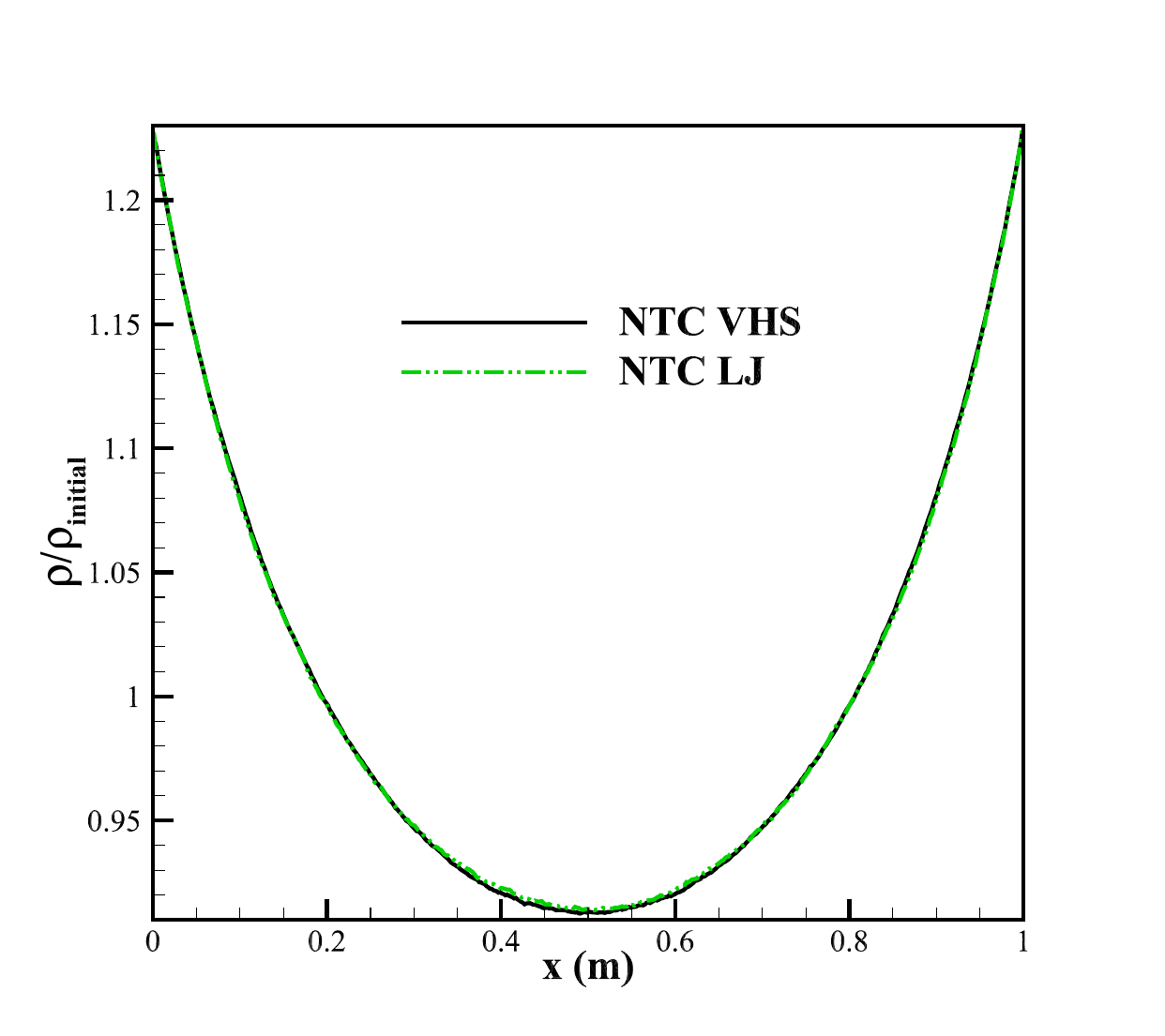}

(a)
\end{minipage} & \begin{minipage}[b]{\linewidth}\raggedright
\includegraphics[width=\linewidth,height=0.32\textheight,keepaspectratio]{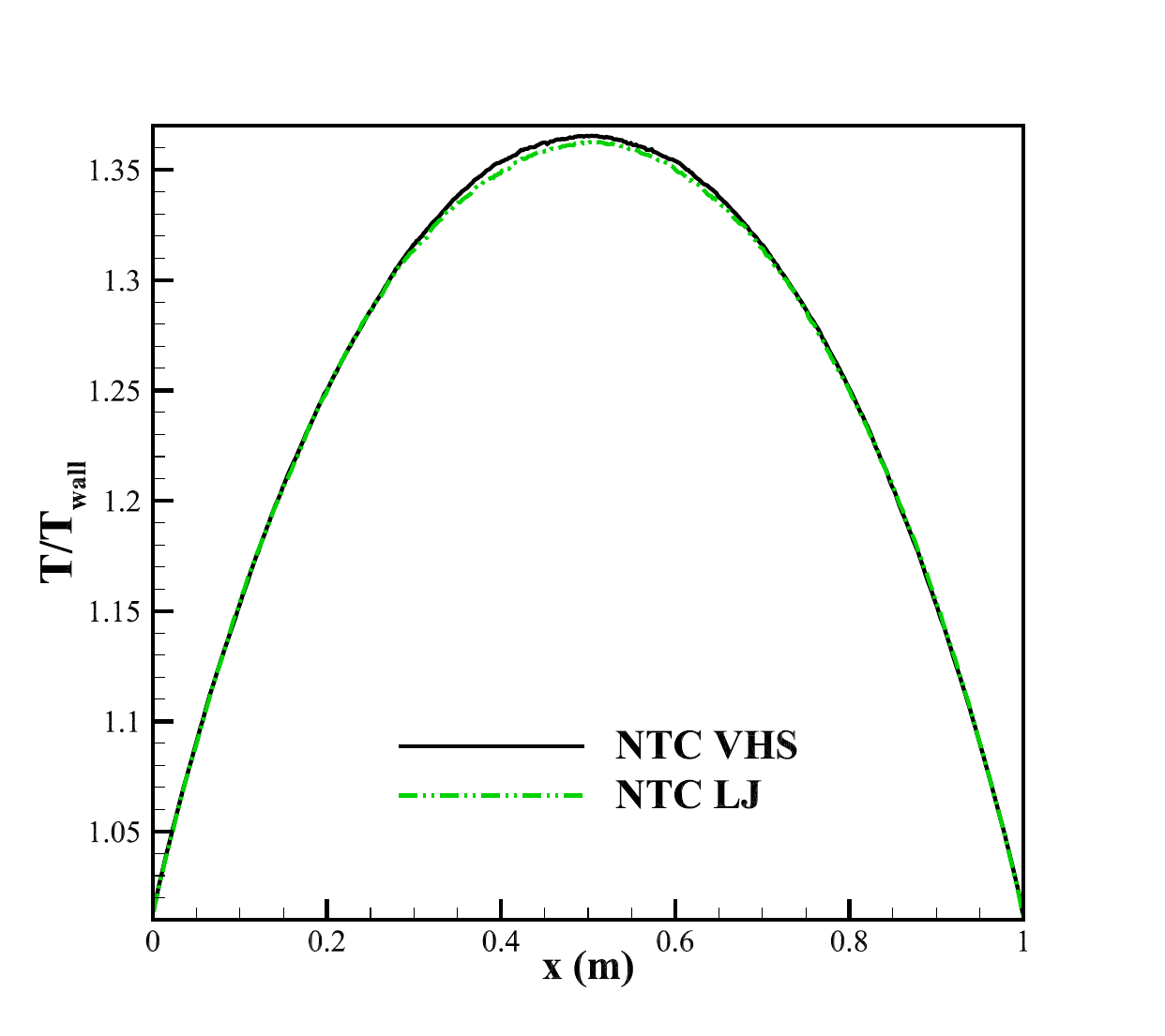}

(b)
\end{minipage} \\

\multicolumn{2}{@{}>{\raggedright\arraybackslash}p{(\columnwidth - 2\tabcolsep) * \real{1.0000} + 2\tabcolsep}@{}}{%
\includegraphics[width=\linewidth,height=0.32\textheight,keepaspectratio]{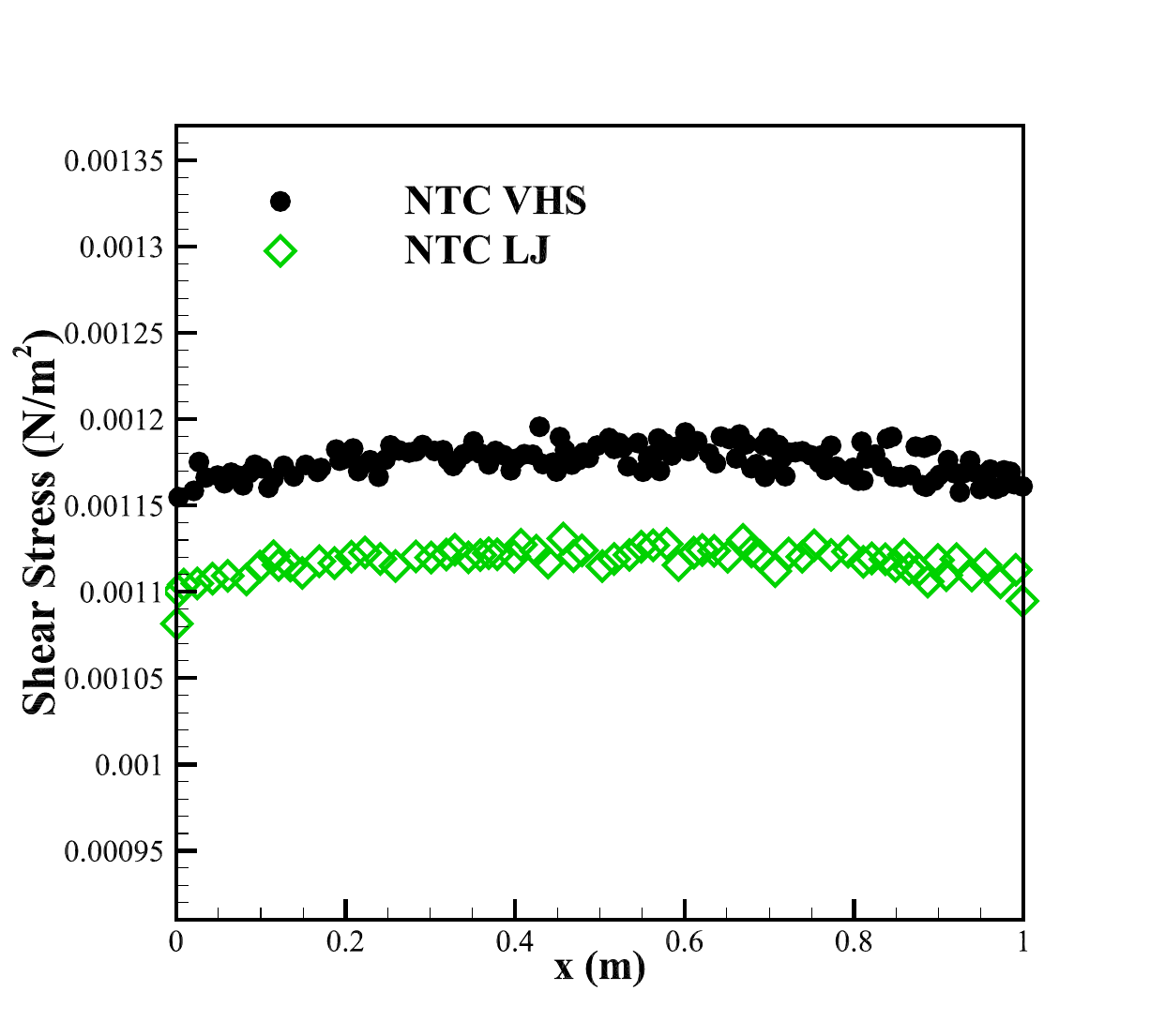}

(c)} \\
\end{longtable}

\autopapercaption{fig:fig11}{Comparison of flow properties obtained using VHS and LJ model and various collision partner selection schemes for supersonic Couette flow of Argon: (a) normalized density, (b) normalized temperature variation, and (c) shear stress.}

\subsubsection{Evaluation of deep-learning surrogates in shear-driven flows}\label{subsubsec:couette-dl-surrogates}
The predictive performance of the trained DeepONet surrogate is further scrutinized by simulating the supersonic Couette flow considered in the previous section. Fig.~\ref{fig:fig12} compares machine learning predictions trained on LJ and LJ-NTC solutions for density and temperature profiles.

As illustrated in Figs.~\ref{fig:fig12}(a) and \ref{fig:fig12}(b), the neural operator demonstrates exceptional fidelity in capturing the non-intuitive density-temperature coupling. The DeepONet effectively reproduces the density surplus in the channel center (\emph{ρ/ρ\textsubscript{initial}} $\approx 0.918$) and the corresponding temperature peak (\emph{T/T\textsubscript{wall}} $\approx 1.365$). These results confirm that the ML model has successfully internalized the complex scattering dynamics of the Lennard--Jones potential.

\begin{longtable}[]{@{}
  >{\raggedright\arraybackslash}p{(\columnwidth - 2\tabcolsep) * \real{0.5136}}
  >{\raggedright\arraybackslash}p{(\columnwidth - 2\tabcolsep) * \real{0.4864}}@{}}

\begin{minipage}[b]{\linewidth}\raggedright
\includegraphics[width=\linewidth,height=0.32\textheight,keepaspectratio]{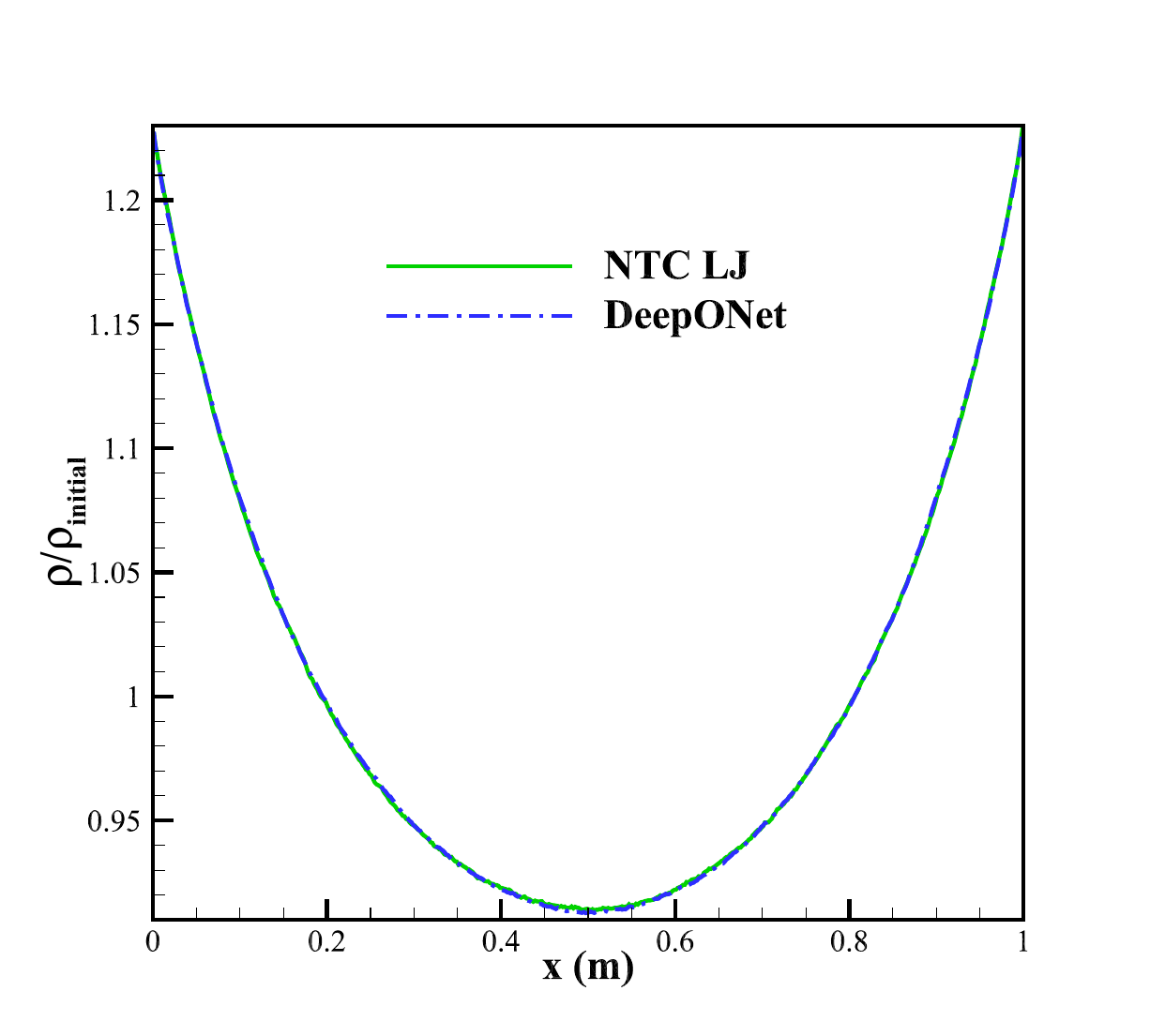}

(a)
\end{minipage} & \begin{minipage}[b]{\linewidth}\raggedright
\includegraphics[width=\linewidth,height=0.32\textheight,keepaspectratio]{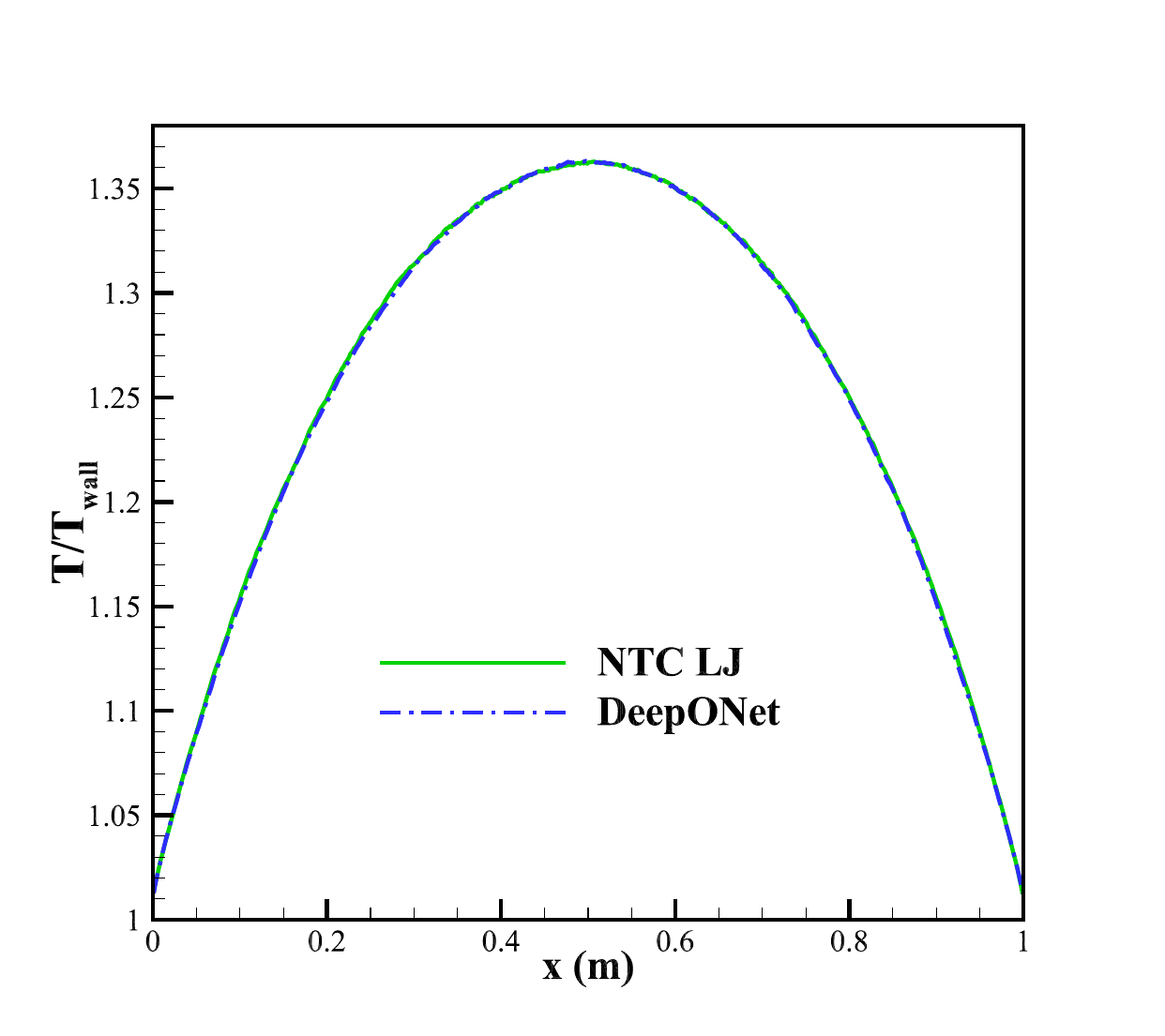}

(b)
\end{minipage} \\

\multicolumn{2}{@{}>{\raggedright\arraybackslash}p{(\columnwidth - 2\tabcolsep) * \real{1.0000} + 2\tabcolsep}@{}}{%
} \\
\end{longtable}

\autopapercaption{fig:fig12}{Comparative evaluation of flow properties in the supersonic argon Couette flow (Kn = 0.0051, \emph{T\textsubscript{w}} = 40 K): (a) normalized density, (b) normalized temperature, (c) normalized velocity with magnified views of boundary slip, and (d) shear stress distribution.}

\subsection{Hypersonic flow over a circular cylinder: validation of geometric robustness}\label{subsec:hypersonic-cylinder}
To evaluate the robustness and numerical stability of the implemented Lennard-Jones (LJ) potential in complex, non-orthogonal geometries, a benchmark simulation of rarefied hypersonic flow over a circular cylinder was performed. This test case is widely recognized in the literature, e.g., \cite{goshayeshi2015ssbt,lofthouse2007continuum,shoja2020gbttas,goshayeshi2015sbttas,shoja2026vacuum}, as it provides a comprehensive scenario for assessing the accuracy of collision algorithms across a wide range of collision frequencies. The flow field inherently encompasses a diverse spectrum of extreme gas dynamic phenomena within a single domain: a strongly compressive, high-temperature detached bow shock at the forebody, a stagnation region characterized by intense thermal excitation, a severe Prandtl-Meyer-type rapid expansion fan accelerating the flow around the cylinder shoulder, and finally, a highly rarefied, ultra-cold, low-speed recirculation zone (the wake) trailing the leeward base.

\subsubsection{Hypersonic flow at Mach 10}\label{subsubsec:hypersonic-mach-10}
\begin{figure}[!htbp]
\centering
\includegraphics[
  width=0.92\linewidth,
  trim={0.5cm 4.0cm 0.5cm 3.2cm},
  clip
]{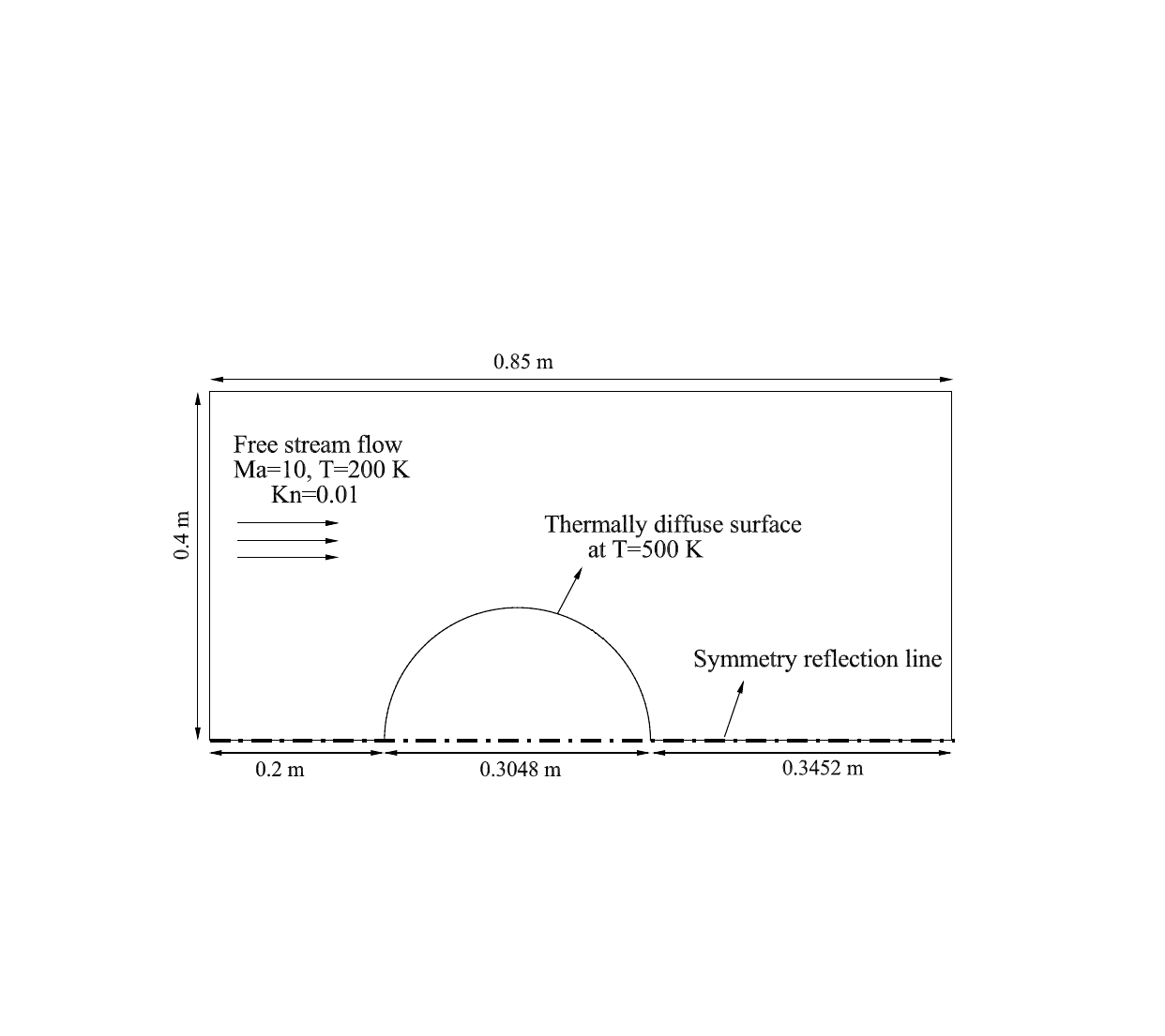}
\caption{Geometry and boundary conditions of the 2D hypersonic cylinder.}
\label{fig:fig13}
\end{figure}

In the first part of the cylinder flow study, a rarefied argon gas flows at Mach 10 (U=2634.1 m/s) with a freestream temperature of 200 K over a 12-inch-diameter cylinder. The cylinder surface is modeled as fully diffusive and is maintained at a constant isothermal temperature of 500 K. Fig.~\ref{fig:fig13} illustrates the computational domain and the applied boundary conditions.

Due to the significant gradients in the mean free path and collision frequency inherent in hypersonic cylinder flows, the transient adaptive subcell (TAS) technique within the DS2V framework \cite{bird2013dsmc} was employed to ensure adequate spatial resolution. Based on our previous grid independence and sensitivity studies \cite{shoja2020gbttas}, the optimal grid resolution was set to 194×100 divisions with 11.5 particles per cell (PPC). This setup ensures that the simulation captures the sharp curved shock and the subsequent flow separation with high fidelity, providing a rigorous environment to verify the integration of the LJ potential in 2D DSMC solvers. We also plan to implement our Collision Hybrid TAS approach \cite{stefanov2022tas} in the DS2V framework, which uses three collision schemes (NTC, SBT, or GBT, as tested in Section~\ref{subsubsec:physical-validation-shocks}) depending on the number of particles per cell. It will ensure uniform accuracy of the method when refining TAS without necessarily increasing the number of particles.

\subsubsection{Physical analysis: flow field and surface characteristics}\label{subsubsec:hypersonic-flow-surface}
\begin{figure}[!htbp]
\centering
\includegraphics[
  width=0.78\linewidth,
  height=0.38\textheight,
  keepaspectratio
]{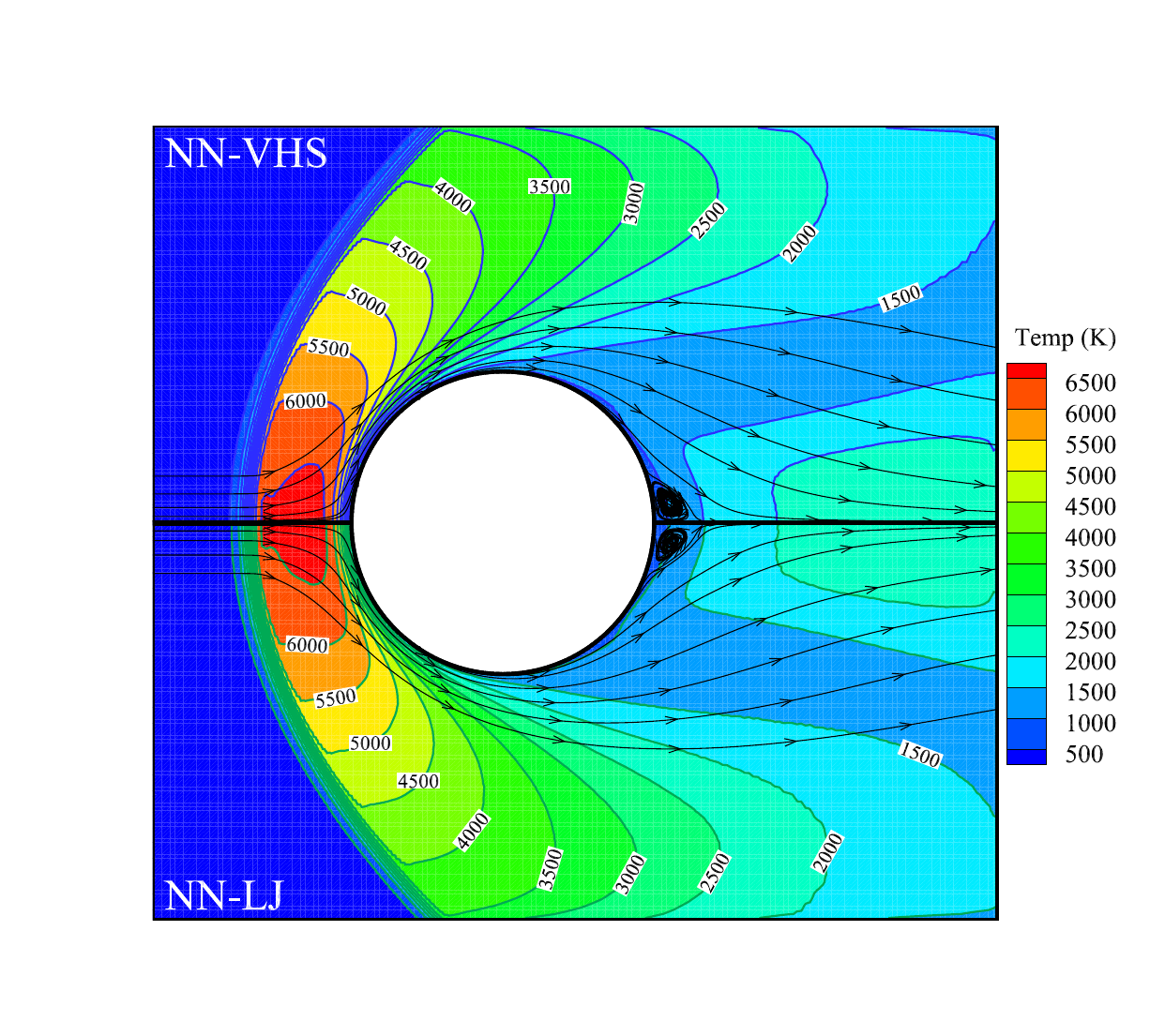}
\caption{Temperature contours and velocity streamlines for the hypersonic cylinder case: upper half, VHS--DSMC; lower half, DeepONet--LJ DSMC.}
\label{fig:fig14}
\end{figure}

The capability of the implemented Lennard-Jones (LJ) potential to resolve complex flow structures is first evaluated through field properties. Fig.~\ref{fig:fig14} shows the temperature contours, comparing DeepONet-Lennard-Jones (DeepONet--LJ) results (lower half) with those of the DeepONet-Variable-Hard-Sphere (DeepONet--VHS) model (upper half). Velocity streamlines are overlaid to examine the global flow structure. The contours show an excellent agreement between the two potentials, accurately capturing the bow shock standoff distance. A notable feature of hypersonic flow is the formation of a small wake vortex downstream. The LJ framework predicts the same shape and size for this vortex as the VHS model, confirming its stability in resolving low-pressure recirculating regions. The observed agreement is attributed to the thermal state of the wake region, where temperatures remain above approximately 800 K. For argon, this corresponds to a reduced temperature (\(T^{*}{= k_{B}T\ /\varepsilon}_{LJ}\)) significantly greater than unity. In this regime, collision dynamics are dominated by high-energy repulsive interactions, rendering the attractive potential well negligible. Consequently, the Lennard-Jones model effectively converges to the purely repulsive VHS behavior.

\begin{figure}[!htbp]
\centering
\includegraphics[
  width=0.78\linewidth,
  height=0.38\textheight,
  keepaspectratio
]{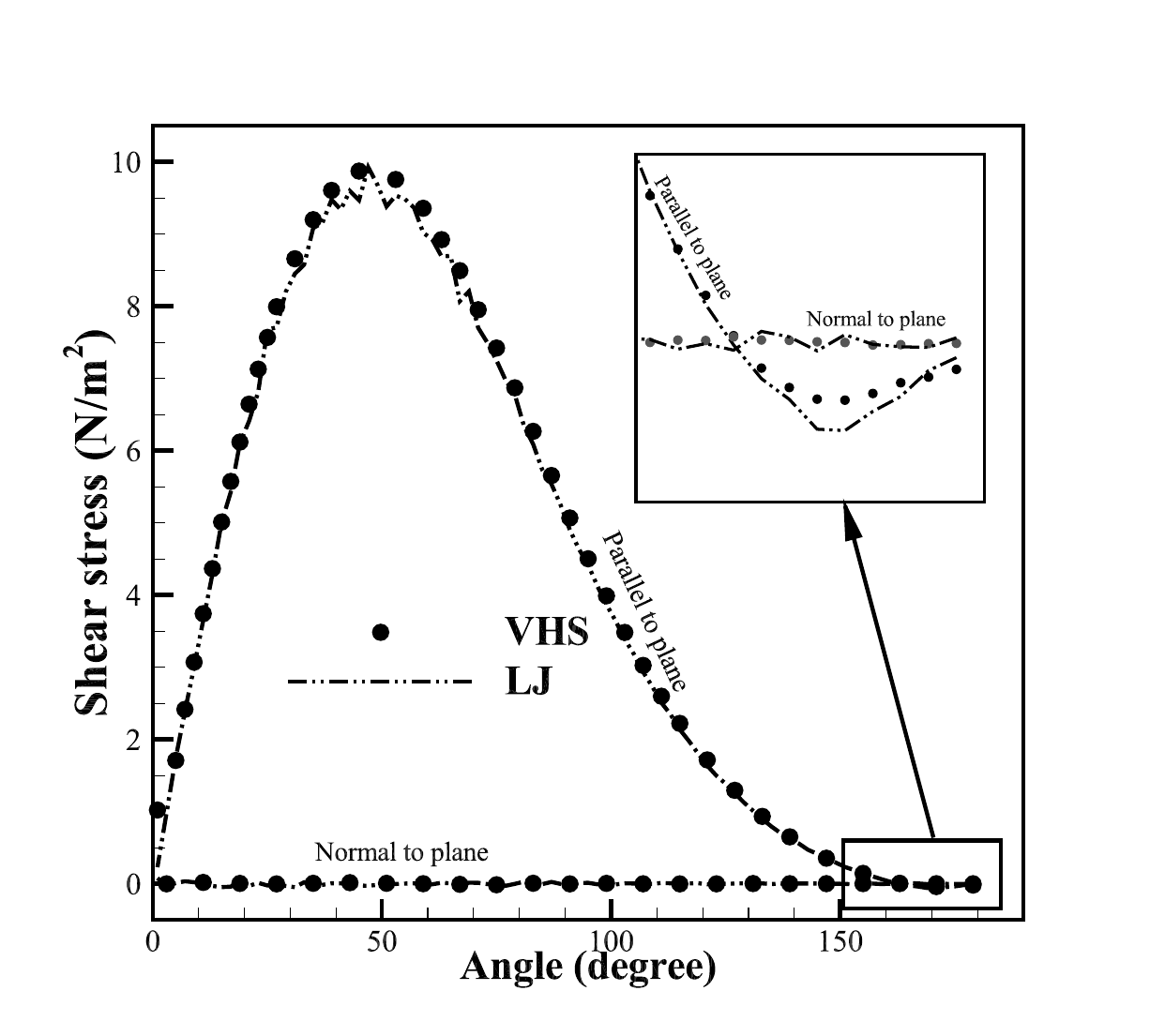}
\caption{Shear-stress distribution over the cylinder surface with a zoomed view of the rear region: LJ versus VHS.}
\label{fig:fig15}
\end{figure}

The surface momentum exchange and flow separation characteristics are scrutinized through the distribution of parallel and normal-to-plane shear stresses in Fig.~\ref{fig:fig15}. Both models predict nearly identical shear stress profiles. A significant observation is made in the magnified view of the cylinder's rear end, where the intersection of the parallel and normal shear-stress components identifies the flow-separation point. The fact that both models predict this point at the same angular location demonstrates that the LJ potential maintains high fidelity in capturing separation and wake structures without introducing numerical artifacts in two-dimensional curved geometries.

The internal structure of the bow shock is scrutinized along the stagnation line ahead of the cylinder in Fig.~\ref{fig:fig16}. Fig.~\ref{fig:fig16}(a) illustrates the normalized Mach number and temperature profiles. The gas undergoes a sharp kinetic-to-thermal energy conversion, and the DeepONet--LJ profiles follow the VHS trends without any spatial lag or numerical thickening of the shock wave. The corresponding normalized density profile is shown in Fig.~\ref{fig:fig16}(b), where a near-perfect overlap between the LJ and VHS models is observed, with the density ratio peaking near the stagnation point. This consistency serves as a critical verification, proving that the LJ potential correctly recovers the repulsive-dominated scattering physics in high-Mach regimes while ensuring numerical robustness in two-dimensional curved geometries.

\begin{longtable}[]{@{}
  >{\raggedright\arraybackslash}p{(\columnwidth - 2\tabcolsep) * \real{0.5314}}
  >{\raggedright\arraybackslash}p{(\columnwidth - 2\tabcolsep) * \real{0.5314}}@{}}

\begin{minipage}[b]{\linewidth}\raggedright
\includegraphics[width=\linewidth,height=0.32\textheight,keepaspectratio]{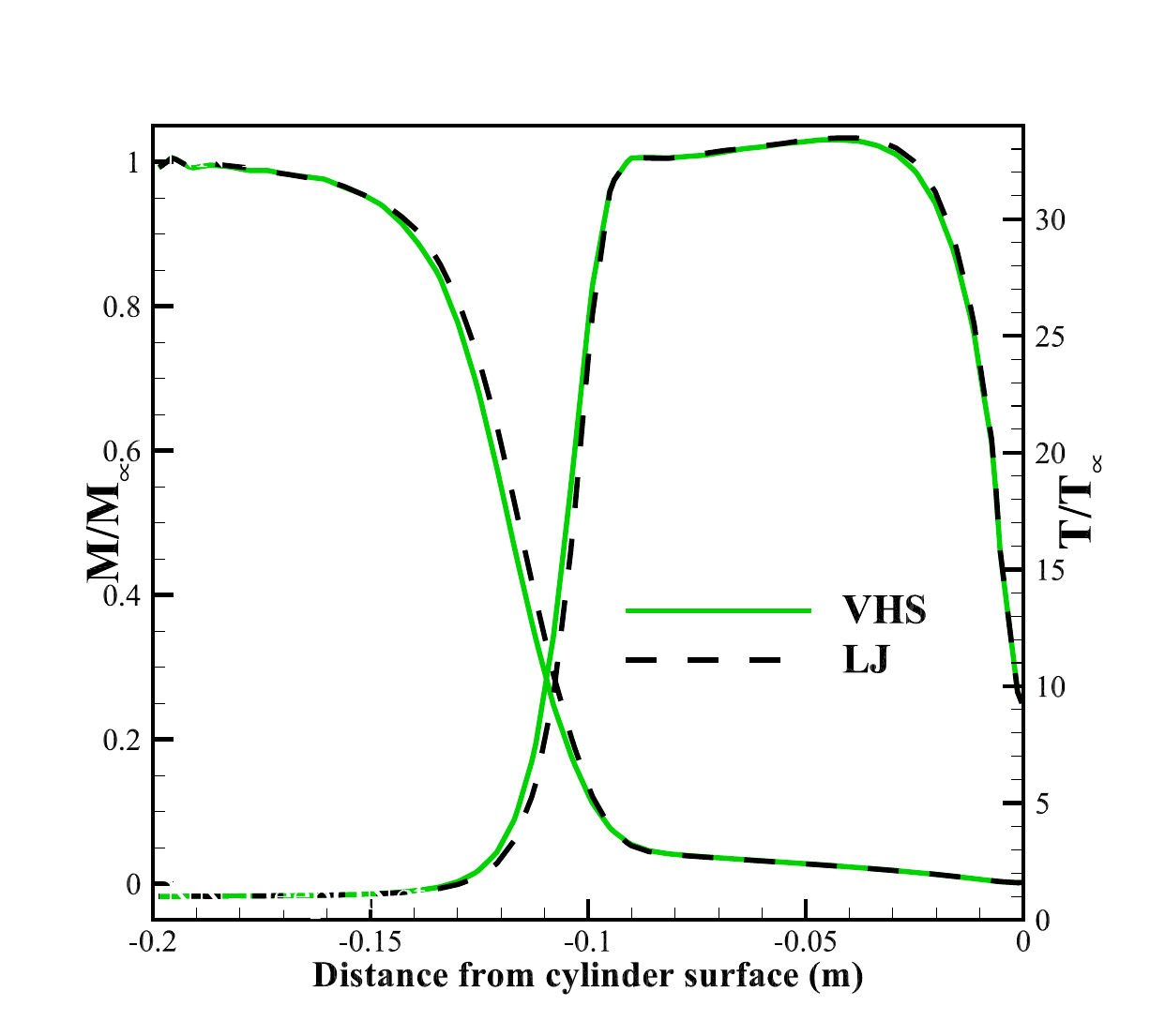}

(a)
\end{minipage} & \begin{minipage}[b]{\linewidth}\raggedright
\includegraphics[width=\linewidth,height=0.32\textheight,keepaspectratio]{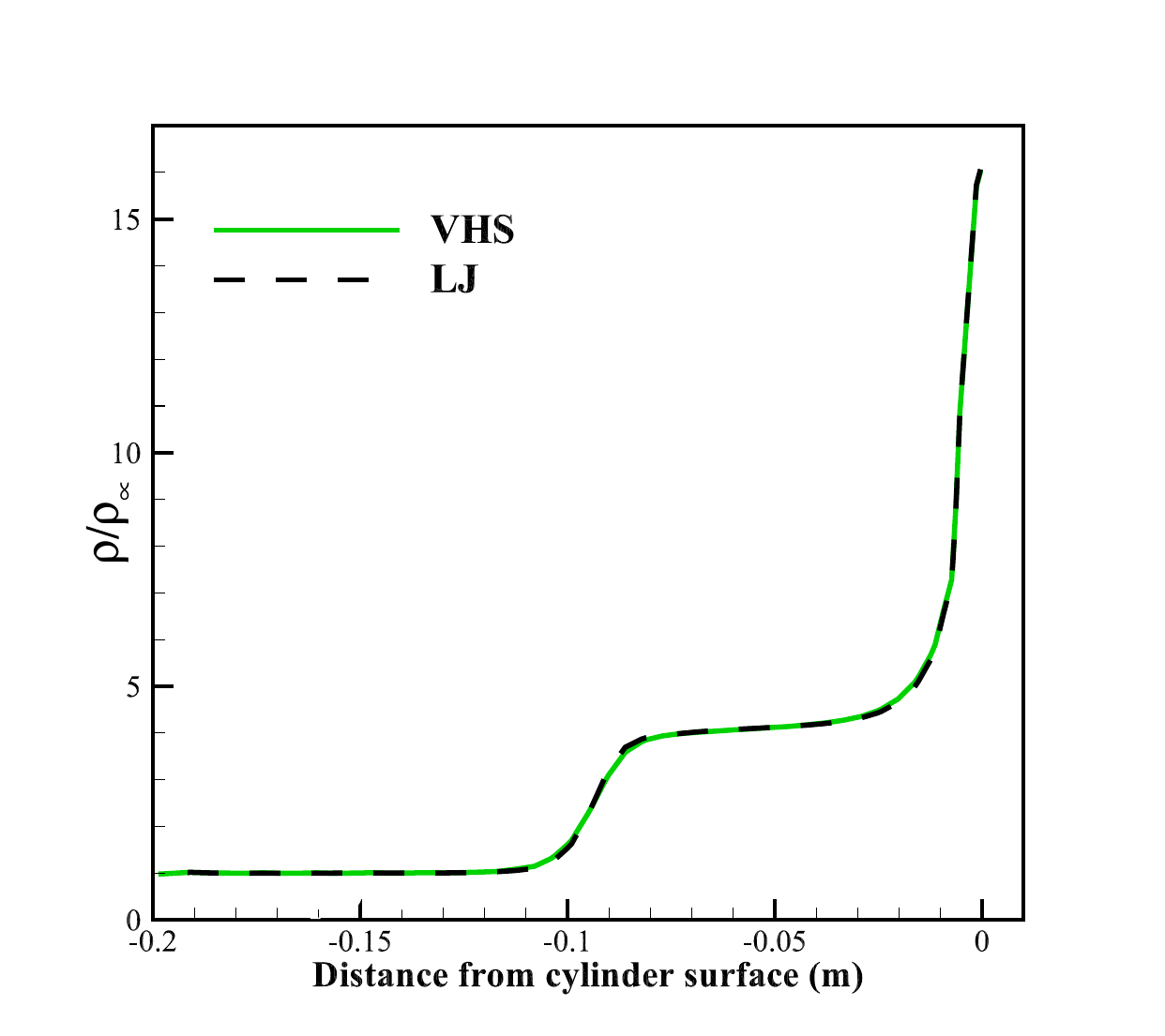}

(b)
\end{minipage} \\

\end{longtable}

\autopapercaption{fig:fig16}{(a) Normalized Mach number and temperature profiles; (b) normalized density profile along the stagnation line in front of the cylinder, LJ vs. VHS.}

The mechanical response of the cylinder surface under hypersonic conditions is presented in Fig.~\ref{fig:fig17}. Fig.~\ref{fig:fig17}(a) depicts the surface velocity distribution. Both the DeepONet--LJ and DeepONet--VHS models predict identical velocity recovery patterns, specifically capturing the flow acceleration and the subsequent deceleration in the vortex wake region. The maximum velocity predicted at the cylinder's shoulder is consistent across all models, confirming that the LJ potential does not alter the macroscopic momentum transport in high-energy regimes. Furthermore, the pressure distribution is illustrated in Fig.~\ref{fig:fig17}(b). Given that pressure varies by several orders of magnitude from the stagnation point to the shadow region, a logarithmic scale is used to examine the model's performance in the wake. The near-perfect overlap between the LJ and VHS results across the entire angular range demonstrates the numerical robustness of the LJ framework, ensuring that the integration of the realistic potential remains artifact-free even in extremely low-pressure environments

\begin{longtable}[]{@{}
  >{\raggedright\arraybackslash}p{(\columnwidth - 2\tabcolsep) * \real{0.5000}}
  >{\raggedright\arraybackslash}p{(\columnwidth - 2\tabcolsep) * \real{0.5000}}@{}}

\begin{minipage}[b]{\linewidth}\raggedright
\includegraphics[width=\linewidth,height=0.32\textheight,keepaspectratio]{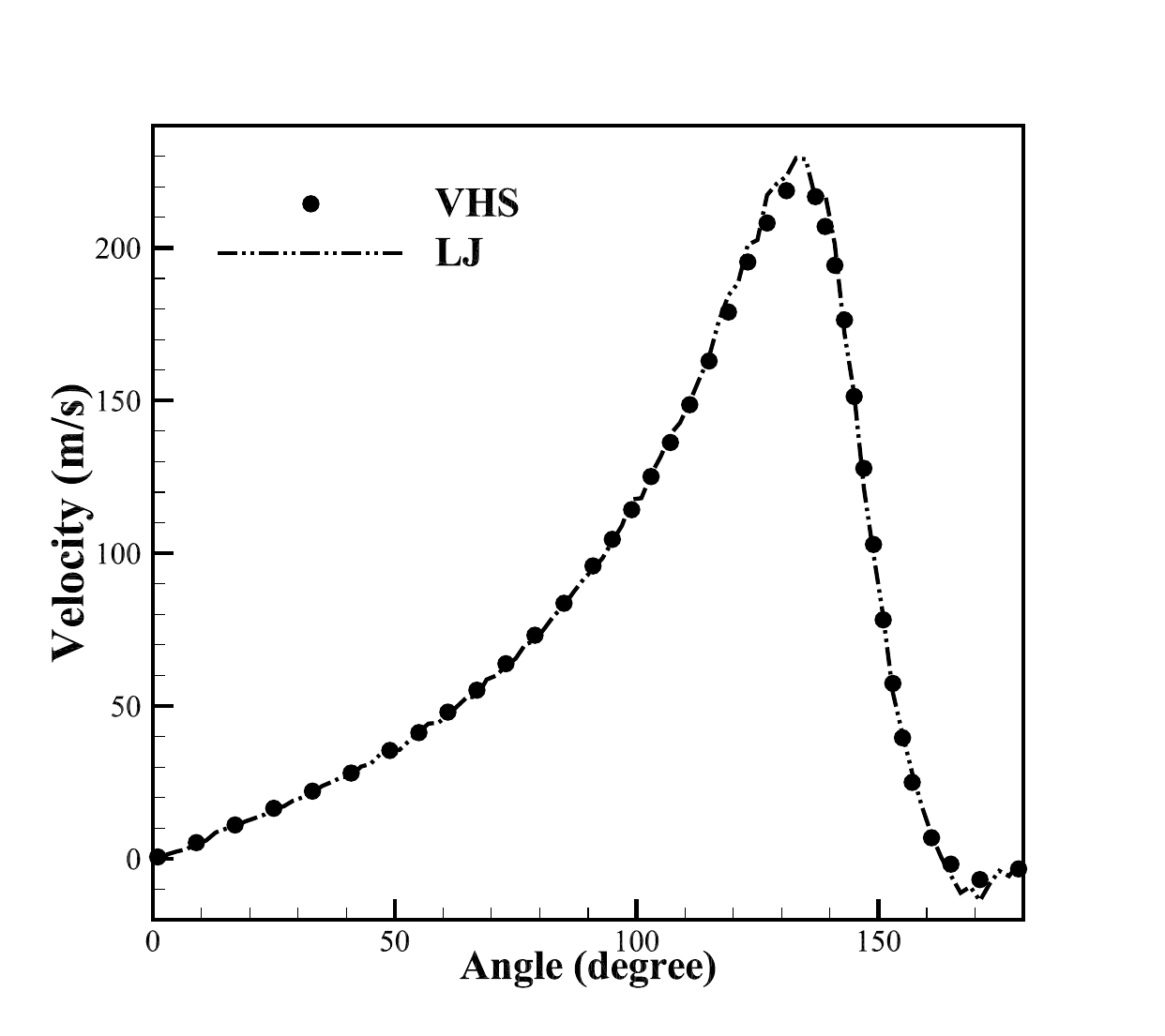}

(a)
\end{minipage} & \begin{minipage}[b]{\linewidth}\raggedright
\includegraphics[width=\linewidth,height=0.32\textheight,keepaspectratio]{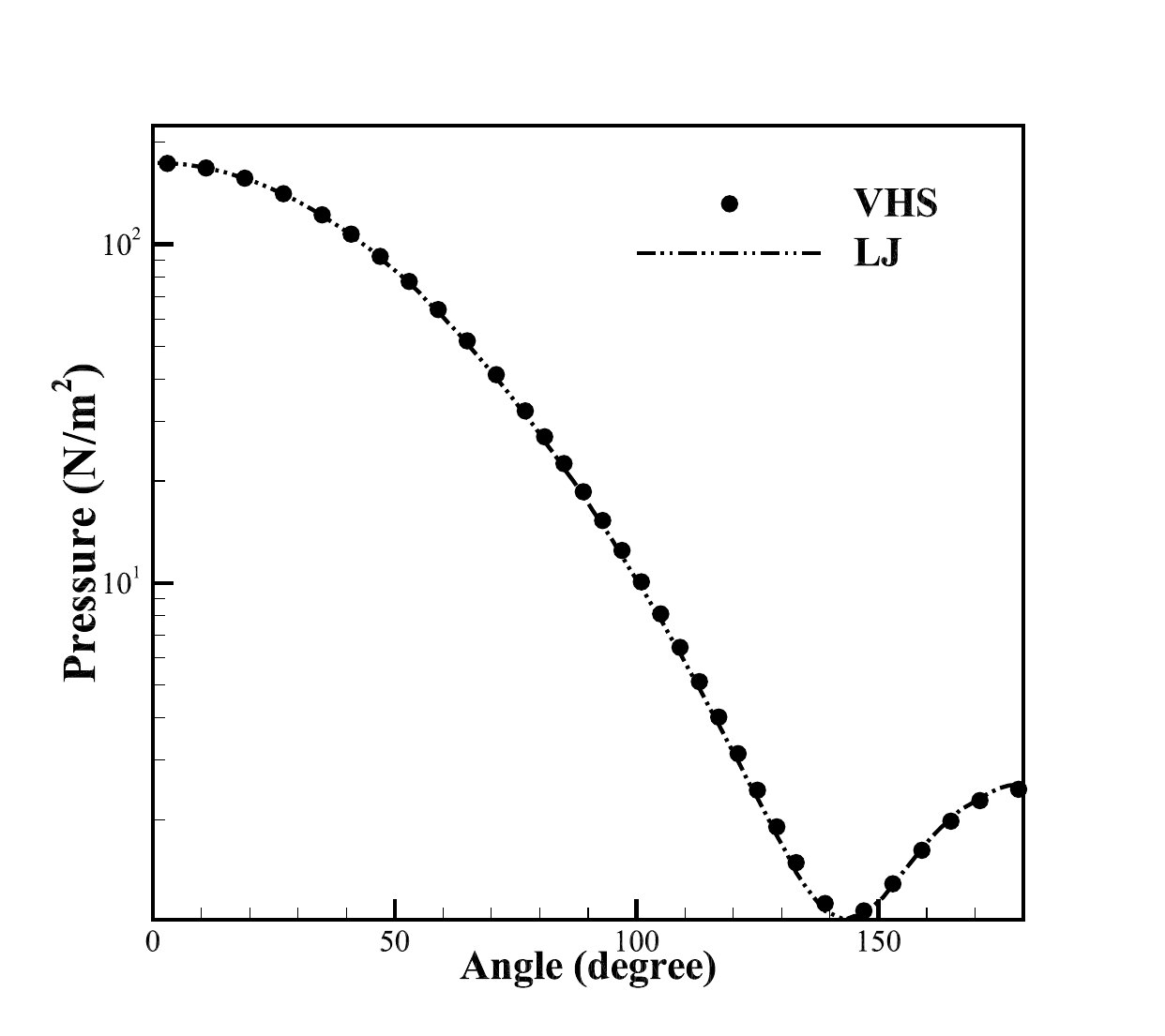}

(b)
\end{minipage} \\

\end{longtable}

\autopapercaption{fig:fig17}{(a) Velocity magnitude and (b) pressure distribution (log scale) on the cylinder surface, LJ vs. VHS.}

The thermal characteristics of the gas-surface interaction are evaluated in Fig.~\ref{fig:fig18}. The normalized surface temperature distribution shown in Fig.~\ref{fig:fig18}(a) is uniform across both potentials, indicating that the energy exchange process is correctly modeled. A more sensitive metric for thermal validation is the surface heat flux, presented in Fig.~\ref{fig:fig18}(b). Because of the massive dissipation of kinetic energy at the stagnation point relative to the rarefied wake, the heat flux is plotted on a logarithmic scale. This representation highlights the high-fidelity performance of the LJ model; it effectively reproduces heat-transfer rates not only in the high-intensity stagnation region but also in the complex, low-energy recirculating flow behind the cylinder. This comprehensive agreement reinforces the conclusion that the LJ-DSMC integration is a reliable tool for predicting both aerodynamic loads and thermal protection requirements in two-dimensional curved geometries.

\begin{longtable}[]{@{}
  >{\raggedright\arraybackslash}p{(\columnwidth - 2\tabcolsep) * \real{0.4974}}
  >{\raggedright\arraybackslash}p{(\columnwidth - 2\tabcolsep) * \real{0.5026}}@{}}

\begin{minipage}[b]{\linewidth}\raggedright
\includegraphics[width=\linewidth,height=0.32\textheight,keepaspectratio]{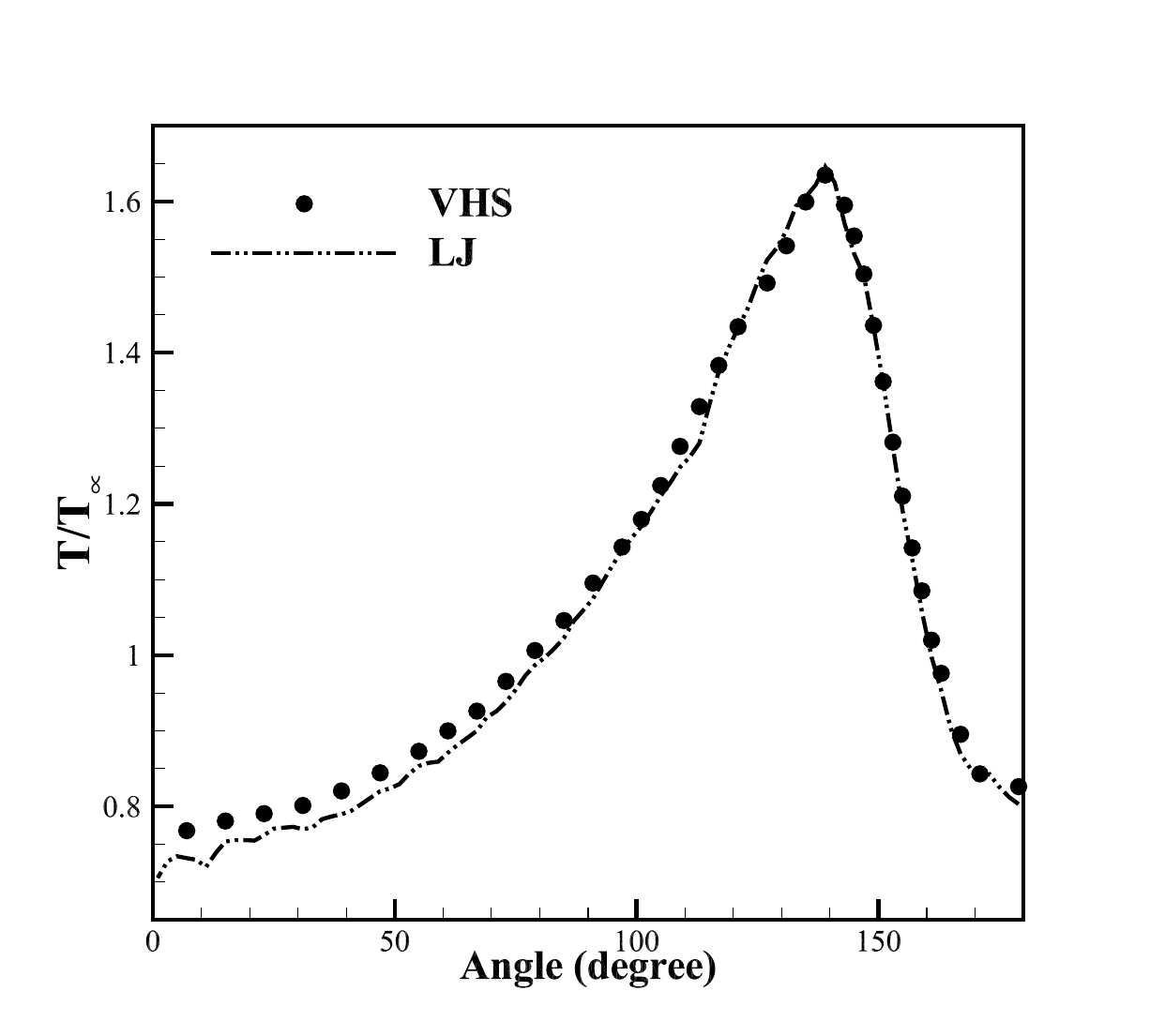}

(a)
\end{minipage} & \begin{minipage}[b]{\linewidth}\raggedright
\includegraphics[width=\linewidth,height=0.32\textheight,keepaspectratio]{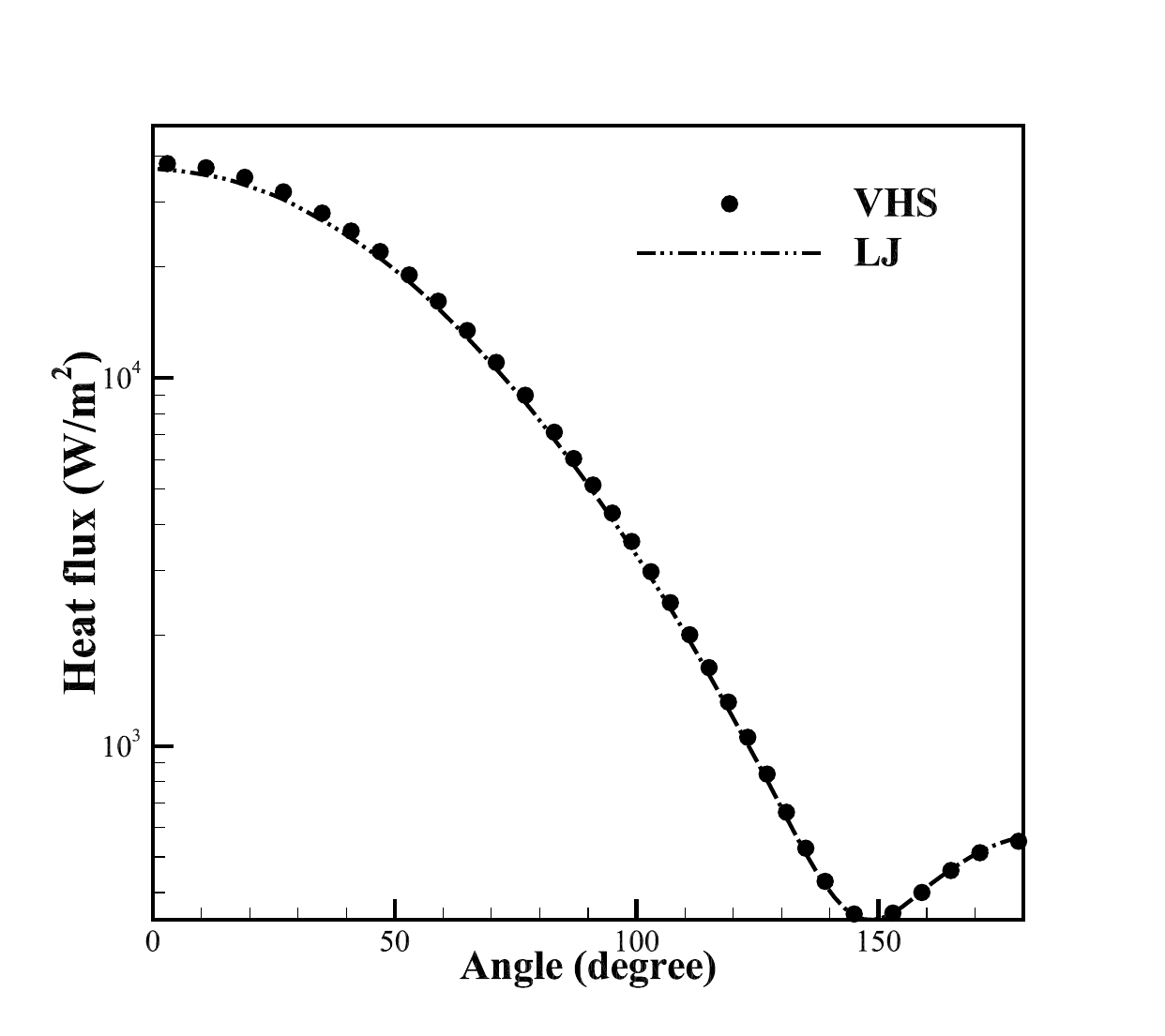}

(b)
\end{minipage} \\

\end{longtable}

\autopapercaption{fig:fig18}{(a) Normalized temperature and (b) heat flux (log scale) on the cylinder surface, LJ vs. VHS.}

\subsubsection{Performance of the DeepONet surrogate in hypersonic flow}\label{subsubsec:hypersonic-deeponet-performance}
\begin{figure}[!htbp]
\centering
\includegraphics[
  width=0.78\linewidth,
  height=0.38\textheight,
  keepaspectratio
]{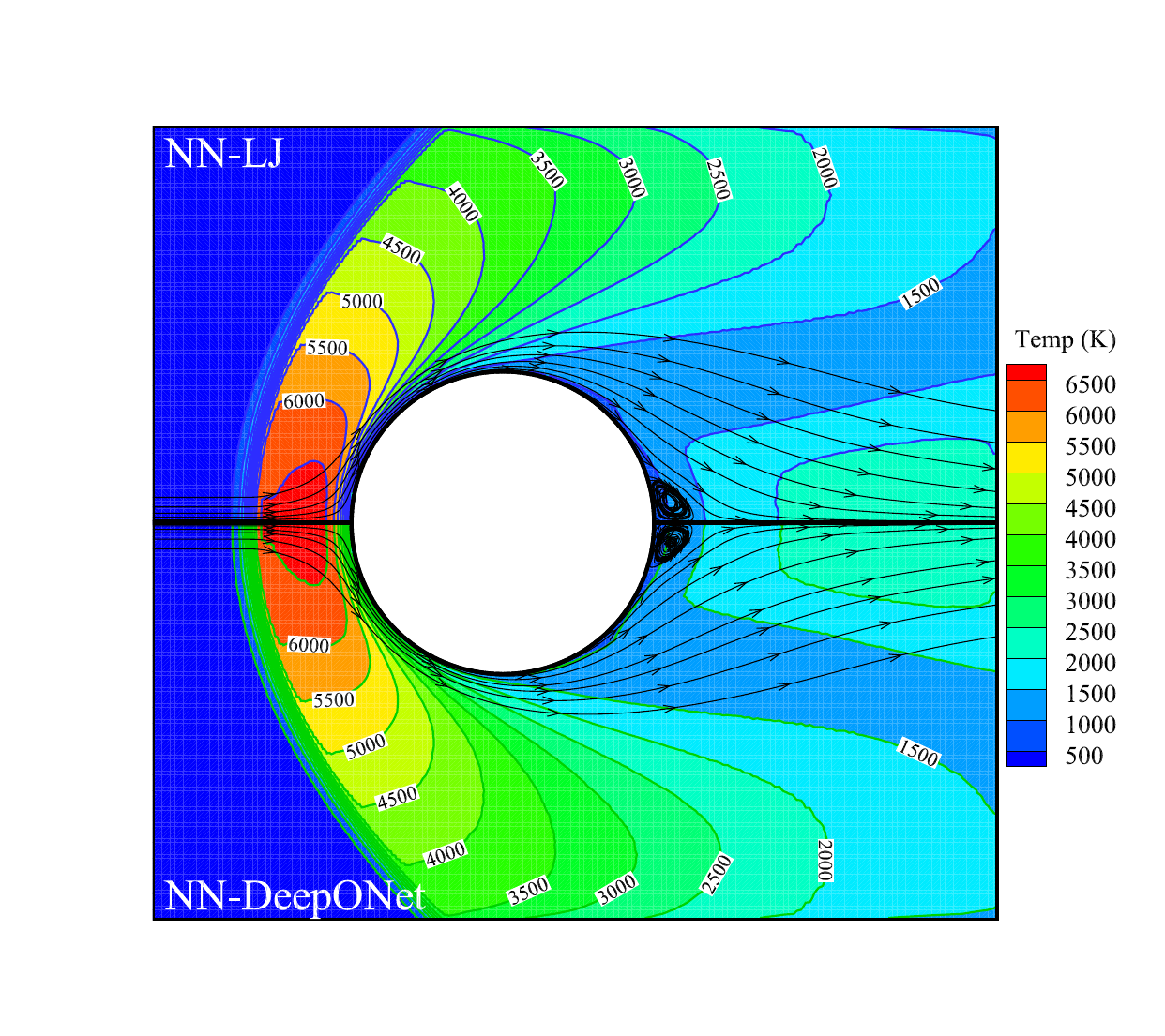}
\caption{Temperature contours and velocity streamlines for the hypersonic cylinder case: upper half, exact LJ--DSMC; lower half, DeepONet--LJ prediction.}
\label{fig:fig19}
\end{figure}

The ultimate objective of this study is to provide a computationally efficient surrogate for the Lennard-Jones (LJ) potential without compromising physical accuracy. To verify this, the performance of the trained DeepONet is evaluated in the hypersonic cylinder flow. Fig.~\ref{fig:fig19} illustrates the temperature contours, with the upper half showing the exact LJ--DSMC result and the lower half showing the DeepONet--LJ prediction. Velocity streamlines are overlaid on both halves to examine the global flow structure. The neural operator reconstructs the thermal field and captures the wake vortex behind the cylinder with high fidelity relative to the exact LJ reference. 

\begin{figure}[!htbp]
\centering
\includegraphics[
  width=0.78\linewidth,
  height=0.38\textheight,
  keepaspectratio
]{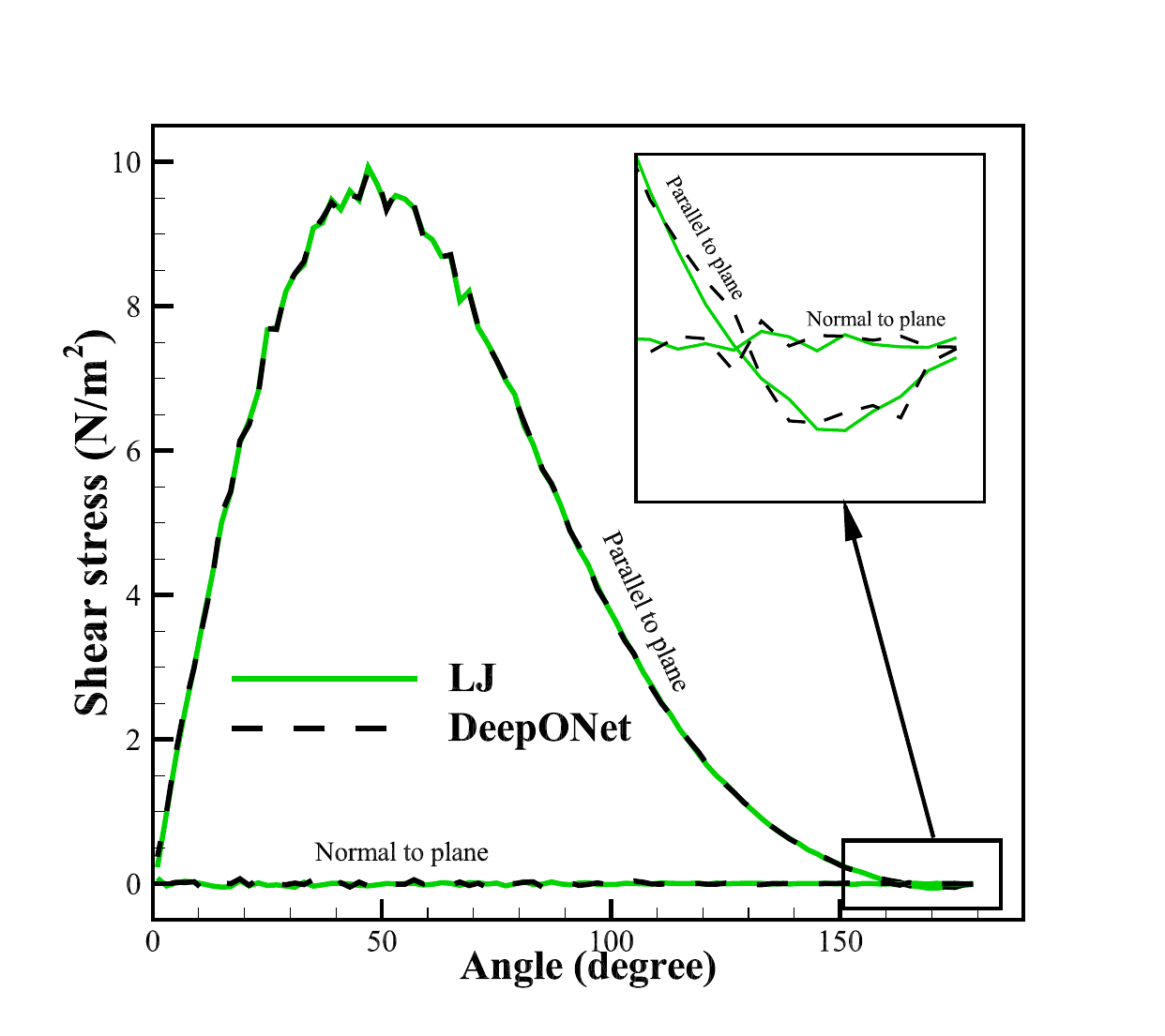}
\caption{Shear-stress distribution over the cylinder surface with a zoomed view of the rear region: DeepONet--LJ versus exact LJ.}
\label{fig:fig20}
\end{figure}

The dynamic accuracy of the surrogate model at the gas-surface interface is further examined in Fig.~\ref{fig:fig20}, which shows the surface shear-stress distribution. The DeepONet results fully agree with the LJ reference in the main view. While a magnified view of the rear-end region reveals minor localized discrepancies between the two models, these variations are negligible relative to the overall stress profile. The fact that the surrogate model predicts the intersection of the parallel and normal shear-stress components---identifying the separation point---at the same location as the LJ solver confirms its robust generalization to 2D curved boundaries.

The internal structure of the bow shock along the stagnation line is analyzed in Fig.~\ref{fig:fig21}. The normalized Mach number, temperature (a), and density (b) profiles predicted by the DeepONet exhibit a near-perfect overlap with the exact LJ-DSMC results. The surrogate's ability to resolve sharp gradients across the bow shock ($T/T_{\infty} \approx 32$ without spatial lag) demonstrates its robustness in high-energy regimes.

\begin{longtable}[]{@{}
  >{\raggedright\arraybackslash}p{(\columnwidth - 2\tabcolsep) * \real{0.5193}}
  >{\raggedright\arraybackslash}p{(\columnwidth - 2\tabcolsep) * \real{0.4807}}@{}}

\begin{minipage}[b]{\linewidth}\raggedright
\includegraphics[width=\linewidth,height=0.32\textheight,keepaspectratio]{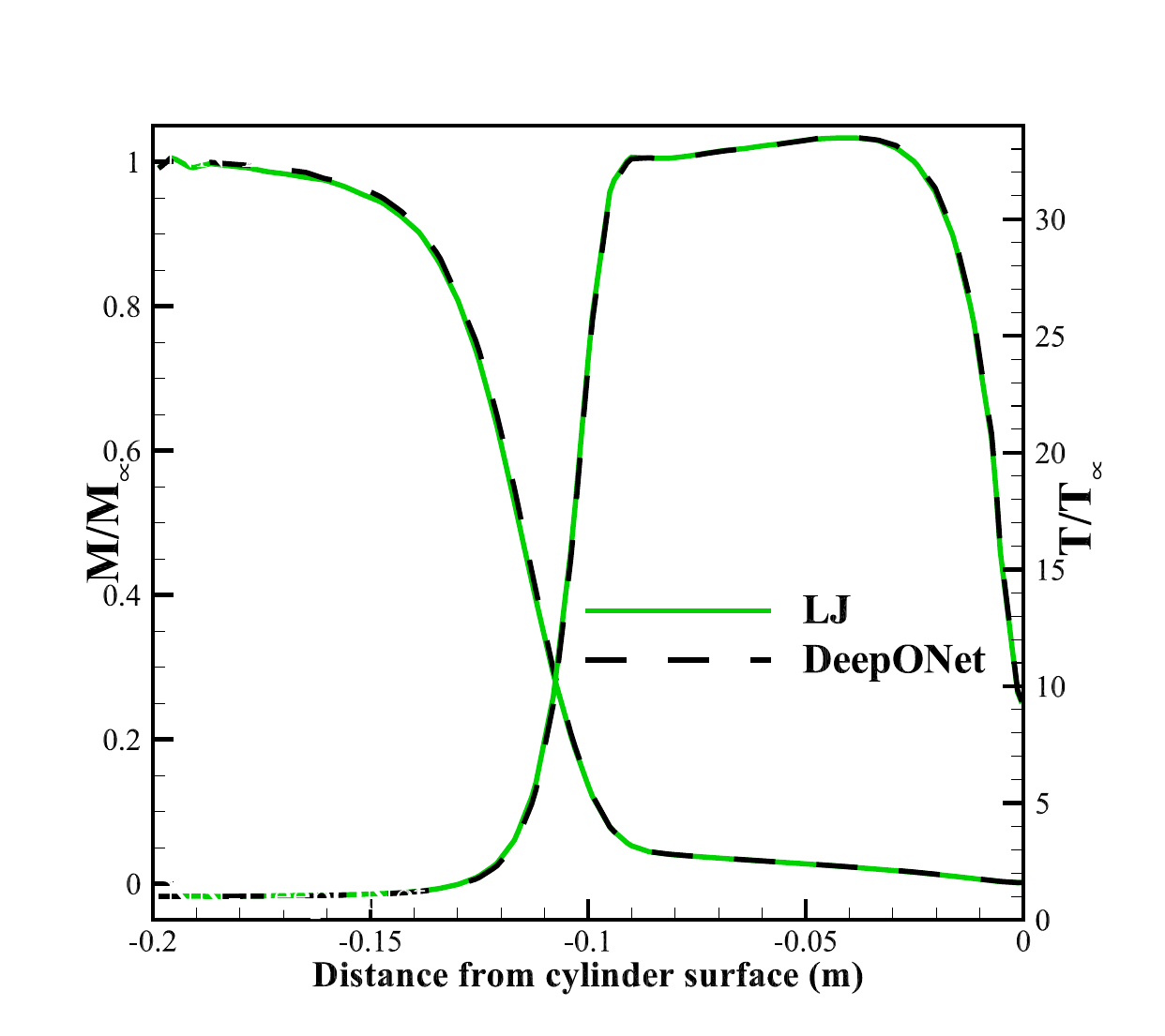}

(a)
\end{minipage} & \begin{minipage}[b]{\linewidth}\raggedright
\includegraphics[width=\linewidth,height=0.32\textheight,keepaspectratio]{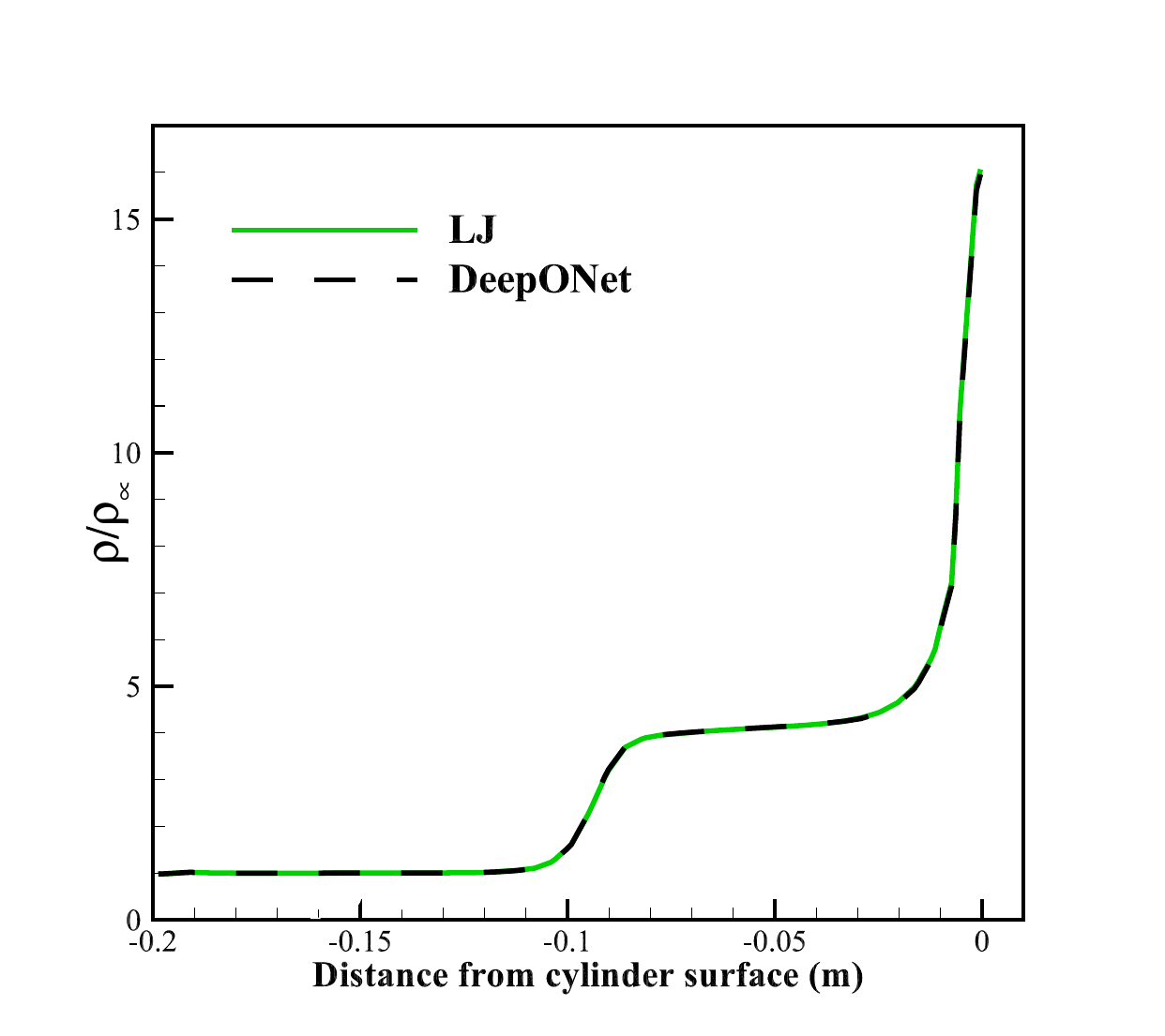}

(b)
\end{minipage} \\

\end{longtable}

\autopapercaption{fig:fig21}{(a) Normalized Mach number and temperature profiles; (b) normalized density profile along the stagnation line in front of the cylinder, DeepONet--LJ versus exact LJ.}

The performance on the cylinder surface is further grouped into mechanical and thermal characteristics. Fig.~\ref{fig:fig22} presents the surface velocity and the pressure coefficient (on a logarithmic scale). For the surface velocity, very subtle differences are observed at the peak velocity and within the wake region; however, these do not affect the overall accuracy of the momentum exchange prediction. Notably, the pressure distribution in the logarithmic plot shows a perfect overlap between the DeepONet and the LJ reference across the entire surface.

\begin{longtable}[]{@{}
  >{\raggedright\arraybackslash}p{(\columnwidth - 2\tabcolsep) * \real{0.5016}}
  >{\raggedright\arraybackslash}p{(\columnwidth - 2\tabcolsep) * \real{0.4984}}@{}}

\begin{minipage}[b]{\linewidth}\raggedright
\includegraphics[width=\linewidth,height=0.32\textheight,keepaspectratio]{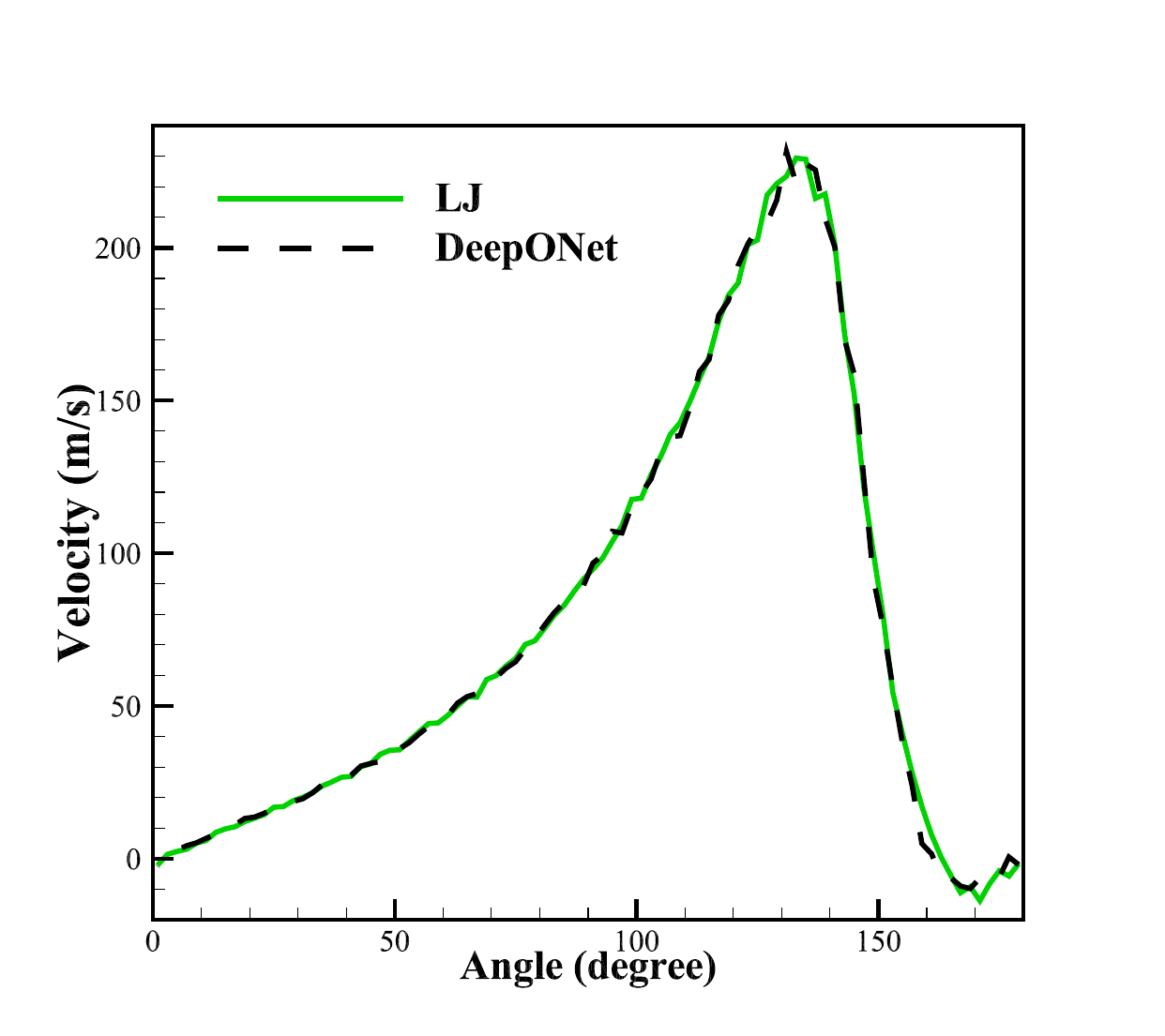}

(a)
\end{minipage} & \begin{minipage}[b]{\linewidth}\raggedright
\includegraphics[width=\linewidth,height=0.32\textheight,keepaspectratio]{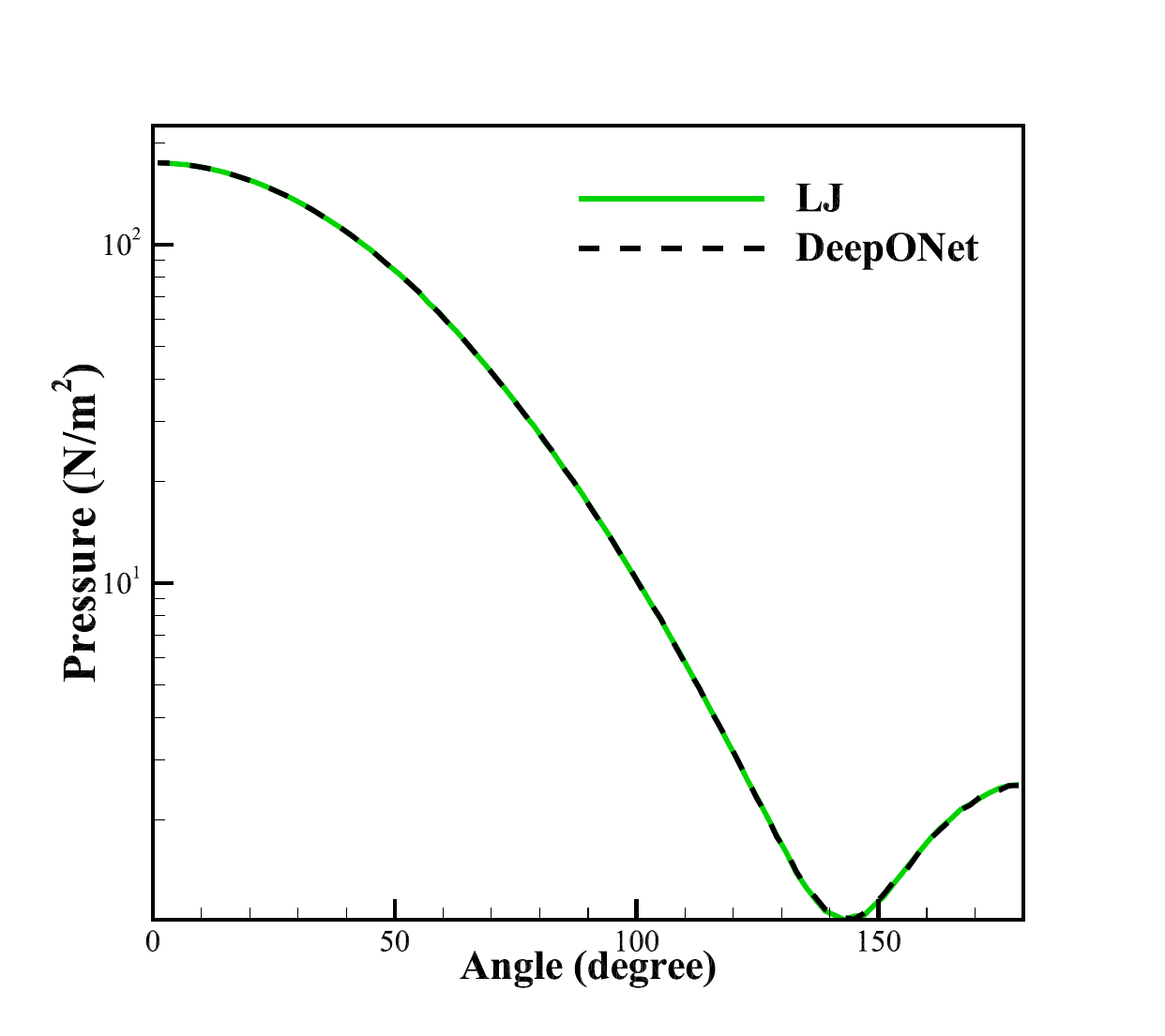}

(b)
\end{minipage} \\

\end{longtable}

\autopapercaption{fig:fig22}{(a) Velocity magnitude and (b) pressure distribution (log scale) on the cylinder surface, DeepONet--LJ versus exact LJ}

Finally, Fig.~\ref{fig:fig23} depicts the surface temperature and heat flux (on a logarithmic scale). Regarding the surface heat flux, both models perform identically, even in the wake's shadow regions. For the surface temperature, while DeepONet captures the general trend, oscillatory behavior is observed around the peak region compared with the smooth LJ curve. Despite these localized oscillations, the agreement near the stagnation point and within the wake remains strong. Smoothed representations of these profiles would show nearly identical behavior, confirming that the DeepONet surrogate is a reliable and highly efficient substitute for expensive LJ potential calculations in complex 2D rarefied flows.

\begin{longtable}[]{@{}
  >{\raggedright\arraybackslash}p{(\columnwidth - 2\tabcolsep) * \real{0.5005}}
  >{\raggedright\arraybackslash}p{(\columnwidth - 2\tabcolsep) * \real{0.4995}}@{}}

\begin{minipage}[b]{\linewidth}\raggedright
\includegraphics[width=\linewidth,height=0.32\textheight,keepaspectratio]{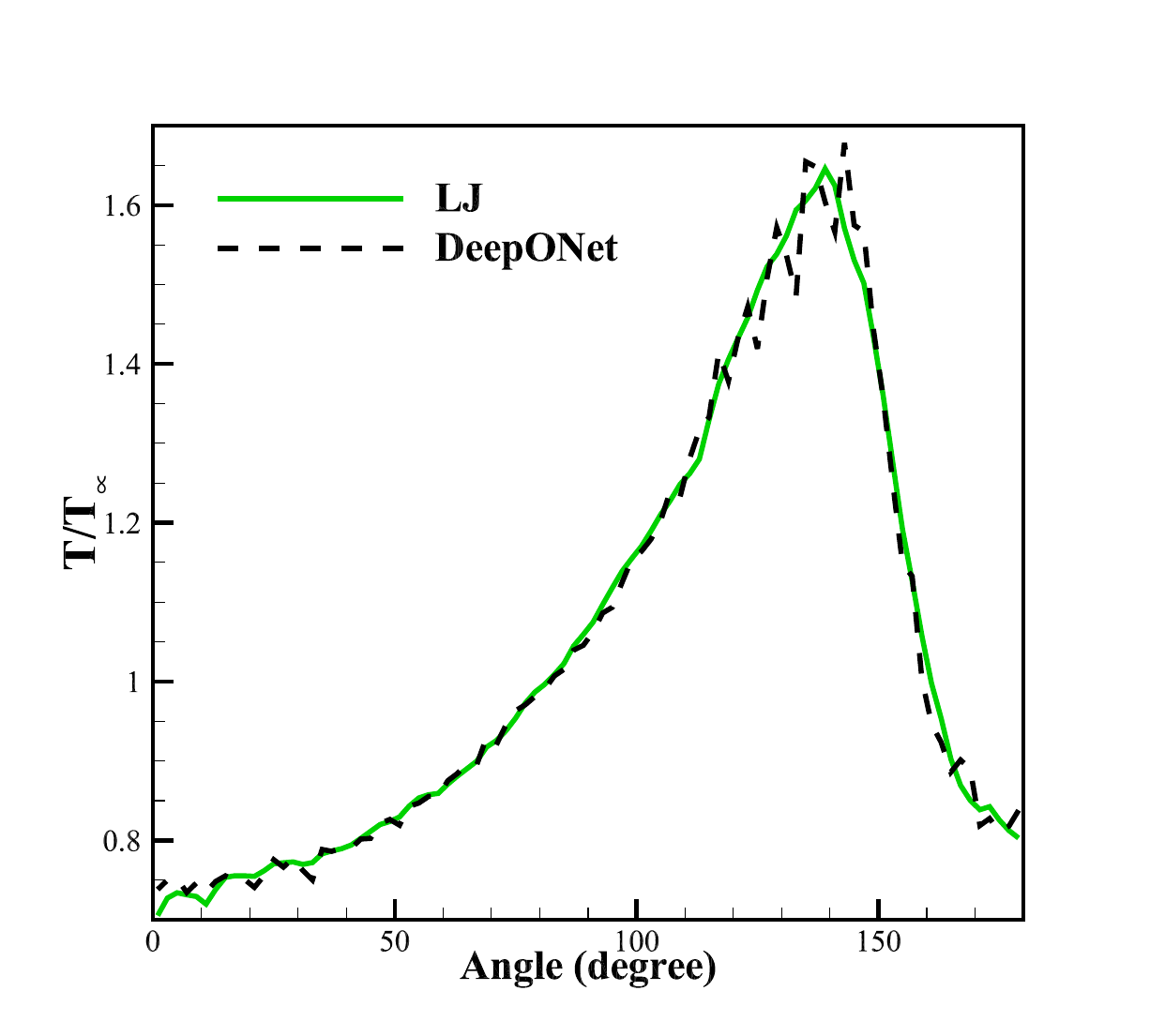}

(a)
\end{minipage} & \begin{minipage}[b]{\linewidth}\raggedright
\includegraphics[width=\linewidth,height=0.32\textheight,keepaspectratio]{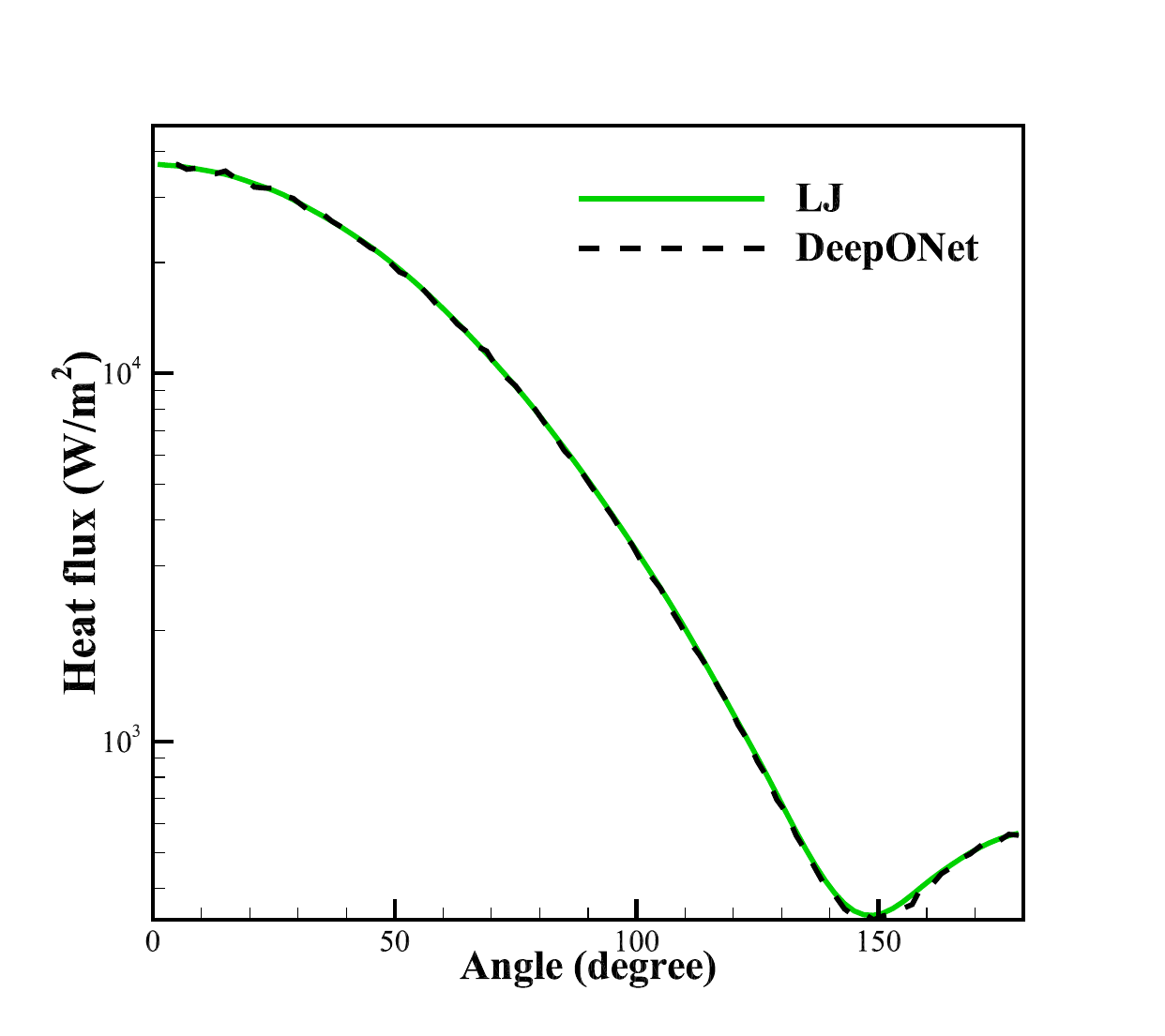}

(b)
\end{minipage} \\

\end{longtable}

\autopapercaption{fig:fig23}{(a) Normalized temperature and (b) heat flux (log scale) on the cylinder surface, DeepONet--LJ versus exact LJ.}

The microscopic fidelity of the DeepONet surrogate was also verified for the high-energy collisions prevalent in this Mach 10 flow. Since the neural operator accurately captures the underlying scattering physics, as originally established and validated in Fig.~\ref{fig:fig10}, the deflection angle distributions are not repeated here for brevity. This cross-scale agreement confirms that the surrogate model is not merely fitting macroscopic trends but is fundamentally grounded in molecular-interaction physics across diverse flow regimes.

To evaluate the computational efficiency of the DeepONet surrogate, a performance test is conducted to measure the time and resources required to reach a predefined target state. Following the methodology established in \cite{shoja2020gbttas}, the simulation starts from an initial condition and continues until a specific 'stop point\textquotesingle{} is reached. This point is defined as the moment when the sum of the normalized differences between the instantaneous cylinder-surface heat flux and a benchmark value falls below 1. It is important to note that this test measures the efficiency of reaching a target physical state rather than the absolute elimination of statistical noise in the final solution.

The results, summarized in Table~\ref{tab:computational-performance-cylinder}, reveal a significant performance gain for the machine learning framework. The exact LJ-DSMC integration required 13 hours and 46 minutes to reach the stop point, while the DeepONet-based solver reached the same stopping criterion in 8 h 48 min on the same CPU, corresponding to a 36\% reduction in wall-clock time. Furthermore, the cumulative sample size required to reach this threshold was analyzed. The exact LJ solver required 3.0513×10\textsuperscript{10} samples, whereas the DeepONet surrogate reached the criteria with 2.4105×10\textsuperscript{10} samples. When normalized against the LJ baseline, the DeepONet shows a 21\% reduction in the required sample size (normalized value of 0.79).

This indicates that the neural operator is more efficient at driving the simulation toward the target macroscopic state. While the surrogate model may exhibit higher localized oscillations in certain properties, such as surface temperature (Fig.~\ref{fig:fig18}(a)), the performance test confirms its ability to resolve the primary physical trends, specifically the surface heat flux, with considerably lower computational overhead and faster state transition compared to the traditional numerical integration of the LJ potential.

\setcounter{table}{4}
\begin{table}[!htbp]
\centering
\caption{Comparison of computational performance between the exact LJ--DSMC and the DeepONet-accelerated framework for the hypersonic cylinder-flow simulation.}
\label{tab:computational-performance-cylinder}
\begin{tabularx}{\linewidth}{lccc}
\toprule
\textbf{Model} &
\textbf{CPU time} &
\textbf{sampled collision events \((\times 10^{10})\)} &
\textbf{Normalized sample size} \\
\midrule
Exact LJ--DSMC & 13 h 46 min & 3.0513 & 1.00 \\
DeepONet--LJ--DSMC & 8 h 48 min & 2.4105 & 0.79 \\
\bottomrule
\end{tabularx}
\end{table}

\textbf{4.3.4. Mach 5 Hypersonic Flow Past a Circular Cylinder: Wake Topology and Viscous Momentum Diffusion}

To rigorously assess the macroscopic aerodynamic implications of the implemented intermolecular potential models in a complex, multi-dimensional, and highly non-equilibrium environment, a hypersonic flow over a two-dimensional macroscopic circular cylinder at Mach 5 was simulated. Because the gas transitions abruptly from a collision-dominated, high-energy state to a sparsely collisional, cryogenic state, the sensitivity of the macroscopic flow topology to the underlying microscopic collision mechanics is profoundly amplified. \RL{}Other simulation conditions are the same as in the previous cylinder case, but the cylinder's surface temperature was reduced to 40 degrees Kelvin, and the flow temperature dropped to 16 degrees Kelvin.

Figs.~\ref{fig:fig24}--\ref{fig:fig26} illustrate the spatial contours of primary macroscopic properties, such as the Mach number, translational temperature, and density fields, along with the velocity streamlines. These figures compare the predictions of the phenomenological VHS model with those of the physically consistent LJ framework. While both collision models predict shock stand-off distances and forebody stagnation properties that are relatively similar (regions where high-energy repulsive-core interactions predominantly govern the scattering kinematics), a critical morphological discrepancy appears in the leeward base region. A meticulous examination of the streamline topologies reveals that the geometrical dimensions of the separated wake, specifically the longitudinal extent and volumetric size of the closed recirculation bubble, are larger and more elongated in the LJ potential simulation, i.e., the choice of inter-molecular potential can change the topology of the flow. The wake stagnation point, which marks the spatial location of flow reattachment and the termination of reverse flow, is displaced considerably farther downstream along the symmetry axis when the LJ model is employed.

\begin{longtable}[]{@{}
  >{\raggedright\arraybackslash}p{(\columnwidth - 2\tabcolsep) * \real{0.5000}}
  >{\raggedright\arraybackslash}p{(\columnwidth - 2\tabcolsep) * \real{0.5000}}@{}}

\begin{minipage}[b]{\linewidth}\raggedright
\includegraphics[width=\linewidth,height=0.32\textheight,keepaspectratio]{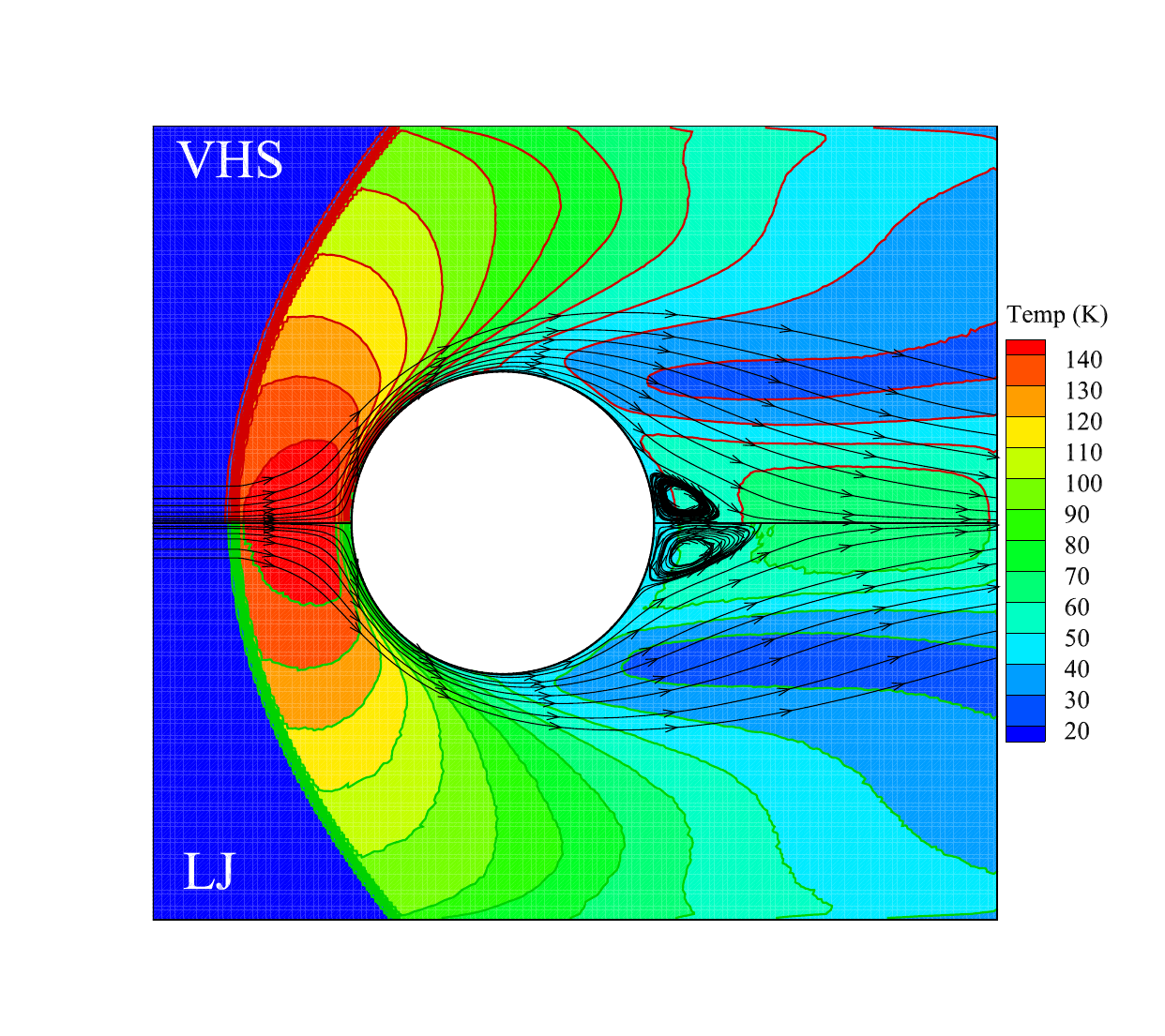}

a)
\end{minipage} & \begin{minipage}[b]{\linewidth}\raggedright
\includegraphics[width=\linewidth,height=0.32\textheight,keepaspectratio]{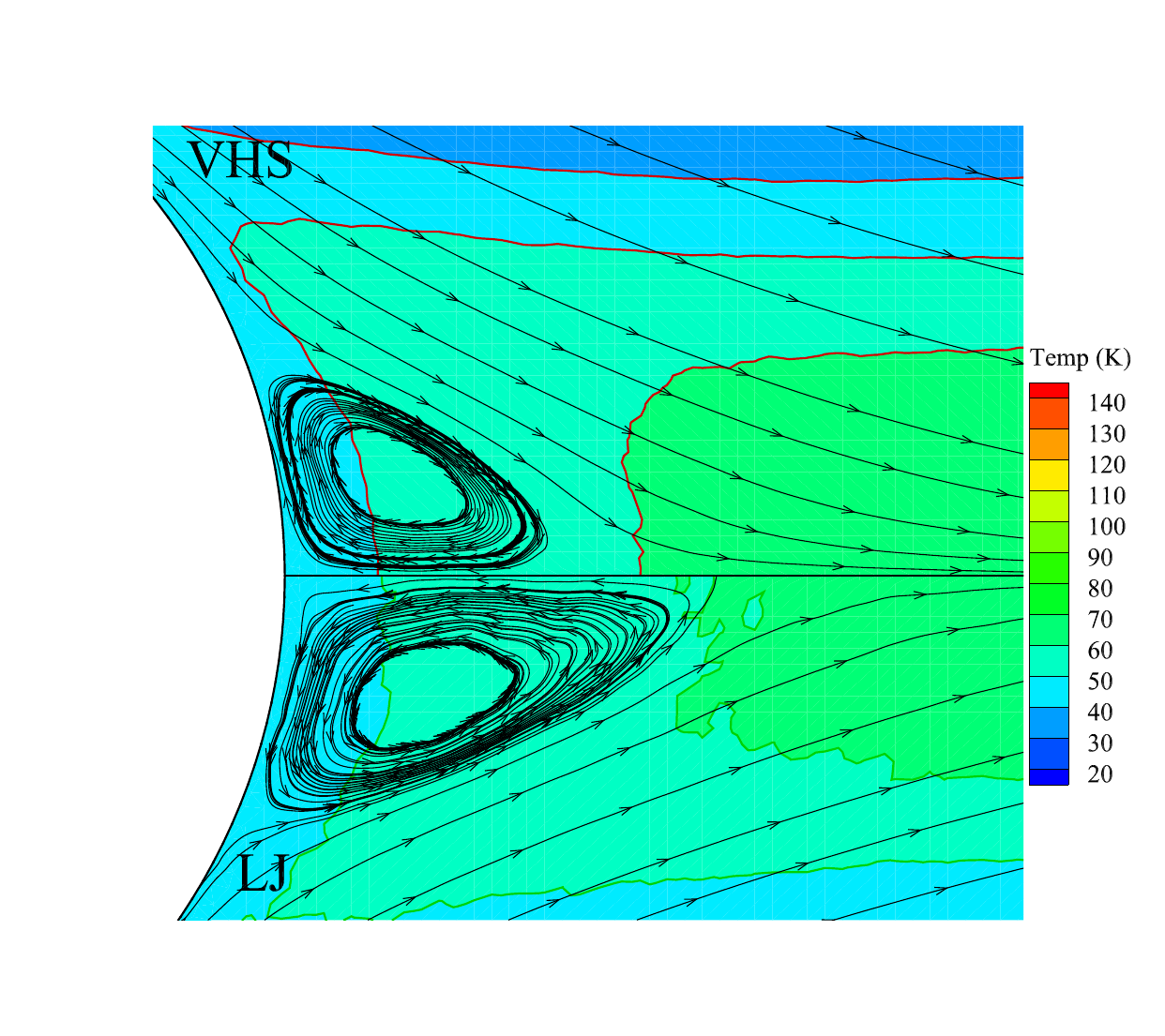}

b)
\end{minipage} \\

\end{longtable}

\autopapercaption{fig:fig24}{Temperature contours and velocity streamlines, upper half VHS and lower half LJ: (a) full domain and (b) zoomed view behind the cylinder to depict the wake, Mach=5.}

The fundamental physical mechanism driving this substantial structural disparity is intimately tied to the gas's macroscopic transport properties, specifically, the local dynamic shear viscosity (\emph{μ}), and to its critical role in governing transverse momentum diffusion. From a fluid-dynamic perspective, the spatial evolution and ultimate closure of a bluff-body wake are governed by the free shear layers that separate from the cylinder surface. Viscosity acts as the primary diffusive agent, transporting kinetic energy and momentum laterally from the high-speed, inviscid outer freestream across the shear layers and into the low-momentum, pressure-depleted wake core. When the gas's effective viscosity is artificially high, lateral momentum transfer is strongly amplified; the external flow rapidly entrains the dead-air fluid within the wake, accelerating the mixing process. This intense viscous coupling forces the separated shear layers to curve inward sharply, merging prematurely and thereby truncating the recirculation bubble into a shorter, more compact configuration.

The macroscopic discrepancies observed in the wake length directly stem from the inherent physical limitations of the VHS model in highly expanded, low-temperature regions. The VHS framework determines the collision cross-section and the resulting gas viscosity via a simple, monotonic power-law relationship (\(\mu \propto T^{\omega}\)). Because the viscosity index (\(\omega\)) is typically calibrated to match transport coefficients at freestream or elevated reference temperatures; this rigid empirical extrapolation severely overpredicts the effective viscosity when the gas is subjected to extreme cooling. As the hypersonic flow expands around the cylinder base, the translational temperature plummets. In this cryogenic, highly rarefied regime, the VHS model, being a purely repulsive hard-sphere derivative, completely fails to capture the true quantum-mechanical nature of low-energy, grazing molecular collisions. This phenomenological blind spot artificially inflates the momentum-exchange efficiency, thereby overestimating the local dynamic viscosity. It is precisely this artificially elevated viscosity in the VHS simulation that over-accelerates momentum diffusion into the wake, thereby forcing the premature shear-layer reattachment observed in the contours.

In stark contrast, the Lennard--Jones potential mathematically integrates both the steep short-range Pauli repulsion and the long-range van der Waals attractive tail (characterized by the potential well depth \emph{ε}). At the extremely low thermal collision energies prevalent in the base region, molecular trajectories are no longer dominated by hard repulsive bounces but are instead overwhelmingly governed by glancing interactions and orbital deflections induced by the attractive tail. The proposed algorithm dynamically couples the microscopic impact parameter directly to the Chapman-Enskog collision integral \emph{W\textsuperscript{(2)}}(2) (\(T_{inv}^{*}\)), which intrinsically captures these complex attractive scattering dynamics. The long-range attractive forces effectively ``pull'' grazing molecules toward each other, significantly increasing the effective collision cross-section. Because dynamic viscosity is inversely proportional to this cross-section (\(\mu \propto T^{0.5}\) \emph{W\textsuperscript{(2)}}(2)), the LJ model correctly registers a distinct, non-linear reduction in momentum transfer efficiency at low temperatures. This yields a demonstrably lower, physically authentic effective viscosity in the cold expansion zones and the wake.

\begin{longtable}[]{@{}
  >{\raggedright\arraybackslash}p{(\columnwidth - 2\tabcolsep) * \real{0.4968}}
  >{\raggedright\arraybackslash}p{(\columnwidth - 2\tabcolsep) * \real{0.5032}}@{}}

\begin{minipage}[b]{\linewidth}\raggedright
\includegraphics[width=\linewidth,height=0.32\textheight,keepaspectratio]{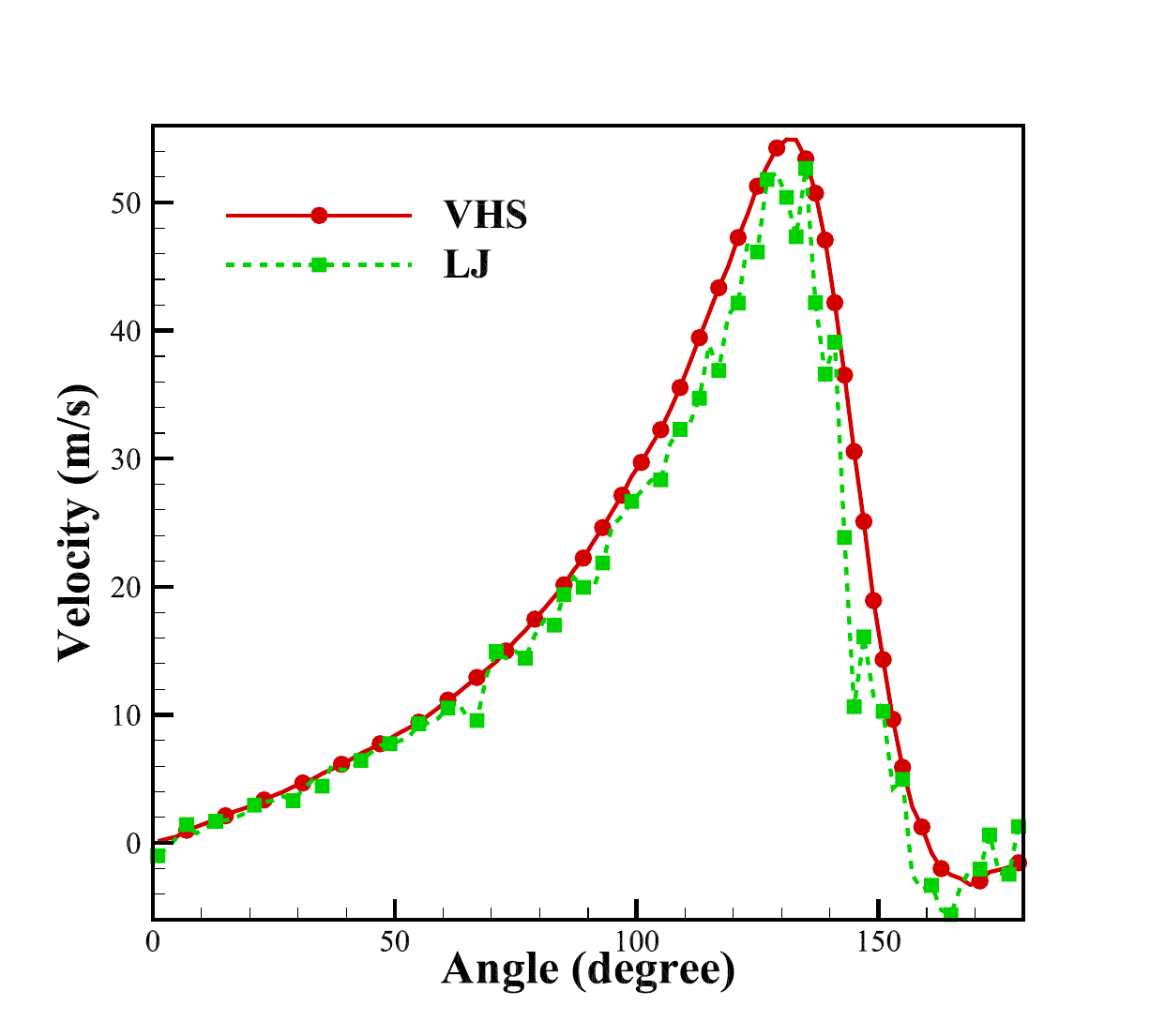}

a)
\end{minipage} & \begin{minipage}[b]{\linewidth}\raggedright
\includegraphics[width=\linewidth,height=0.32\textheight,keepaspectratio]{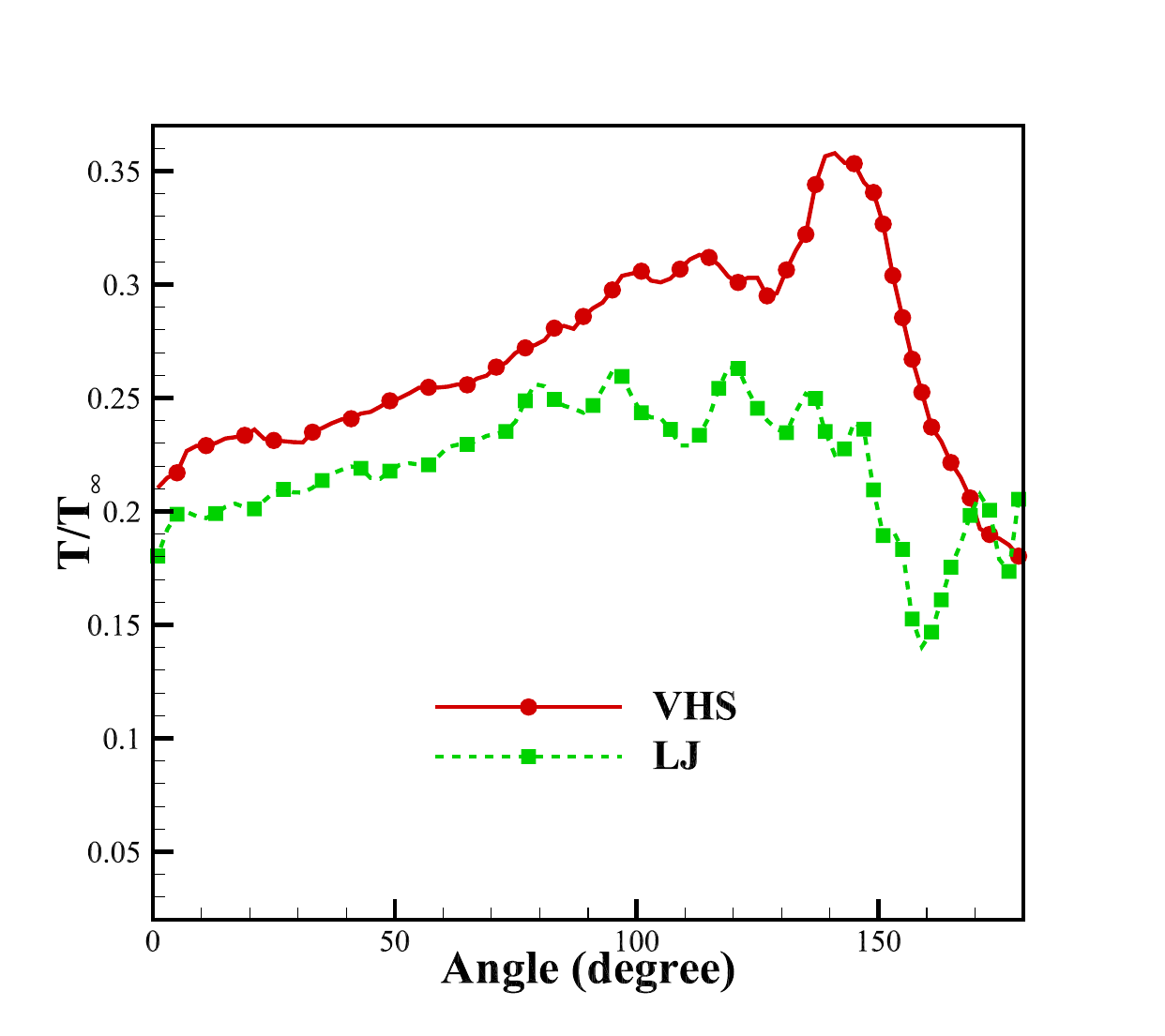}

b)
\end{minipage} \\

\end{longtable}

\autopapercaption{fig:fig25}{(a) Slip velocity and (b) normalized temperature over the cylinder surface, Mach=5.}

\begin{longtable}[]{@{}
  >{\raggedright\arraybackslash}p{(\columnwidth - 2\tabcolsep) * \real{0.5047}}
  >{\raggedright\arraybackslash}p{(\columnwidth - 2\tabcolsep) * \real{0.4953}}@{}}

\begin{minipage}[b]{\linewidth}\raggedright
\includegraphics[width=\linewidth,height=0.32\textheight,keepaspectratio]{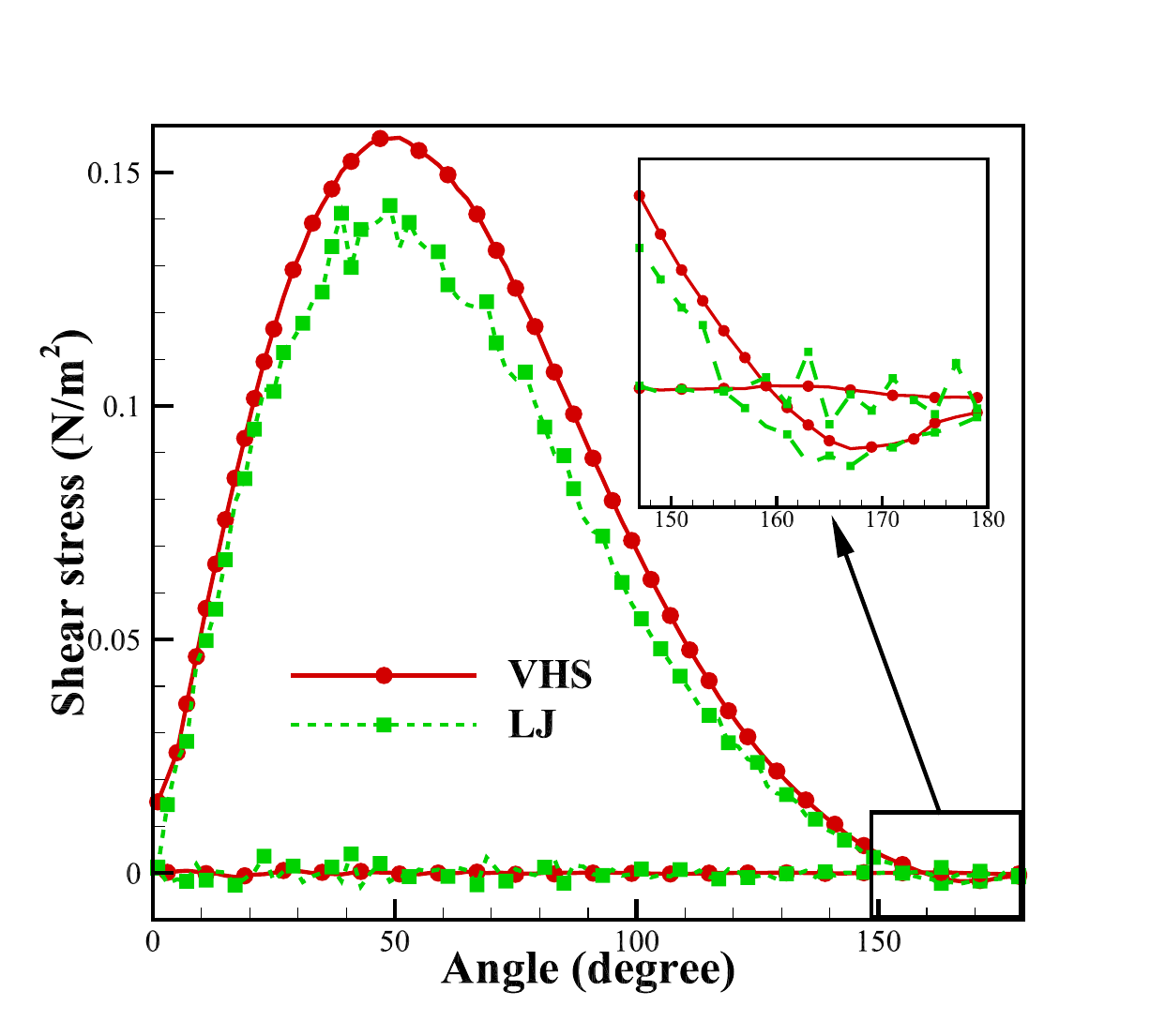}

\begin{enumerate}
\def\labelenumi{\alph{enumi})}
\tightlist
\item
\end{enumerate}
\end{minipage} & \begin{minipage}[b]{\linewidth}\raggedright
\includegraphics[width=\linewidth,height=0.32\textheight,keepaspectratio]{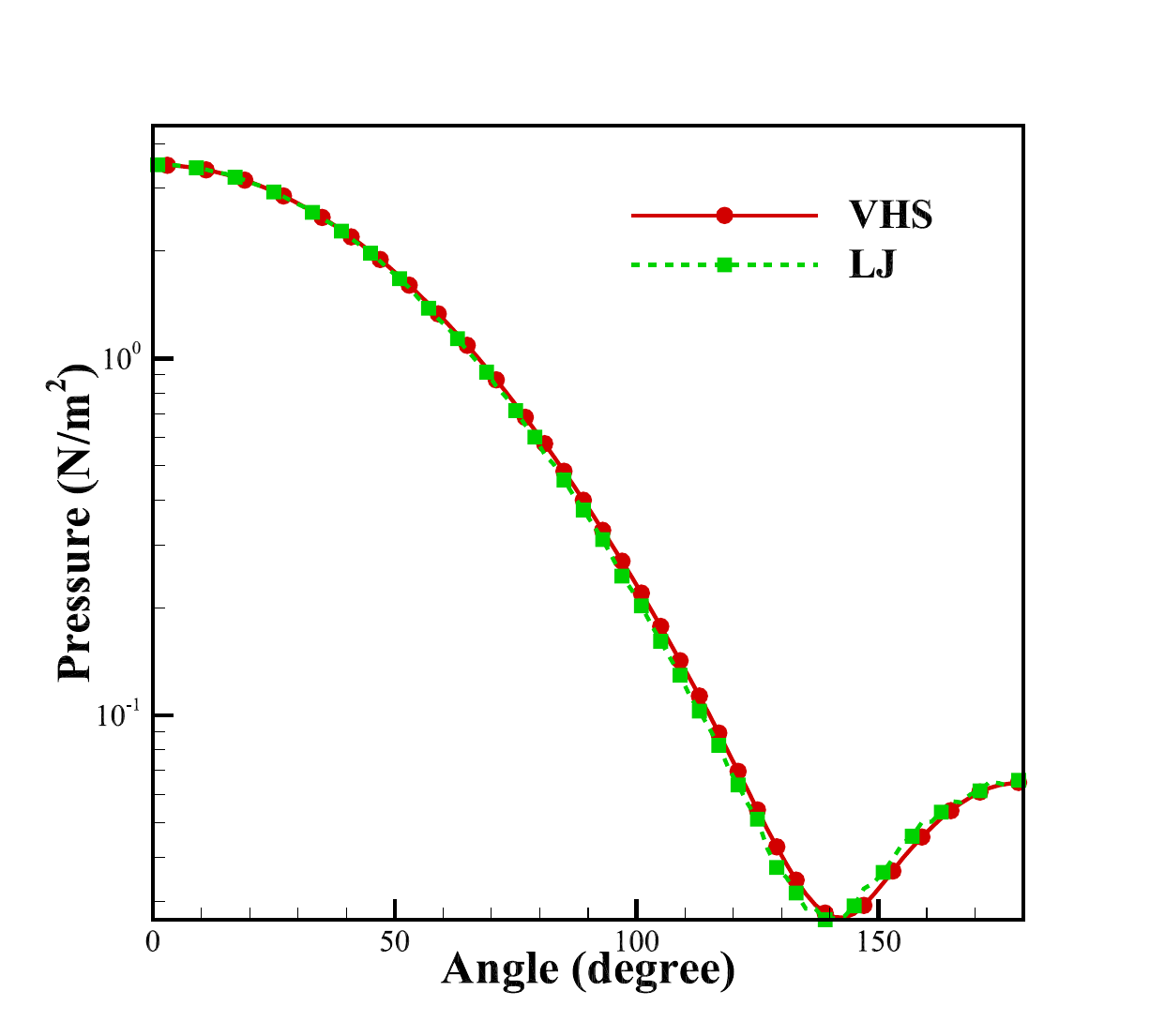}

\begin{enumerate}
\def\labelenumi{\alph{enumi})}
\setcounter{enumi}{1}
\tightlist
\item
\end{enumerate}
\end{minipage} \\

\end{longtable}

\begin{longtable}[]{@{}
  >{\raggedright\arraybackslash}p{(\columnwidth - 0\tabcolsep) * \real{1.0000}}@{}}

\begin{minipage}[b]{\linewidth}\raggedright
\includegraphics[width=\linewidth,height=0.32\textheight,keepaspectratio]{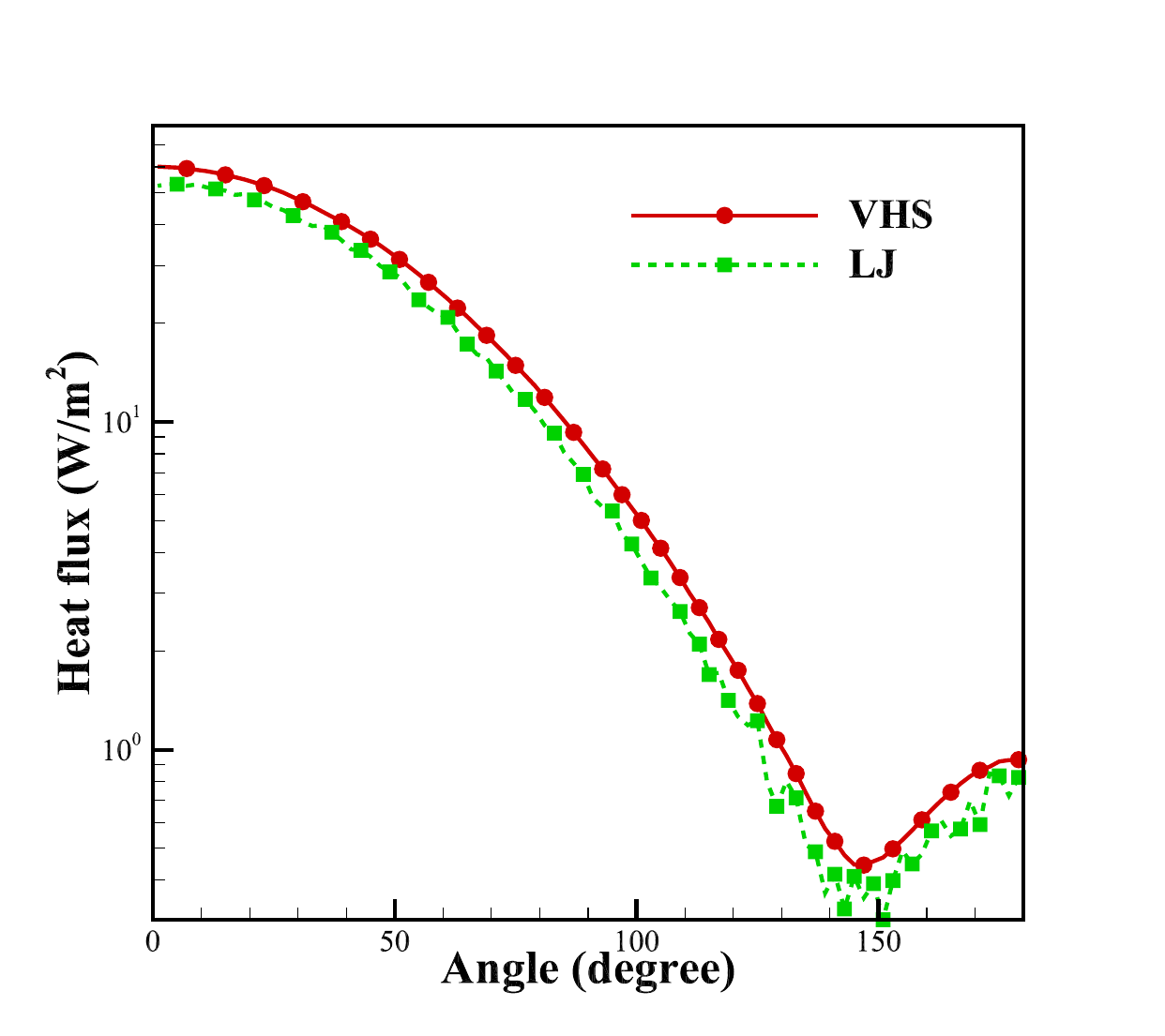}

c)
\end{minipage} \\

\end{longtable}

\autopapercaption{fig:fig26}{(a) Shear stress, (b) pressure, and (c) heat flux over the cylinder surface, Mach=5.}

Because the authentic LJ viscosity is markedly lower in the wake shear layers compared to the VHS overprediction, the transverse diffusion of momentum from the high-speed bypass flow into the recirculation zone is considerably attenuated. Lacking the artificial viscous entrainment present in the VHS model, the free shear layers in the LJ simulation sustain their downstream trajectory over a substantially longer spatial distance before accumulating sufficient cross-stream momentum to induce reattachment. This delayed momentum mixing perfectly justifies the significantly larger, more elongated, and physically precise wake topology visualized in Figs.~\ref{fig:fig24}--\ref{fig:fig26}.

The quantitative implications of these viscous momentum-diffusion phenomena are explicitly corroborated by the spatial profiles extracted along the symmetry axis, as shown in Fig.~\ref{fig:fig27}. These profiles, which detail the spatial recovery of flow properties downstream of the cylinder base, clearly show a much more gradual spatial gradient for the LJ model. The slower rate of macroscopic property recovery in the LJ results perfectly mirrors the restricted shear-driven momentum transfer, mathematically confirming a deeper, more extended aerodynamic wake. Interestingly, Fig.~\ref{fig:fig27}(b) illustrates the spatial variation of the Separation of Free Paths (SOF) along the wake centerline. The SOF is a fundamental quality metric in DSMC, defined as the mean physical distance between accepted collision pairs normalized by the local molecular mean free path. As observed, the SOF for the LJ potential is consistently higher than that of the VHS model throughout the expanded wake region, despite both cases utilizing identical grid resolutions and the same nominal number of simulator particles.

Rather than implying a deficiency in spatial resolution, this behavior is a profound manifestation of the distinct interaction physics in low-temperature environments and their physical coupling with the DSMC selection algorithm. The elevated SOF for the LJ potential is inherently driven by three synergistic factors affecting both the numerator and the denominator of the SOF ratio:

1. Mean Free Path Shrinkage (Denominator Effect)

In the highly expanded, ultra-cold recirculation zone trailing the cylinder, the relative translational velocities between interacting molecules drop significantly. Under these cryogenic conditions, the long-range van der Waals attractive tail of the LJ potential becomes kinematically dominant, drastically inflating the effective collision cross-section \({(\sigma}_{LJ}\) $\gg$ \(\sigma_{VHS}\)). Because the local mean free path is strictly inversely proportional to the collision cross-section (\(\lambda \propto \frac{1}{n\sigma}\)), the highly collisional nature of the LJ attractive well yields a substantially diminished local \(\lambda\). Normalizing the collision separation distance by a severely reduced \(\lambda\)\textsubscript{LJ} mathematically shifts the non-dimensional SOF ratio to a higher baseline.

2. Macroscopic Wake Expansion (Numerator Effect)

As established in the preceding macroscopic contour analyses, the LJ potential induces a significantly more expanded and elongated aerodynamic wake. This profound expansion results in a lower gas density in the leeward region than in the VHS model. Because the global particle weight \emph{F\textsubscript{num}} is fixed, this density deficit results in a proportionally lower number of simulated Particles Per Cell (PPC). Geometrically, when fewer particles populate a constant cell volume, the absolute physical distance between any randomly distributed particles inherently increases, thereby enlarging the absolute collision separation distance.

3. Elevated Collision Frequency and Sub-cell Depletion

From an algorithmic standpoint, the enlarged effective cross-section in the LJ model directly translates into a considerably higher instantaneous pair-acceptance probability. To accommodate this increased physical collision frequency, the stochastic selection algorithm must process and accept more candidate pairs. In DSMC routines that use spatial sub-cells, accommodating more successful collisions in a highly rarefied region quickly depletes the available nearest-neighbor candidates in the immediate sub-cell. Consequently, the algorithm is dynamically forced to systematically "reach out" to adjacent sub-cells to pair the remaining molecules, directly increasing the mean physical distance between the accepted collision partners.

\begin{longtable}[]{@{}
  >{\raggedright\arraybackslash}p{(\columnwidth - 6\tabcolsep) * \real{0.5022}}
  >{\raggedright\arraybackslash}p{(\columnwidth - 6\tabcolsep) * \real{0.0017}}
  >{\raggedright\arraybackslash}p{(\columnwidth - 6\tabcolsep) * \real{0.4939}}
  >{\raggedright\arraybackslash}p{(\columnwidth - 6\tabcolsep) * \real{0.0021}}@{}}

\multicolumn{2}{@{}>{\raggedright\arraybackslash}p{(\columnwidth - 6\tabcolsep) * \real{0.5040} + 2\tabcolsep}}{%
\begin{minipage}[b]{\linewidth}\raggedright
\includegraphics[width=\linewidth,height=0.32\textheight,keepaspectratio]{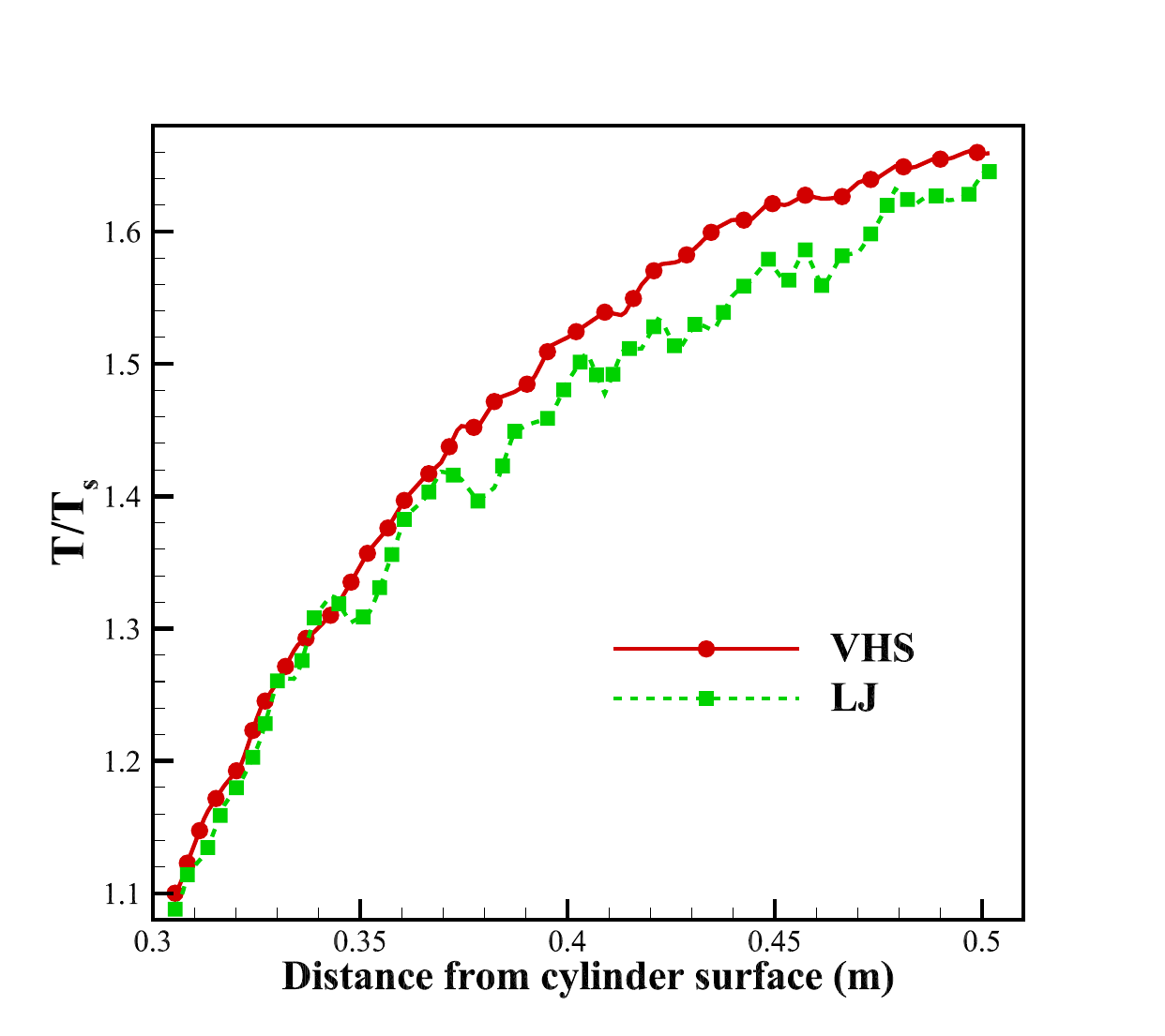}

\begin{enumerate}
\def\labelenumi{\alph{enumi})}
\item
  Normalized temperature
\end{enumerate}
\end{minipage}} & \multicolumn{2}{>{\raggedright\arraybackslash}p{(\columnwidth - 6\tabcolsep) * \real{0.4960} + 2\tabcolsep}@{}}{%
\begin{minipage}[b]{\linewidth}\raggedright
\includegraphics[width=\linewidth,height=0.32\textheight,keepaspectratio]{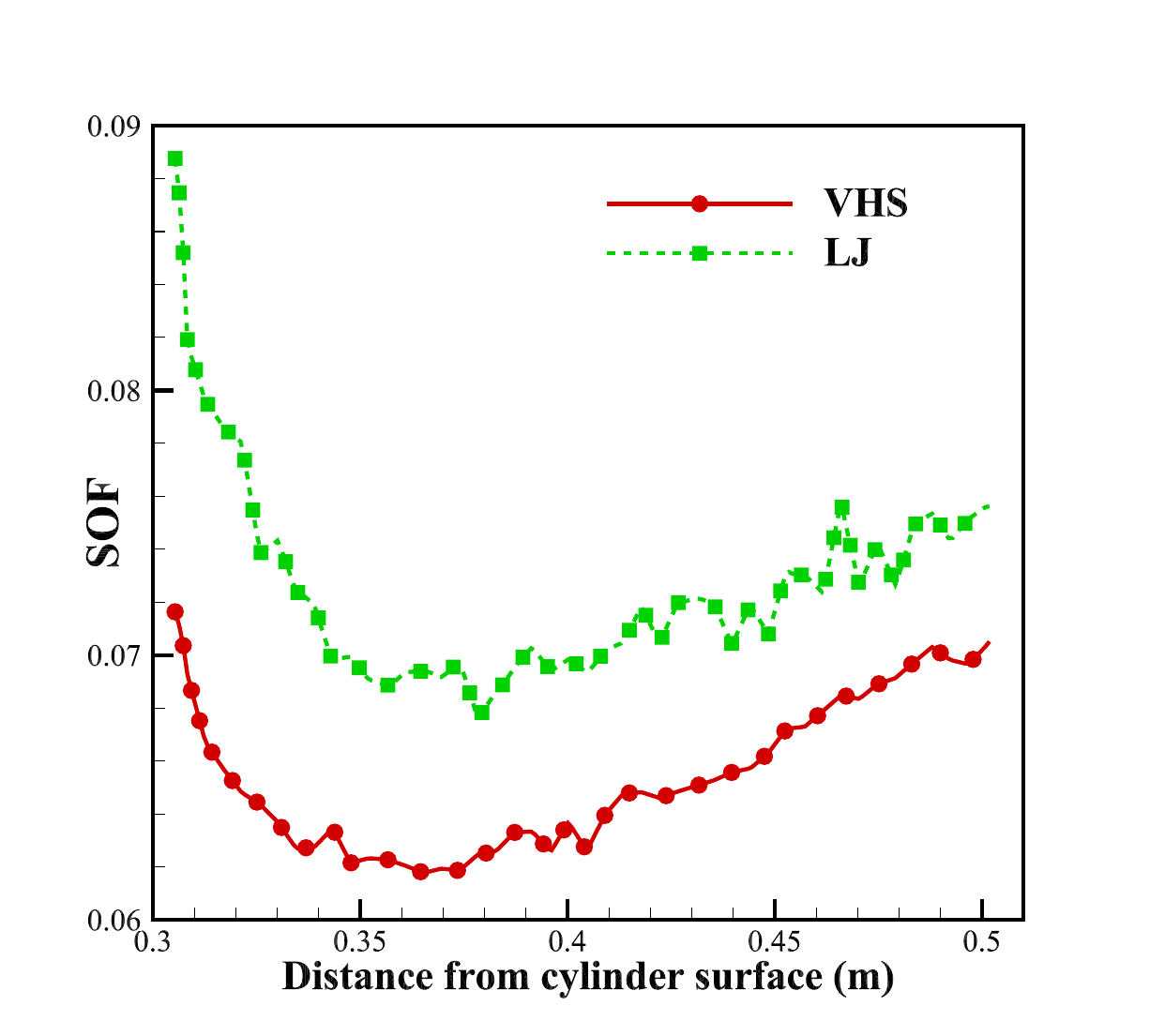}

\begin{enumerate}
\def\labelenumi{\alph{enumi})}
\setcounter{enumi}{1}
\item
  Separation of free paths (SOF)
\end{enumerate}
\end{minipage}} \\

\begin{minipage}[t]{\linewidth}\raggedright
\includegraphics[width=\linewidth,height=0.32\textheight,keepaspectratio]{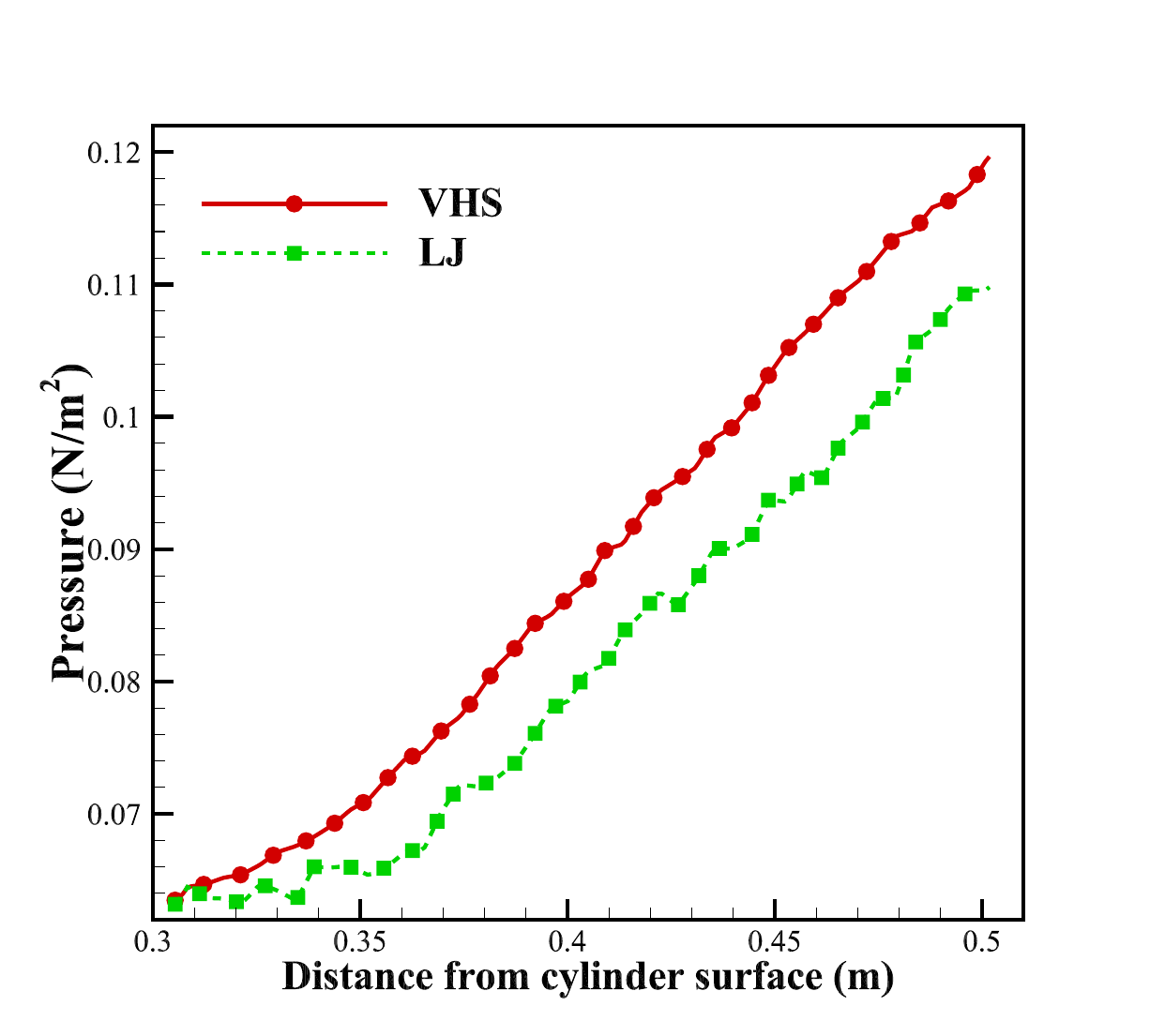}

\begin{enumerate}
\def\labelenumi{\alph{enumi})}
\setcounter{enumi}{2}
\item
  Pressure
\end{enumerate}
\end{minipage} & \multicolumn{2}{>{\raggedright\arraybackslash}p{(\columnwidth - 6\tabcolsep) * \real{0.4956} + 2\tabcolsep}}{%
\begin{minipage}[t]{\linewidth}\raggedright
\includegraphics[width=\linewidth,height=0.32\textheight,keepaspectratio]{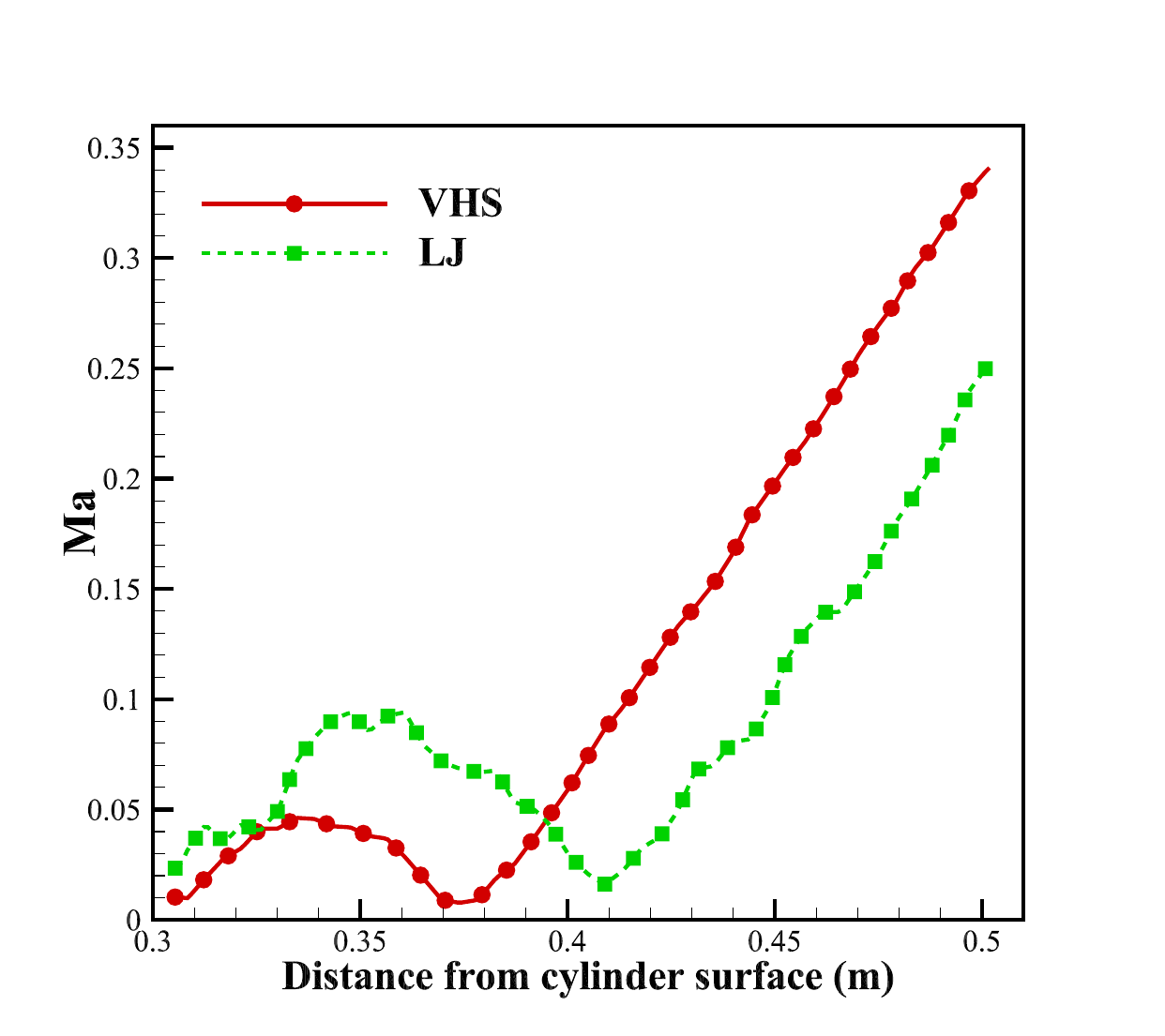}

\begin{enumerate}
\def\labelenumi{\alph{enumi})}
\setcounter{enumi}{3}
\item
  Mach number
\end{enumerate}
\end{minipage}} & \\
\end{longtable}

\autopapercaption{fig:fig27}{Flow properties behind the cylinder, Ma=5.}

Ultimately, the elevated SOF curve for the LJ model perfectly reflects the physical reality of attractive intermolecular forces: the cold recirculation zone becomes far more "collisional," inherently demanding collision pairings over larger physical intra-cell distances. Importantly, despite this relative increase, the maximum SOF for the LJ model strictly remains below the conventional reliability threshold of 0.1 across the entire domain. This unequivocally confirms that the employed grid resolution is sufficiently fine, and the spatial discretization error remains well within the standard acceptable limits for high-fidelity DSMC simulations.

To quantitatively evaluate the impact of the chosen intermolecular potential on the global aerodynamic forces, the total time-averaged drag force parallel to the freestream was computed. The total drag evaluates to 0.396 N for the conventional VHS model and 0.391 N for the physically rigorous LJ potential. This represents a marginal relative reduction of approximately 1.2\%. At first glance, this negligible discrepancy in the total integrated drag appears paradoxical when juxtaposed with the local wall shear stress profiles presented in Fig.~\ref{fig:fig26}(a). As observed, the LJ potential predicts a distinctly lower shear stress magnitude across the attached windward forebody, with the peak shear stress near 50° reduced by over 10\%-15\% compared to the VHS model.

The fundamental physical reasoning that resolves this apparent aerodynamic paradox lies in the classical decomposition of total drag into normal pressure (form) drag and tangential skin friction (viscous) drag. For aerodynamically blunt bodies, such as a circular cylinder operating in a Mach 5 hypersonic regime, the total drag is overwhelmingly dominated by the immense normal pressure generated immediately behind the strong, detached bow shock. In such bluff-body configurations, pressure drag typically accounts for more than 90\%-95\% of the total aerodynamic resistance. As shown in Fig.~\ref{fig:fig26}(b), the pressure distributions predicted by VHS and LJ are almost identical.

In the highly compressed stagnation region over the windward face, thermal collision energies are exceptionally high. Under these hyper-energetic conditions, interacting molecules deeply penetrate their respective potential fields, rendering the long-range attractive van der Waals tail of the LJ potential kinematically negligible. Consequently, molecular scattering is dictated almost entirely by the steep, short-range Pauli repulsive core of the potential. Because both the phenomenological VHS model and the LJ model fundamentally replicate this strong repulsive behavior during high-energy collisions, the resulting macroscopic surface pressure distribution and thus the dominant pressure drag component remains virtually identical across both simulations.

Conversely, the tangential momentum accommodation that dictates the surface shear stress (and the subsequent skin-friction drag) is acutely sensitive to local dynamic gas viscosity and velocity gradients within the boundary layer. As the flow accelerates and expands around the cylinder's shoulder, it cools significantly. In this lower-temperature regime, the dynamically coupled cross-section of the LJ potential accurately captures the increased probability of grazing collisions induced by attractive forces, leading to a physically realistic, lower effective viscosity compared to the artificially extrapolated VHS model. This localized reduction in viscosity directly attenuates the transverse momentum diffusion from the gas to the solid wall, resulting in the suppressed shear-stress profile observed for the LJ model.

However, because the integrated skin friction drag constitutes a minuscule fraction of the overall force budget on a hypersonic cylinder, a pronounced relative reduction in the local tangential shear stress mathematically translates to a fractional change restricted exclusively to the third decimal place of the total drag coefficient. Furthermore, the inset in Fig.~\ref{fig:fig26}(a) highlights the shear stress distribution in the severely rarefied, separated leeward wake. While the distinct negative shear stress fluctuations exhibited by the LJ model corroborate a more pronounced flow reversal and an elongated wake topology, the absolute magnitude of the stresses in this base region O(10\textsuperscript{-2}) is intrinsically minute compared to the windward peak, rendering its contribution to the global drag mathematically imperceptible.

Ultimately, these results underscore a critical insight for high-fidelity non-equilibrium aerodynamics: while gross macroscopic loads (such as total drag) on bluff bodies may appear highly insensitive to the choice of the intermolecular collision model due to pressure dominance, the underlying local viscous transport mechanisms, boundary layer friction, and separated wake topologies are profoundly sensitive and require the physical fidelity of advanced models like the Lennard--Jones potential.

\subsection{Diffusion validation and surrogate baseline comparison}\label{subsec:diffusion-validation-baselines}
\begin{figure}[!htbp]
\centering
\includegraphics[width=0.82\textwidth,trim={0 0 0 20pt},clip]{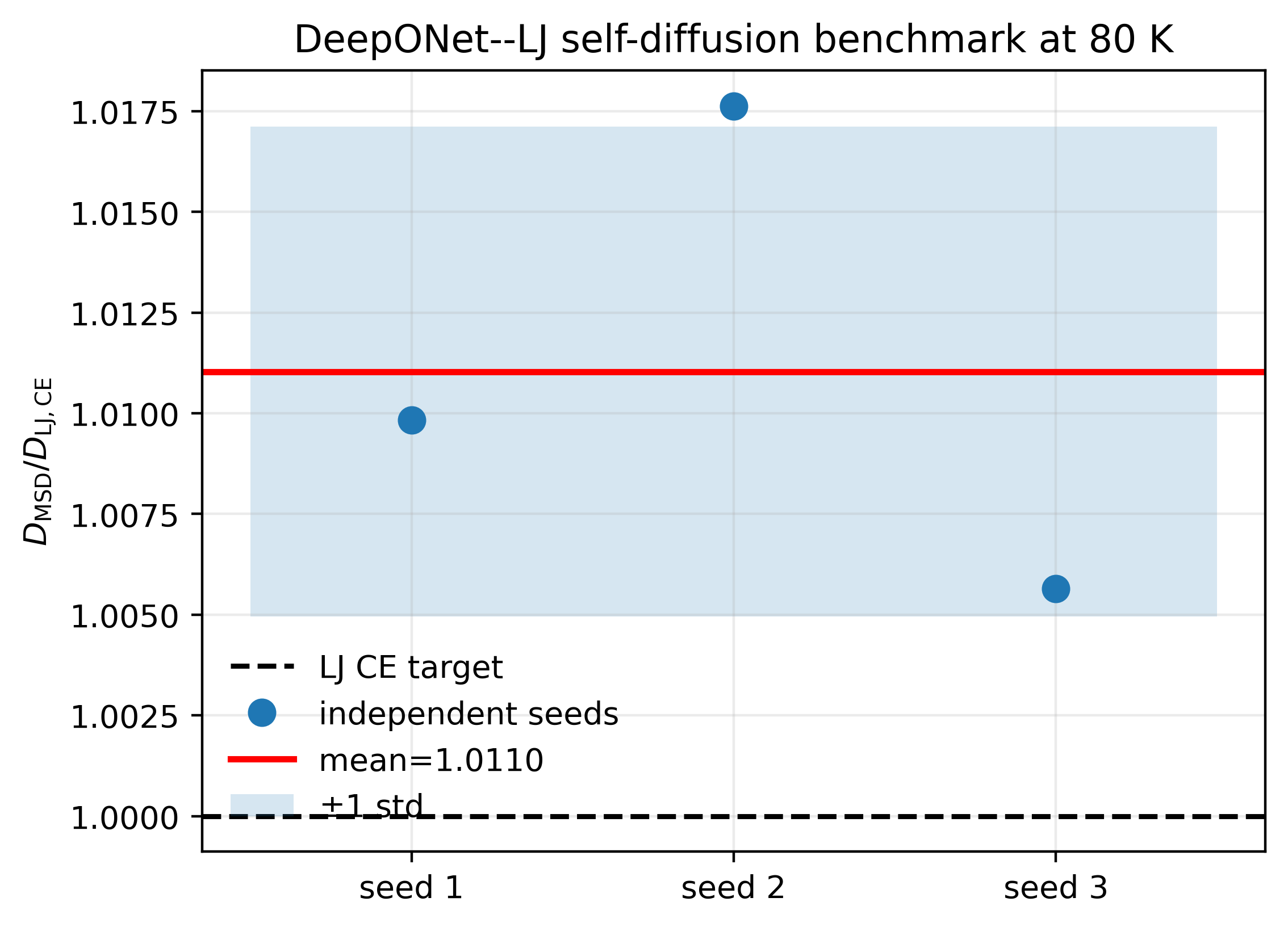}
\caption{Low-temperature self-diffusion validation of the DeepONet--LJ DSMC scattering kernel at \(T=80\,\mathrm{K}\). Three independent MSD realizations are compared with the Chapman--Enskog LJ diffusion coefficient. The mean ratio is \(D_{\mathrm{MSD}}/D_{\mathrm{LJ,CE}}=1.0110\), corresponding to a mean deviation of approximately \(1.10\%\).}
\label{fig:msd_low_temperature_diffusion}
\end{figure}

\begin{figure}[!htbp]
\centering
\includegraphics[width=0.99\textwidth,trim={0 0 0 48pt},clip]{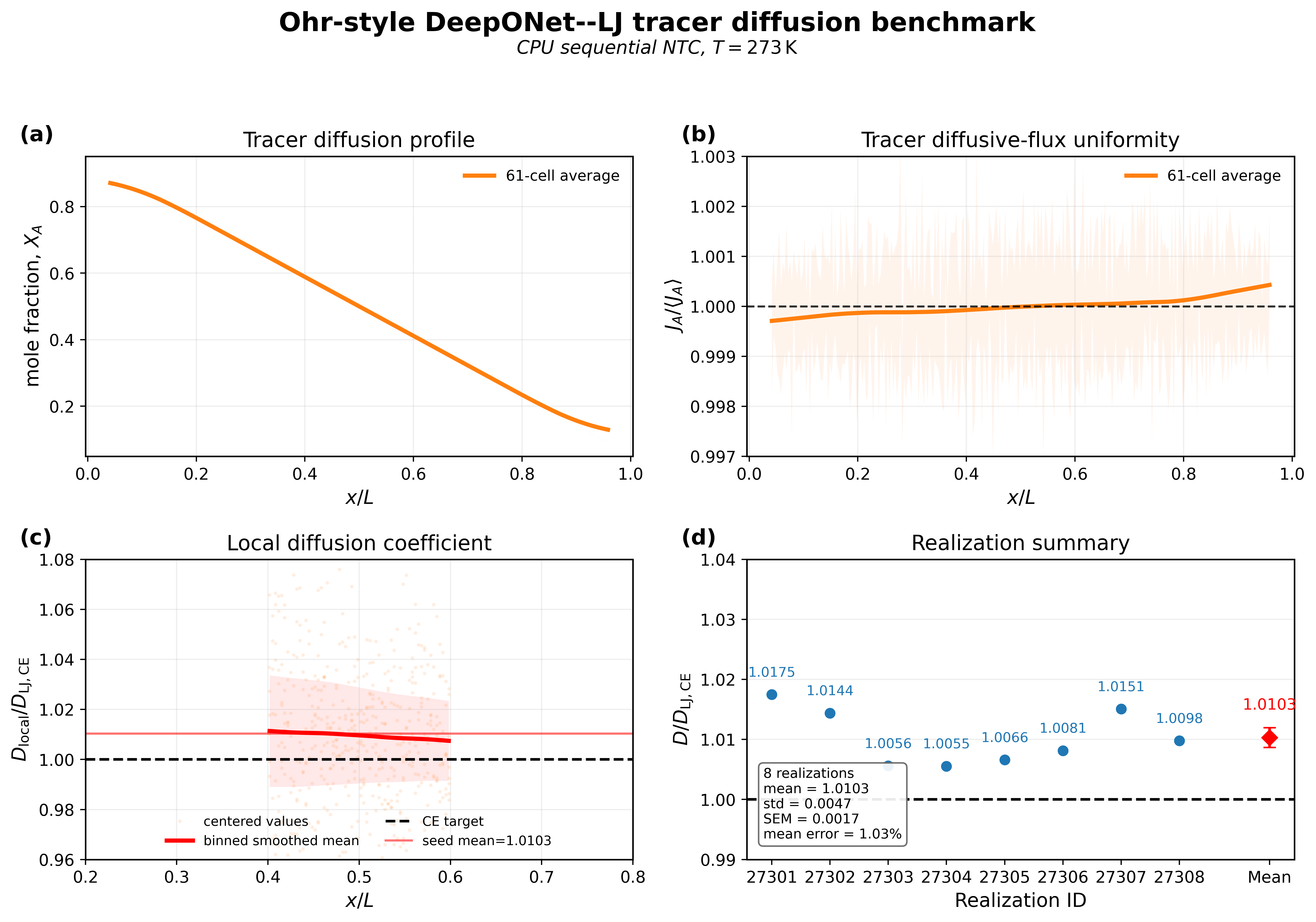}
\caption{Two-label tracer-diffusion validation of the DeepONet--LJ DSMC kernel at \(T=273\,\mathrm{K}\). Panel (a) shows the tracer mole-fraction profile, panel (b) shows the normalized diffusive-flux uniformity, panel (c) shows the local diffusion coefficient normalized by the Chapman--Enskog LJ target, and panel (d) summarizes the independent realizations. The reported mean is \(D/D_{\mathrm{LJ,CE}}=1.0103\). Curves are spatially coarse-grained only for visualization; the reported diffusion coefficient is computed from the centered local estimates.}
\label{fig:ohr_tracer_diffusion}
\end{figure}

\begin{figure}[!htbp]
\centering
\includegraphics[width=0.99\textwidth,trim={0 0 0 48pt},clip]{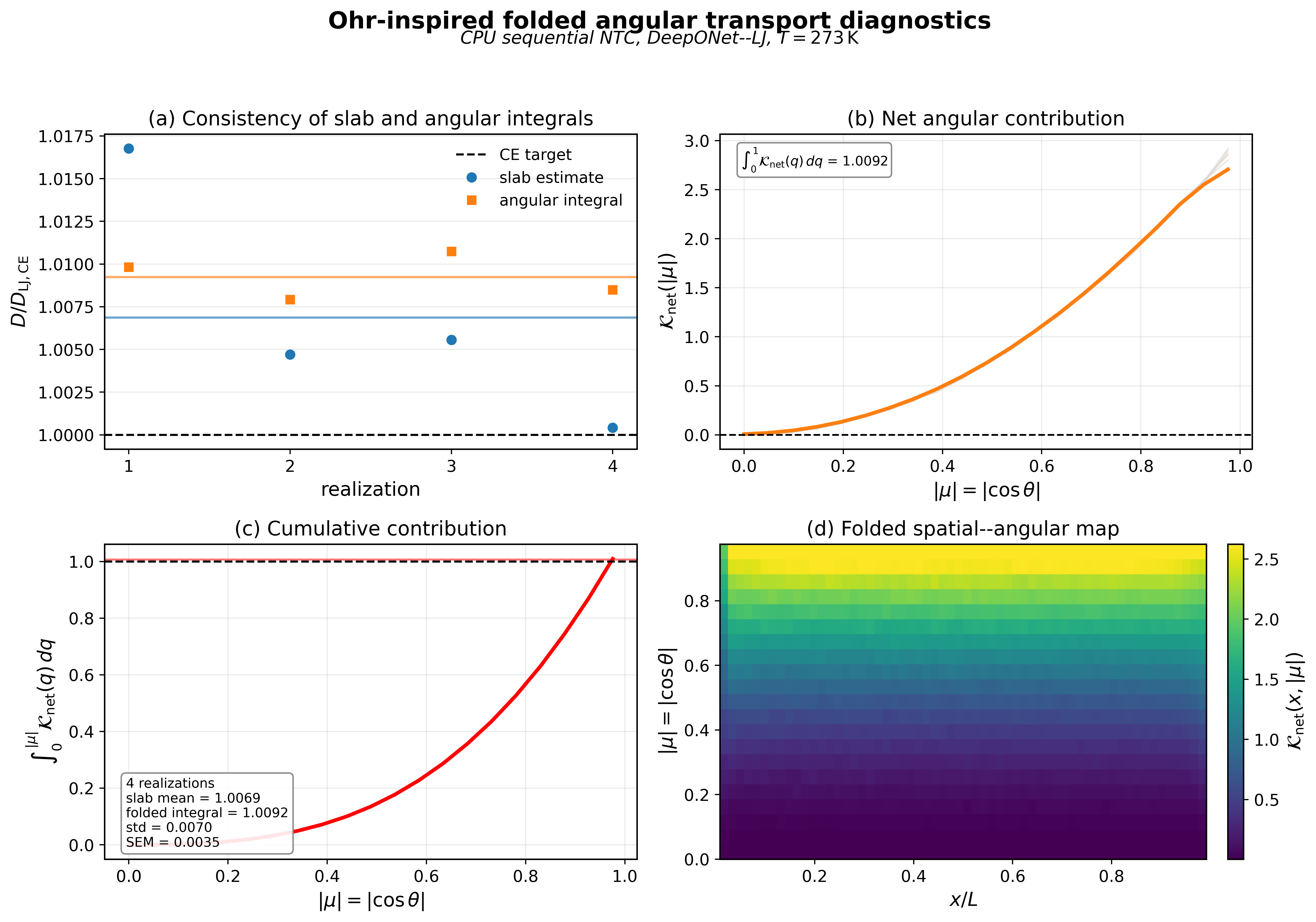}
\caption{Folded angular decomposition of the tracer-diffusion coefficient. Panel (a) compares the normalized diffusion coefficient obtained from the slab-flux estimate with that obtained by integrating the folded angular contribution over four independent realizations. Panel (b) shows the net folded angular contribution density, \(\mathcal{K}_{\mathrm{net}}(|\mu|)\), where \(|\mu|=|\cos\theta|\) measures the streamwise alignment of the molecular velocity; the raw binned estimate and the smoothed/integrated representation differ slightly near \(|\mu|\to 1\) because of finite-sample endpoint effects. Panel (c) reports the cumulative angular integral, demonstrating how the diffusion coefficient is accumulated from increasingly streamwise molecular directions. Panel (d) shows the folded spatial--angular contribution map over the central slab, confirming that the contribution is nearly uniform in the physical coordinate and is therefore not a localized sampling artifact.}
\label{fig:folded_angular_transport}
\end{figure}

The direct scattering-angle validation in Section~\ref{subsec:training-dataset-sampling}, including Fig.~\ref{fig:scattering_angle_surrogate_validation}, confirms that the trained DeepONet accurately reproduces the LJ angular map used in the DSMC collision step. A remaining question is whether the same LJ scattering kernel also preserves diffusion-controlled transport, particularly at low temperature where the attractive tail of the potential dominates low-energy and grazing encounters. To address this point, two independent diffusion benchmarks were added to the validation set. The corresponding diffusion diagnostics are reported in Figs.~\ref{fig:msd_low_temperature_diffusion}--\ref{fig:folded_angular_transport}, progressing from a homogeneous MSD-based self-diffusion test to an Ohr-style tracer-slab benchmark and a final angular-transport decomposition. The first is a homogeneous self-diffusion benchmark based on the mean-square displacement (MSD), and the second is a two-label tracer-diffusion slab inspired by the diffusion tests used in DSMC verification and by the recent angular-collision diagnostics of Ohr~\cite{ohr2023representative}. In both cases, the reference value is the first-order Chapman--Enskog (CE) LJ self-diffusion coefficient computed with the same collision-integral data used in the transport closure. For a monatomic gas of identical particles, this reference coefficient is written as~\cite{hirschfelder1948transport,chapman1970nonuniform}
\setcounter{equation}{26}
\begin{equation}
D_{\mathrm{LJ,CE}}(T,n)=\frac{3}{8 n d_{\mathrm{LJ}}^{2}\Omega_{D}^{(1,1)}(T^*)}
\left(\frac{k_B T}{\pi m}\right)^{1/2},
\label{eq:ce_lj_diffusion}
\end{equation}
where \(n\) is the number density, \(m\) is the molecular mass, \(d_{\mathrm{LJ}}\) is the LJ diameter, \(T^*=k_B T/\varepsilon_{\mathrm{LJ}}\), and \(\Omega_{D}^{(1,1)}\) is the reduced LJ diffusion collision integral. The corresponding tracer-slab measurement uses Fick's law,
\begin{equation}
J_A=-nD_{\mathrm{LJ,CE}}\frac{dX_A}{dx},
\label{eq:fick_diffusion_reference}
\end{equation}
with \(J_A=n_A(u_A-u)\), so that the same CE target can be compared with both the MSD and the tracer-flux measurements.

The first benchmark is deliberately performed at \(T=80\,\mathrm{K}\), where the argon reduced temperature is close to the regime in which attractive LJ interactions are most important. A homogeneous periodic argon system with approximately \(10^5\) simulator particles was evolved using the same DeepONet--LJ collision kernel and the NTC pair-selection procedure. The time step was kept small relative to the mean collision time, and the MSD was computed from unwrapped particle trajectories after the initial ballistic transient. Three independent heavy realizations with seeds 8101, 8102, and 8103 were used. The resulting MSD-based diffusion coefficients were \(1.3063\), \(1.3164\), and \(1.3009\,\mathrm{m^2/s}\), compared with the Chapman--Enskog target \(D_{\mathrm{LJ,CE}}=1.2936\,\mathrm{m^2/s}\). The mean ratio is
\begin{equation}
D_{\mathrm{MSD}}/D_{\mathrm{LJ,CE}}=1.0110
\tag{29}
\end{equation}
with a standard deviation of \(0.0061\) and a standard error of \(0.0035\). Thus, the low-temperature self-diffusion coefficient is recovered to within approximately \(1.10\%\), confirming that the neural LJ scattering kernel preserves diffusion-relevant angular statistics in a regime where VHS-type purely repulsive models are least reliable. This agreement is visualized in Fig.~\ref{fig:msd_low_temperature_diffusion}, where the three independent MSD realizations remain clustered around the Chapman--Enskog LJ reference, indicating that the DeepONet--LJ kernel reproduces the diffusive long-time slope rather than merely matching an averaged scalar coefficient.

\FloatBarrier

The second diffusion benchmark follows an Ohr-style tracer-slab construction~\cite{ohr2023representative}. Two labels of the same LJ gas are maintained at different nominal mole fractions at the two boundaries while the mixture temperature is fixed at \(273\,\mathrm{K}\). Since the two labels have identical molecular properties, the benchmark isolates self-diffusion without introducing mixture-specific mass or diameter effects. The slab length is \(L=0.3883\,\mathrm{m}\), the number density is \(n_0=10^{20}\,\mathrm{m^{-3}}\), and the inverse Knudsen number is \(\mathrm{Kn}^{-1}=30\). The one-dimensional domain was discretized with 300 cells and 240 simulator particles per cell on average, giving approximately \(7.2\times10^4\) particles. The time step was \(\Delta t/\tau_c=0.01\), where \(\tau_c\) is the reference mean collision time; each realization used 360000 steps, a 90000-step burn-in interval, and sampling every 8 steps. The local tracer flux \(J_A=n_A(u_A-u)\), the fitted mole-fraction gradient, and the local Fickian diffusion coefficient were computed over the central region of the slab to avoid boundary-layer contamination. Eight independent realizations give
\begin{equation}
D_{\mathrm{Ohr}}/D_{\mathrm{LJ,CE}}=1.0103
\tag{30}
\end{equation}
with a standard deviation of \(0.0047\) and a standard error of \(0.0017\). The spatially averaged tracer flux is uniform to within small stochastic fluctuations, and the local diffusion coefficient remains centered on the Chapman--Enskog target. This benchmark directly addresses the diffusion-cross-section concern: replacing the exact LJ scattering integral with the DeepONet surrogate does not degrade the diffusion coefficient. Figure~\ref{fig:ohr_tracer_diffusion} supports this conclusion by showing that the tracer mole-fraction profile is nearly linear in the sampled central region, the normalized diffusive flux remains spatially uniform, and the local diffusion coefficient fluctuates around the Chapman--Enskog LJ target across independent realizations.

\FloatBarrier

The remaining scatter in the local diffusion coefficient should not be interpreted as a systematic failure of the DeepONet--LJ collision kernel. This scatter is expected in a finite-particle DSMC tracer-diffusion measurement because the diffusive flux is obtained from relatively small differences between two labeled populations and is sensitive to rare collision events associated with the attractive LJ tail. The important point is that the spatially averaged diffusion coefficient and the ensemble of independent realizations remain centered around the Chapman--Enskog LJ target. Therefore, the tracer-slab benchmark confirms that the global Fickian transport coefficient is recovered even though the local cellwise estimates retain finite-sampling fluctuations.

For further consistency, the tracer-slab calculation was decomposed into a folded angular contribution density. Defining \(\mu=\cos\theta=v_x/|\mathbf{v}|\), the signed contribution \(\mathcal{K}(\mu)=d(D/D_{\mathrm{LJ,CE}})/d\mu\) was paired as \(\mathcal{K}_{\mathrm{net}}(|\mu|)=\mathcal{K}(+|\mu|)+\mathcal{K}(-|\mu|)\). This decomposition is useful because diffusion is not controlled only by the total number of accepted collisions, but also by how the post-collision velocity directions redistribute streamwise molecular displacement. Therefore, a correct LJ scattering surrogate should recover not only the scalar diffusion coefficient, but also the angular structure through which that coefficient is produced.

The four panels in Fig.~\ref{fig:folded_angular_transport} provide complementary checks of this requirement. Panel (a) compares the diffusion coefficient obtained from the slab flux with the value obtained by integrating the folded angular contribution for each realization. The close clustering of both estimates around the Chapman--Enskog target shows that the angular diagnostic is consistent with the macroscopic tracer-diffusion measurement. Panel (b) shows the net folded angular contribution density, \(\mathcal{K}_{\mathrm{net}}(|\mu|)\). The increase toward \(|\mu|\approx 1\) indicates that particles with larger streamwise velocity components contribute most strongly to the diffusive flux, as expected for one-dimensional tracer diffusion. The gray curve represents the raw binned angular contribution, while the orange curve represents the smoothed/integrated representation used for the cumulative angular integral. Their small separation near \(|\mu|\to 1\) is an endpoint-bin effect caused by finite sampling and smoothing of a sharply increasing contribution near the streamwise direction; it does not indicate a systematic bias because the integrated value remains consistent with the slab estimate. Panel (c) gives the cumulative angular integral and shows how the normalized diffusion coefficient is built up as increasingly streamwise molecular directions are included. Finally, panel (d) maps the folded spatial--angular contribution over the central slab and confirms that the angular contribution is nearly uniform in \(x/L\), so the recovered diffusion coefficient is not produced by a localized spatial artifact. Together, these diagnostics show that the DeepONet--LJ kernel preserves the angular transport structure underlying diffusion, not merely the final scalar value of \(D\).

\FloatBarrier

A final question is whether the operator-based DeepONet architecture is genuinely needed for the LJ scattering surrogate, or whether a simpler pointwise regression model would be sufficient. This distinction is important because the LJ deflection map is not a smooth single-regime function: it contains rapid variations near the attractive--repulsive transition, low-angle grazing events, and strong dependence on both reduced collision energy and impact parameter. A surrogate used inside DSMC must therefore reproduce not only the mean scattering response, but also the angular-error tails, since rare but systematic errors in the deflection angle can accumulate over millions of collision events and affect transport-level observables such as diffusion, shear stress, and shock structure. To isolate the architectural effect, the trained DeepONet was compared with two simpler alternatives evaluated on the same held-out LJ scattering samples: a standard multilayer perceptron (MLP) and a bilinear lookup table.

\begin{figure}[!htbp]
\centering
\includegraphics[width=0.99\textwidth,trim={0 0 0 42pt},clip]{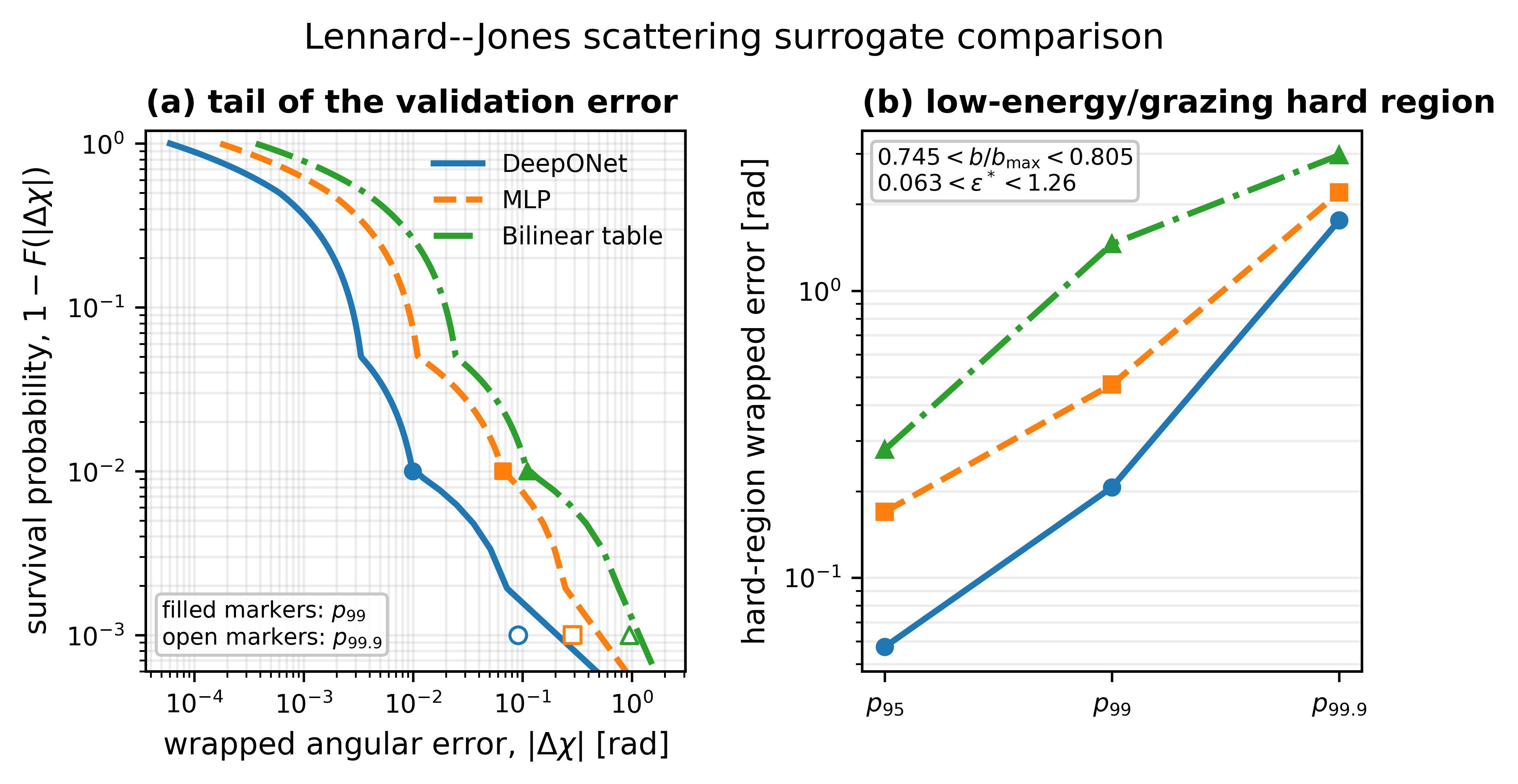}
\caption{Comparison of the DeepONet--LJ scattering surrogate with a conventional MLP and a bilinear lookup table on held-out scattering-angle samples. Panel (a) reports the survival probability of the wrapped angular error, while panel (b) focuses on the low-energy/grazing region, \(0.745<b/b_{\max}<0.805\) and \(0.063<\epsilon^*<1.26\), where the LJ scattering map is most difficult to represent.}
\label{fig:deeponet_baseline_comparison}
\end{figure}

As shown in Fig.~\ref{fig:deeponet_baseline_comparison}, DeepONet gives the narrowest error distribution and the smallest high-percentile wrapped-angle errors. The improvement is most important in the non-smooth regions of the scattering map, where a pointwise MLP tends to smooth sharp angular variations and a lookup table is limited by grid resolution. The branch--trunk decomposition of DeepONet therefore provides a more stable representation of the LJ scattering operator over the full collision-parameter space. This comparison justifies the use of DeepONet in the production DSMC calculations rather than a simpler surrogate, particularly for strongly non-equilibrium flows where the collision-energy and impact-parameter distributions are broad.

\FloatBarrier

\section{Concluding remarks}\label{sec:concluding-remarks}

This study established a generalized and computationally efficient framework for integrating the physically rigorous Lennard--Jones (LJ) potential into the Direct Simulation Monte Carlo (DSMC) method. The framework is built on two methodological advances. The first is a Variable Effective Diameter (VED) model that provides a viscosity-consistent collision-rate closure for selecting collision pairs in the presence of a long-range LJ potential. The second is a Deep Operator Network (DeepONet) surrogate that replaces the computationally expensive numerical evaluation of the LJ scattering angle while preserving the standard elastic post-collision update. In addition to validating these two components in canonical rarefied-flow simulations, we performed direct scattering-angle and diffusion benchmarks to verify that the learned LJ angular kernel preserves transport behavior beyond viscosity matching.

\begin{enumerate}
\def\labelenumi{\arabic{enumi}.}
\item
  Physical fidelity in strong non-equilibrium shock waves: Simulations of normal shock waves in helium and argon demonstrated that the LJ-based framework accurately reproduces experimental density and temperature profiles. The method correctly distinguishes between the repulsive-dominated behavior of helium and the stronger attractive--repulsive dynamics of argon, validating the viscosity-consistent VED approach as a robust bridge between Chapman--Enskog transport closure and DSMC collision selection.

\item
  Dominance of attractive forces in cryogenic shear flows: A key physical finding emerged from the supersonic Couette flow simulation with cryogenic walls at \(40\,\mathrm{K}\). In this regime, where the reduced temperature drops below unity, the attractive well of the LJ potential plays a governing role. The LJ model predicts a smaller shear stress than the standard VHS model, highlighting the inability of purely repulsive models to capture the correct collision behavior in low-temperature shear layers.

\item
  Robustness in high-energy hypersonic regimes: In Mach 10 flow over a cylinder, the LJ and VHS models produced nearly identical wake structures and surface properties. This agreement is physically consistent with the high thermal state of the wake, where temperatures remain sufficiently large for repulsive collisions to dominate. The result confirms that the proposed LJ framework smoothly approaches the standard repulsive limit in high-temperature engineering flows without introducing numerical artifacts.

\item
  Cryogenic wake sensitivity at lower hypersonic temperature: For hypersonic flow over a cylinder at Mach 5 with cryogenic wall and freestream temperatures, the LJ and VHS predictions diverge more strongly. The VHS model predicts a more viscously dissipated and prematurely closed wake, whereas the LJ model yields a larger and more elongated vortex. This difference demonstrates that realistic attractive interactions are essential when high-speed compression is coupled with cold wake expansion.

\item
  DeepONet acceleration of LJ scattering dynamics: The DeepONet surrogate was trained to reproduce the LJ scattering-angle map generated from the exact Matsumoto--Koura integral. This surrogate does not replace the VED model and does not alter the DSMC collision structure; it only accelerates the deflection-angle evaluation for accepted collision pairs. Because the standard elastic velocity transformation is retained, momentum and kinetic energy conservation are preserved. The surrogate accelerates the collision subroutine by 40\% and reduces the total wall-clock time by 36\% in the multidimensional cylinder simulations.

\item
  Diffusion preservation by the same DeepONet--LJ scattering kernel: Direct held-out scattering-angle validation confirms that the trained DeepONet reproduces the exact LJ angular map with small wrapped-angle error. More importantly, two diffusion benchmarks show that the same learned scattering kernel preserves transport statistics beyond viscosity-controlled flows. At \(80\,\mathrm{K}\), the homogeneous MSD test recovers the Chapman--Enskog LJ self-diffusion coefficient with a mean deviation of approximately \(1.10\%\). In the Ohr-style tracer-slab benchmark at \(273\,\mathrm{K}\), the measured Fickian diffusion coefficient gives \(D/D_{\mathrm{LJ,CE}}=1.0103\), while the folded angular decomposition independently recovers the same transport coefficient. These results demonstrate that the learned LJ scattering angle preserves diffusion-relevant angular statistics in addition to supporting viscosity-consistent DSMC simulations.
\end{enumerate}

In summary, this work provides a scalable pathway to incorporate realistic molecular physics into engineering-scale DSMC simulations. The VED model supplies a local viscosity-consistent collision-rate closure, the DeepONet surrogate removes the dominant cost of LJ scattering-angle evaluation, and the diffusion benchmarks verify that the same learned angular kernel retains the transport information required for diffusion. Future work will consider extending the LJ--ML framework to gas mixtures, reacting flows, and complex three-dimensional configurations. For mixtures, the framework can be generalized by using species-dependent LJ parameters and cross-collision rules in the pair-selection stage, while the scattering surrogate can be trained on the corresponding cross-potential collision data. Extension to inelastic or reacting collisions will require augmenting the present elastic scattering kernel with state-resolved internal-energy and reaction-probability models, but the separation between local transport closure and pairwise scattering dynamics remains directly applicable.

\section*{Acknowledgments}\phantomsection\addcontentsline{toc}{section}{Acknowledgments}\label{sec:acknowledgments}

Stefan Stefanov was partially supported by the MES under the Grant No. D01-325/01.12.2023 for NCDSC -- part of the Bulgarian National Roadmap on RIs.

\section*{Declaration of interests}\phantomsection\addcontentsline{toc}{section}{Declaration of interests}\label{sec:declaration-of-interests}
The authors report no conflict of interest. % Bibliography embedded to avoid BibTeX/file-name issues on Editorial Manager

\end{document}